\documentclass[aps,prx,twocolumn,floatfix,showpacs,superscriptaddress]{revtex4-2}

\usepackage{amsfonts}
\usepackage{amsmath,bbm, amsfonts}
\usepackage{amssymb}
\usepackage{amsthm}
\usepackage{amsbsy}
\usepackage{bm}
\usepackage{comment}
\usepackage{dcolumn}
\usepackage{epstopdf}
\usepackage{graphicx}
\usepackage{graphbox}
\usepackage{enumerate}
\usepackage{latexsym}
\usepackage{multirow}
\usepackage{mathtools}
\usepackage{subfig}
\usepackage{tabularx}
\usepackage{times}
\usepackage{tikz}
\usepackage{cancel}
\usepackage{verbatim}
\usepackage{appendix}
\usepackage{ragged2e}
\usepackage[normalem]{ulem}

\makeatletter
\long\def\@makecaption#1#2{%
  \vskip\abovecaptionskip
  {\small
  \justifying
  \hyphenpenalty=50 
  \exhyphenpenalty=50
  \sloppy              
  \noindent #1. #2\par}
  \vskip\belowcaptionskip}
\makeatother

\usepackage[unicode=true,
bookmarks=true,bookmarksnumbered=false,bookmarksopen=false, breaklinks=false,pdfborder={0 0 1},backref=false,colorlinks=true]{hyperref}
\hypersetup{linkcolor=magenta, citecolor=blue,urlcolor=blue}   
\makeatletter

\newcommand{\ket}[1]{|#1\rangle}
\newcommand{\bra}[1]{\langle #1 |}

\newcommand{\Tr}[1]{\textnormal{Tr}\left[#1\right]}
\tikzset{every picture/.style={line width=0.75pt}} 

\begin{document}

\title{Temporal Complexity Hierarchies in Solvable Quantum Many-Body Dynamics}

\author{He-Ran Wang}
\thanks{These authors contributed equally to this work.}
\affiliation{Chen-Ning Yang Institute for Advanced Study, Tsinghua University, Beijing 100084,
People's Republic of China}
\author{Ilya Vilkoviskiy}
\thanks{These authors contributed equally to this work.}
\affiliation{Department of Physics, Princeton University, Princeton, New Jersey 08544, USA}
\author{Dmitry A. Abanin}
\email{dabanin@princeton.edu}
\affiliation{Department of Physics, Princeton University, Princeton, New Jersey 08544, USA}
\affiliation{\'Ecole Polytechnique F\'ed\'erale de Lausanne (EPFL), 1015 Lausanne, Switzerland}
\affiliation{Google Research, Brandschenkestrasse 150, 8002 Zürich, Switzerland}
\begin{abstract} 

The influence matrix (IM) provides a powerful framework for characterizing nonequilibrium quantum many-body dynamics by encoding multitime correlations into tensor-network states. Understanding how its computational complexity relates to underlying dynamics is crucial for both theoretical insight and practical utility yet remains largely unexplored despite a few case studies. Here, we address this question for a family of brickwork quantum circuits ranging from integrable to chaotic regimes. Using tools from geometric group theory, we identify three qualitatively distinct scalings of temporal entanglement entropy, establishing a hierarchy of computational resources required for accurate tensor-network representations of the IM for these models. We further analyze the memory structure of the IM and distinguish between classical and quantum temporal correlations. In particular, for certain examples, we identify effectively classical IMs that admit an efficient Monte Carlo algorithm for computing multitime correlations. In more generic settings without an explicit classical description of the IM, we introduce an operational measure of quantum memory with an experimental protocol, and discuss examples exhibiting long-time genuinely quantum correlations. Our results establish a new connection between quantum many-body dynamics and group theory, providing fresh insights into the complexity of the IM and its intricate connection to the physical characteristics of the dynamics.

\end{abstract}

\maketitle
\section{Introduction}

Describing and classifying nonequilibrium quantum matter has attracted considerable attention for several decades \cite{Rigol2008Thermalization,Polkovnikov2011Noneq,Kosloff2013quantum,Abanin2019Colloquium}. Recent advances in experimental techniques, particularly quantum simulation platforms, now enable probing quantum many-body dynamics with remarkable precision, thereby calling for parallel theoretical investigations to identify and analyze dynamical signatures. Despite this progress, efficiently characterizing nonequilibrium quantum states remains a formidable challenge, primarily due to the substantial computational complexity of simulating quantum dynamics in many-body Hilbert space \cite{Daley2022Practical}. This difficulty is slightly mitigated in low dimensions, where celebrated tensor-network techniques allow for representations of some physically relevant quantum states with the cost of computational resources that scale only polynomially with system size \cite{Vidal2004Efficient,White2004Real,Vidal2007Classical,Orus2008Infinite,Wolf2008Area,Haegeman2011Time,Haegeman2017Diag,Paeckel2019Time,Cirac2021Matrix}. However, the generally rapid growth of entanglement degrades the accuracy of such representations as time grows \cite{Calabrese2005Evlution,Kim2013Ballistic}, thus hindering both numerical simulations and analytical approaches to long time evolution along this direction.

Recently, a novel tensor-network method has been developed that encodes multitime correlations and dynamical characteristics into a temporal matrix-product state (MPS), referred to as the \textit{influence matrix} (IM) \cite{Lerose2021Influence, Sonner2021Influence,Ye2021Constructing, Cerezo2025Spatio}.
Following the idea of the Feynman-Vernon influence functional in the path integral formulation \cite{Feynman1963Theory}, the IM is defined as the tensor network obtained by performing trace operations on the full time-evolved system but a local subsystem. 
Therefore, this construction provides a direct approach to efficiently calculating spatially local quantities that characterize global nonequilibrium dynamics, such as the relaxation of local observables toward equilibrium values \cite{Berges2004Prethermalization, Manmana2007Strongly, Rigol2008Thermalization} and the asymptotic decay of autocorrelation functions \cite{Schreiber2015Observation, Schiulaz2019Thouless, Garratt2021Local}.
This method is particularly attainable and tractable in (1+1) D quantum-circuit dynamics, where it can be implemented as the transversal tensor-network contraction along the spatial direction \cite{Banuls2009Matrix, Hermes2012Tensor, Hastings2015Connecting, Perez2022Light}. Notably, closely related tensorial objects have also been introduced in the quantum information and open system communities as the quantum comb \cite{Chiribella2008Quantum, Chiribella2009Theoretical} and the process tensor \cite{Pollock2018non, Pollock2018Operational, Milz2021Tutorial, Keeling2025Process, Taranto2025Higher}, respectively.

By compressing the full time-evolved system into an MPS, the effective non-Markovian memory that impacts local dynamics is encoded in the inner bonds \cite{Strathearn2018Efficient, Luchnikov2019Simulation, Cygorek2025Understanding}. The efficacy of such an approach is supported by its moderate computational complexity for increasingly large times. 
This complexity is quantified by the scaling of the dimension of inner bonds and the \textit{temporal entanglement entropy} (TEE).
To be concrete, MPS representations of IM with polynomial (exponential) scaling of bond dimension with time fall into the polynomial (exponential) complexity class, and usually manifest as sublinear (linear) growth of TEE. Numerical results further suggest a slower growth than conventional spatial entanglement, even in regimes of exponential complexity \cite{Banuls2009Matrix, Lerose2021Influence}. 
However, it remains elusive how the computational complexity of the IM corresponds to the nature of the underlying dynamics, despite a few examples.

So far, only a handful of analytically tractable examples of the IM are known. Below, we present an overview of existing results in increasing order of ergodicity, i.e., from noninteracting to chaotic dynamics. First, for quantum impurity models in noninteracting fermionic environments, the polynomial scaling of IM computational complexity has been rigorously established in Refs.~\cite{Lerose2021Scaling, Thoenniss2023Efficient, Thoenniss2023non, Thoenniss2024Efficient}, with similar results for free bosons \cite{Strathearn2018Efficient, Jorgensen2019Exploiting, Cygorek2024Sublinear, Link2024Open, Vilkoviskiy2024Bound}.
For interacting integrable dynamics, exact solutions of IM are obtained only in a few isolated examples: (i) area-law entangled solutions for the ``rule 54'' cellular automaton \cite{Klobas2020Matrix, Klobas2020Space, Klobas2021Exact, Klobas2021ExactII} and (ii) at most logarithmic growth of TEE in the dispersionless Trotterized $XXZ$ Heisenberg model as demonstrated in Ref.~\cite{Giudice2022Temporal}. It was further conjectured that IMs in integrable models generally exhibit sublinear TEE scaling, also implied by the later work \cite{Carignano2024Temporal}.
Finally, in dual-unitary circuits \cite{Bertini2019Exact, Gopalakrishnan2019Unitary, Bertini2025Exactly}, which are typically maximally chaotic and scrambled according to various diagnostic tools \cite{Bertini2018Exact, Bertini2020Scrambling, Bertini2021Random, Zhou2022Maximal, Foligno2023Growth, Rampp2024Haydenpreskill, Chen2025Free, Claeys2025Fock, Fritzsch2025Eigenstate}, the scaling of TEE exhibits qualitatively distinct behavior depending on the choice of initial states: when prepared in so-called ``solvable initial states,'' the IM reduces to a simple product state representing a completely thermalizing bath and perfect dephaser (or depolarizer, depending on the circuit structure) acting on the subsystem \cite{Piroli2020Exact, Lerose2021Influence, Claeys2023From}, while for generic initial states, asymptotic linear growth of TEE has been rigorously shown in Ref.~\cite{Foligno2023Temporal}, which furthermore applies to Haar random brickwork circuits. Moreover, various generalizations of dual unitarity with compatible initial states lead to area-law TEE \cite{Bertini2023Exact, Yu2024Hierarchical, Wang2024Exact, Foligno2023Quantum, Liu2025Solvable, Foligno2025Non}. 

Overall, these previous results have established area-law, logarithmic, and linear scaling of TEE in separate examples, while the broader picture regarding connections between IM complexity and dynamical properties has been fragmented so far.

In this work, we make progress toward tackling the aforementioned question by studying a novel family of brickwork quantum circuits that provides a unified and flexible setting for exploring the complexity of nonequilibrium quantum dynamics, placing three regimes of TEE scalings within a common analytically tractable framework.
For this class of models, we provide analytical MPS representations of IMs for generic MPS initial states, where the inner bonds, surprisingly, are associated with a group algebra.
We establish a rigorous correspondence between bond dimension scaling and the \textit{growth function}, a concept from geometric group theory that quantifies how fast a group is generated from a set of generators, which are directly tied to the parametrization of local gates in our models.
Through this framework, we reproduce known solutions for integrable circuits and obtain new exact results for nonintegrable and chaotic cases, supported by numerical simulations. 
Hence, our construction leads to a systematic landscape in which the full hierarchy of classical simulation complexities of the IM -- belonging to constant, polynomial, or exponential computational resources -- can be realized and compared on equal footing. We refer to the corresponding models as class I, II, and III models, respectively. We complement our results with numerical evaluations of the TEE, finding agreement between its saturation, logarithmic growth, and linear growth, and the complexity landscape identified above.

The different scaling regimes of temporal complexity are interesting from complementary physical perspectives. Area-law TEE suggests that an extended many-body environment can be effectively compressed into a low-complexity dynamical description, as in finite-dimensional Markovian embeddings developed, for example, for free-fermionic environments \cite{Thoenniss2023Efficient, Thoenniss2023non, Thoenniss2024Efficient, Sonner2025Semigroup}. Our class I models show how such simple reduced dynamics can arise even in interacting Floquet systems with signatures of quantum chaos. Logarithmic TEE represents an intermediate regime with unbounded temporal memory but a polynomially tractable IM. Previously observed mainly in integrable systems, here we find such behavior in both integrable and nonintegrable examples. Finally, linear TEE represents the high temporal complexity expected in generic chaotic systems. In our class III models, although IM cannot be represented with polynomial resources, two-point functions remain accessible, whereas higher-order multitime correlations retain the complexity of the full temporal state. This provides an explicit mechanism underlying earlier observations that selected observables can be substantially easier to compute than suggested by the TEE~\cite{Banuls2009Matrix, Lerose2021Influence, Vilkoviskiy2025TEE}.

As a second main result, we explore another notion of complexity that focuses on the interplay between the classical and quantum nature of temporal correlations encoded in the IM.
More broadly, identifying the quantumness of a physical state, process, or correlation is a fundamental problem in quantum physics, with direct relevance to the characterization of quantum resources and their applications in quantum technologies.
While the full circuit exhibits genuine quantum dynamics, we identify special cases in which the IM can nevertheless be described exactly by purely classical memory. 
This enables efficient classical simulations for finite-order temporal correlation functions, even in regimes that are otherwise difficult to access with tensor-network approaches due to exponentially growing bond dimension. Specifically, we show that the full multitime statistics encoded in the IM can be reproduced by classical stochastic processes, thereby mapping quantum correlation functions onto averages of observables in random walks. According to the central limit theorem, simulations of the latter object require computational resources scaling only polynomially with the desired accuracy.

This is analogous to the situation for spatial states, where faithful matrix-product operator (MPO) representations of low-temperature Gibbs states may cost exponential computational resources \cite{Verstraete2008Matrix, Kliesch2014Matrix}, but several sampling algorithms can circumvent the difficulty and yield efficient estimates of observables \cite{White2009Minimally, Stoudenmire2010Minimally, Iwaki2024Sample}.
There is a general intuition developed in Ref.~\cite{Kudler2025Optimal} that quantum correlations are often suppressed on large scales compared to the classical ones. This avenue remains largely unexplored in the context of IMs \cite{Dowling2023Equilibration}, and our findings provide one of the first analytically tractable examples.

Beyond these special cases of ``classical'' IM, quantum memory is expected to be ubiquitous. 
To probe it, we propose an original quantum-teleportation protocol that extracts quantum memory hidden in the IM. 
While our protocol applies broadly to generic nonequilibrium dynamics, we find concrete realizations within our model that exhibit persistent quantum memory at long times.
These findings open an additional dimension of temporal complexity, enriching the framework for characterizing IMs and nonequilibrium quantum dynamics.

The rest of the paper is organized as follows. 
In Sec.~\ref{sec:MPO}, we introduce parametrizations of the local gates constituting the brickwork quantum circuits considered in this article. We then derive an exact and compact MPS representation of the IM by recasting it as a tensor-network operator acting on the initial state and expressing this operator as a matrix-product operator with inner bonds associated with a group algebra.
In Sec.~\ref{sec:growth} we demonstrate how the temporal scaling of bond dimension is related to the group-theoretic concept -- growth function -- which, in turn, gives rise to the full hierarchy of computational complexities with mathematical rigor. Each class is then illustrated by explicit realizations of quantum circuits, supported by numerical studies of TEE: we (i) reproduce results in Ref.~\cite{Giudice2022Temporal} by showing sublinear growth of TEE in an integrable model (Sec.~\ref{sec:Abelian}), (ii) demonstrate the same complexity class in a nonintegrable model in Sec.~\ref{sec:nonAbelian}, and (iii) show linear TEE growth for generic circuit realizations within our model (Sec.~\ref{sec:TEE_generic_group}). 

Next, we turn to the perspective of classical and quantum memory in Sec.~\ref{sec:Hiererchy_nMarkovian}. After clarifying our notions of memory from the viewpoint of open system and non-Markovianity (Sec.~\ref{sec:q_and_cl}), we explain the mapping from quantum dynamics to stochastic processes in the case of classical IM (Sec.~\ref{sec:stochastic}). We also perform Monte Carlo simulations to compute local observables, which we benchmark by comparing with tensor-network calculations.
In Sec.~\ref{sec:hybrid}, we characterize quantum IM, followed by introducing an operational measure of quantum memory computable from the IM in Sec.~\ref{sec:quantum_memory}.
We propose how this quantity could, in principle, be extracted from experiments. The validity of this approach is demonstrated by explicit examples that preserve long-lived quantum memory. Our conclusions and outlook are reported in Sec.~\ref{sec:conclusion}. Details of derivations and various generalizations are presented in the appendixes. 

\section{Setting and exact solutions}\label{sec:MPO}

\begin{figure*}[ht]
\hspace*{-1\textwidth}
\includegraphics[width=1\linewidth]{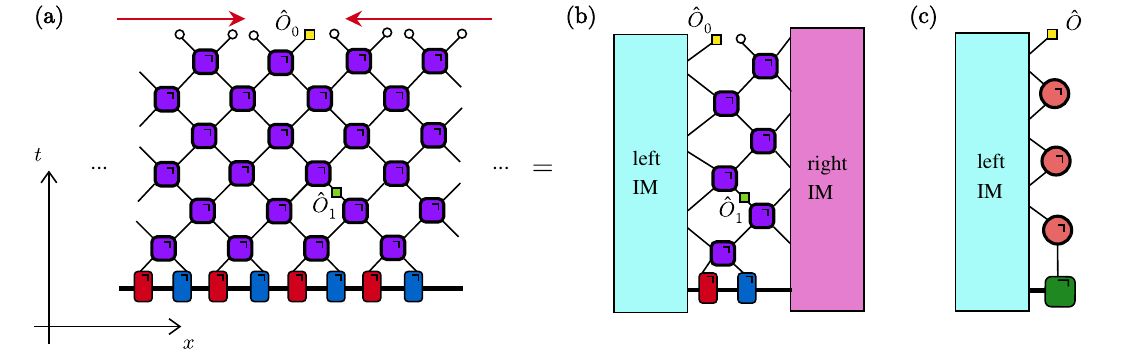} 
\caption{Graphical representations of the quantum-circuit dynamics and the influence matrix. Total number of time steps $T=3$. 
(a) Folded representation of the spatiotemporal correlation function in (1+1) D quantum circuits. The evolution of a matrix-product initial state [Eq.~\eqref{eq:MPSinitial}] is generated by two-site gates [Eq.~\eqref{eq:swap_control}] arranged in the brickwork architecture. Local operators $\hat{O}_0$ and $\hat{O}_1$ are inserted at different time steps. Red arrows indicate the direction of tensor-network contractions leading to influence matrices.
(b) Illustration of the left and right influence matrices. The partially traced components are compressed into MPS extending along the time direction.
(c) Representation of the observable expectation value in impurity dynamics. An impurity site (green tensor) is coupled to the quantum-circuit dynamics on the left. Red round tensors represent sequential quantum operations acting only on the impurity site.
}
\label{fig:setting}
\end{figure*}

In this work, we focus on quantum circuits on an infinite-size 1D chain of qudits, representing discrete-time quantum many-body dynamics with local interactions.
Each lattice site is labeled by an integer $x$ and associated with a $q$-dimensional Hilbert space:
\begin{equation}
    \mathcal{H}_x = \{\ket{a},a=0,1,\dots,q-1\}.
\end{equation}
One step of time evolution is generated by a global Floquet operator $\mathbb{U}$ composed of two-qudit unitaries $U$ arranged in a brickwork architecture:
\begin{align}
    &\mathbb{U}=\mathbb{U}_{\text{odd}}\mathbb{U}_{\text{even}},\nonumber\\
    & \mathbb{U}_{\text{odd(even)}}=\otimes_{x\in {\text{odd(even)}}} ~U_{x,x+1},
\end{align}
where $U_{x,x+1}$ acts locally on neighboring sites $x$ and $x+1$. 

We consider circuits with spacetime shift invariance, where the local gate is taken to be 
\begin{equation}\label{eq:swap_control}
    U=S(\sum_{a=0}^{q-1}u_a\otimes \ket{a}\bra{a}).
\end{equation}
Here, $u_a$ are single-site unitaries drawn from the unitary group $\mathrm{U}(q)$, and $S$ denotes the {\scriptsize${\mathrm{SWAP}}$} gate exchanging quantum states on two qudits. Although this construction belongs to the dual-unitary class \cite{Bertini2019Exact, Claeys2021Ergodic, Wang2025Hopf}, our exact solutions go beyond the scope of product-state IM, which requires special solvable initial states \cite{Piroli2020Exact, Lerose2021Influence, Foligno2023Quantum, Foligno2025Non}.

Since observables and correlation functions always involve unitaries paired with their Hermitian conjugates, we adopt the folded representation, in which the forward and backward branches of time evolution are folded together. Therein, the folded two-qudit unitary can be depicted in two equivalent ways:
\begin{equation}\label{eq:swap_control_tensor}
\tikzset{every picture/.style={line width=0.75pt}} 
\begin{tikzpicture}[x=0.75pt,y=0.75pt,yscale=-1,xscale=1]

\draw  [fill={rgb, 255:red, 144; green, 19; blue, 254 }  ,fill opacity=1 ][line width=1.5]  (123.5,118.96) .. controls (123.5,117.05) and (125.05,115.5) .. (126.96,115.5) -- (135.04,115.5) .. controls (136.95,115.5) and (138.5,117.05) .. (138.5,118.96) -- (138.5,127.04) .. controls (138.5,128.95) and (136.95,130.5) .. (135.04,130.5) -- (126.96,130.5) .. controls (125.05,130.5) and (123.5,128.95) .. (123.5,127.04) -- cycle ;
\draw  [fill={rgb, 255:red, 144; green, 19; blue, 254 }  ,fill opacity=1 ][line width=0.75]  (131,119) -- (135,119) -- (135,123) ;
\draw [fill={rgb, 255:red, 144; green, 19; blue, 254 }  ,fill opacity=1 ][line width=0.75]    (138.29,130.12) -- (148.5,140.5) ;
\draw [fill={rgb, 255:red, 144; green, 19; blue, 254 }  ,fill opacity=1 ][line width=0.75]    (113.5,105.5) -- (124,116) ;
\draw [fill={rgb, 255:red, 144; green, 19; blue, 254 }  ,fill opacity=1 ][line width=0.75]    (149.17,105.5) -- (138.28,116.44) ;
\draw [fill={rgb, 255:red, 144; green, 19; blue, 254 }  ,fill opacity=1 ][line width=0.75]    (124.17,129.17) -- (113.5,140.5) ;

\draw [fill={rgb, 255:red, 144; green, 19; blue, 254 }  ,fill opacity=1 ][line width=0.75]    (239.39,123.41) -- (251.48,139.39) ;
\draw [fill={rgb, 255:red, 144; green, 19; blue, 254 }  ,fill opacity=1 ][line width=0.75]    (218.39,102.91) -- (234.39,118.41) ;
\draw [fill={rgb, 255:red, 144; green, 19; blue, 254 }  ,fill opacity=1 ][line width=0.75]    (252.91,105.27) -- (217.5,140.5) ;
\draw    (228.52,128.8) -- (243.39,128.91) ;
\draw  [fill={rgb, 255:red, 255; green, 255; blue, 255 }  ,fill opacity=1 ] (223.52,128.26) .. controls (223.52,125.5) and (225.76,123.26) .. (228.52,123.26) .. controls (231.29,123.26) and (233.52,125.5) .. (233.52,128.26) .. controls (233.52,131.02) and (231.29,133.26) .. (228.52,133.26) .. controls (225.76,133.26) and (223.52,131.02) .. (223.52,128.26) -- cycle ;
\draw  [fill={rgb, 255:red, 0; green, 0; blue, 0 }  ,fill opacity=1 ] (241.89,128.91) .. controls (241.89,128.08) and (242.57,127.41) .. (243.39,127.41) .. controls (244.22,127.41) and (244.89,128.08) .. (244.89,128.91) .. controls (244.89,129.74) and (244.22,130.41) .. (243.39,130.41) .. controls (242.57,130.41) and (241.89,129.74) .. (241.89,128.91) -- cycle ;

\draw (282.17,112.9) node [anchor=north west][inner sep=0.75pt]    {$=U_{ab}^{cd}\left( U_{a'b'}^{c'd'}\right)^{*}$};
\draw (150,137.4) node [anchor=north west][inner sep=0.75pt]  [font=\scriptsize]  {$b,b'$};
\draw (92,96.4) node [anchor=north west][inner sep=0.75pt]  [font=\scriptsize]  {$c,c'$};
\draw (150,96.4) node [anchor=north west][inner sep=0.75pt]  [font=\scriptsize]  {$d,d'$};
\draw (92,137.4) node [anchor=north west][inner sep=0.75pt]  [font=\scriptsize]  {$a,a'$};
\draw (254,137.4) node [anchor=north west][inner sep=0.75pt]  [font=\scriptsize]  {$b,b'$};
\draw (196,96.4) node [anchor=north west][inner sep=0.75pt]  [font=\scriptsize]  {$c,c'$};
\draw (254,96.4) node [anchor=north west][inner sep=0.75pt]  [font=\scriptsize]  {$d,d'$};
\draw (196,137.4) node [anchor=north west][inner sep=0.75pt]  [font=\scriptsize]  {$a,a'$};
\draw (223.6,125.8) node [anchor=north west][inner sep=0.75pt]  [font=\tiny]  {$u$};
\draw (177.17,123.9) node [anchor=north west][inner sep=0.75pt]    {$=$};

\end{tikzpicture}.
\end{equation}
The first diagram has been widely applied to represent generic brickwork quantum circuits. In the second one, the three-leg tensor centered by the solid dot denotes the Kronecker delta that forces basis states on three legs to be the same.
Accordingly, the identity operator in the physical Hilbert space is represented as the hollow bullet:
\begin{equation}
\tikzset{every picture/.style={line width=0.75pt}} 
\begin{tikzpicture}[x=0.75pt,y=0.75pt,yscale=-1,xscale=1]

\draw [fill={rgb, 255:red, 144; green, 19; blue, 254 }  ,fill opacity=1 ][line width=0.75]    (114.9,83.5) -- (114.8,94.07) ;
\draw  [fill={rgb, 255:red, 255; green, 255; blue, 255 }  ,fill opacity=1 ] (112.1,80.7) .. controls (112.1,79.15) and (113.35,77.9) .. (114.9,77.9) .. controls (116.45,77.9) and (117.7,79.15) .. (117.7,80.7) .. controls (117.7,82.25) and (116.45,83.5) .. (114.9,83.5) .. controls (113.35,83.5) and (112.1,82.25) .. (112.1,80.7) -- cycle ;

\draw (105.4,96.28) node [anchor=north west][inner sep=0.75pt]  [font=\footnotesize]  {$a,a'$};
\draw (123.6,80.8) node [anchor=north west][inner sep=0.75pt]    {$\ =\delta _{a,a'}$};
\end{tikzpicture},
\end{equation}
which appears when performing partial trace operations. It follows that we can present the unitarity condition of $U$ as
\begin{equation}\label{eq:time_unitary}
\tikzset{every picture/.style={line width=0.75pt}} 
\begin{tikzpicture}[x=0.75pt,y=0.75pt,yscale=-1,xscale=1]

\draw  [fill={rgb, 255:red, 144; green, 19; blue, 254 }  ,fill opacity=1 ][line width=1.5]  (119.2,104.66) .. controls (119.2,102.75) and (120.75,101.2) .. (122.66,101.2) -- (130.74,101.2) .. controls (132.65,101.2) and (134.2,102.75) .. (134.2,104.66) -- (134.2,112.74) .. controls (134.2,114.65) and (132.65,116.2) .. (130.74,116.2) -- (122.66,116.2) .. controls (120.75,116.2) and (119.2,114.65) .. (119.2,112.74) -- cycle ;
\draw  [fill={rgb, 255:red, 144; green, 19; blue, 254 }  ,fill opacity=1 ][line width=0.75]  (126.7,104.7) -- (130.7,104.7) -- (130.7,108.7) ;
\draw [fill={rgb, 255:red, 144; green, 19; blue, 254 }  ,fill opacity=1 ][line width=0.75]    (133.99,115.82) -- (144.2,126.2) ;
\draw [fill={rgb, 255:red, 144; green, 19; blue, 254 }  ,fill opacity=1 ][line width=0.75]    (109.2,91.2) -- (119.7,101.7) ;
\draw [fill={rgb, 255:red, 144; green, 19; blue, 254 }  ,fill opacity=1 ][line width=0.75]    (144.87,91.2) -- (133.98,102.14) ;
\draw [fill={rgb, 255:red, 144; green, 19; blue, 254 }  ,fill opacity=1 ][line width=0.75]    (119.87,114.87) -- (109.2,126.2) ;

\draw  [fill={rgb, 255:red, 255; green, 255; blue, 255 }  ,fill opacity=1 ] (106.4,91.2) .. controls (106.4,89.65) and (107.65,88.4) .. (109.2,88.4) .. controls (110.75,88.4) and (112,89.65) .. (112,91.2) .. controls (112,92.75) and (110.75,94) .. (109.2,94) .. controls (107.65,94) and (106.4,92.75) .. (106.4,91.2) -- cycle ;
\draw    (174.4,94.18) -- (174.2,114.44) -- (174.2,125.4) ;
\draw   (171.6,91.29) .. controls (171.6,89.69) and (172.85,88.4) .. (174.4,88.4) .. controls (175.95,88.4) and (177.2,89.69) .. (177.2,91.29) .. controls (177.2,92.89) and (175.95,94.18) .. (174.4,94.18) .. controls (172.85,94.18) and (171.6,92.89) .. (171.6,91.29) -- cycle ;

\draw  [fill={rgb, 255:red, 255; green, 255; blue, 255 }  ,fill opacity=1 ] (142.07,91.2) .. controls (142.07,89.65) and (143.32,88.4) .. (144.87,88.4) .. controls (146.41,88.4) and (147.67,89.65) .. (147.67,91.2) .. controls (147.67,92.75) and (146.41,94) .. (144.87,94) .. controls (143.32,94) and (142.07,92.75) .. (142.07,91.2) -- cycle ;
\draw    (194.2,94.18) -- (194,114.44) -- (194,125.4) ;
\draw   (191.4,91.29) .. controls (191.4,89.69) and (192.65,88.4) .. (194.2,88.4) .. controls (195.75,88.4) and (197,89.69) .. (197,91.29) .. controls (197,92.89) and (195.75,94.18) .. (194.2,94.18) .. controls (192.65,94.18) and (191.4,92.89) .. (191.4,91.29) -- cycle ;

\draw (147.6,99.4) node [anchor=north west][inner sep=0.75pt]    {$=$};
\end{tikzpicture}.
\end{equation}

We consider pure normalized initial states $\ket{\Psi_\text{in}}$ in MPS form defined in terms of two $q\times D\times D$ tensors: $A^{a}_{jk},B^{b}_{jk}$, with $D$ the dimension of the auxiliary Hilbert space, such that
\begin{align}\label{eq:MPSinitial}
    \ket{\Psi_\text{in}} = &\sum_{\dots,a_{-1},a_0,a_1,\dots=0}^{q-1}\Tr{\cdots B^{a_{-1}}A^{a_0}B^{a_1}A^{a_2}\cdots}\nonumber\\
   & \ket{\dots,a_{-1},a_0,a_1,a_2,\dots}.
\end{align}
The folded three-leg tensors have graphical representations as
\begin{equation}\label{eq:MPS_tensor}
\tikzset{every picture/.style={line width=0.75pt}} 
\begin{tikzpicture}[x=0.75pt,y=0.75pt,yscale=-1,xscale=1]

\draw [line width=1.5]    (120,143.04) -- (149,143.04) ;
\draw  [fill={rgb, 255:red, 208; green, 2; blue, 27 }  ,fill opacity=1 ] (128.47,135.68) .. controls (128.47,134.42) and (129.49,133.4) .. (130.75,133.4) -- (137.59,133.4) .. controls (138.85,133.4) and (139.87,134.42) .. (139.87,135.68) -- (139.87,150.39) .. controls (139.87,151.65) and (138.85,152.67) .. (137.59,152.67) -- (130.75,152.67) .. controls (129.49,152.67) and (128.47,151.65) .. (128.47,150.39) -- cycle ;
\draw  [line width=0.75]  (133.3,135.63) -- (137.3,135.63) -- (137.3,139.63) ;

\draw [fill={rgb, 255:red, 144; green, 19; blue, 254 }  ,fill opacity=1 ][line width=0.75]    (134.2,124) -- (134.25,132.9) ;
\draw [line width=1.5]    (302,143.04) -- (331,143.04) ;
\draw  [fill={rgb, 255:red, 2; green, 100; blue, 200 }  ,fill opacity=1 ] (310.47,135.68) .. controls (310.47,134.42) and (311.49,133.4) .. (312.75,133.4) -- (319.59,133.4) .. controls (320.85,133.4) and (321.87,134.42) .. (321.87,135.68) -- (321.87,150.39) .. controls (321.87,151.65) and (320.85,152.67) .. (319.59,152.67) -- (312.75,152.67) .. controls (311.49,152.67) and (310.47,151.65) .. (310.47,150.39) -- cycle ;
\draw  [fill={rgb, 255:red, 2; green, 100; blue, 200 }  ,fill opacity=1 ][line width=0.75]  (315.3,135.63) -- (319.3,135.63) -- (319.3,139.63) ;

\draw [fill={rgb, 255:red, 144; green, 19; blue, 254 }  ,fill opacity=1 ][line width=0.75]    (316.2,124) -- (316.25,132.9) ;

\draw (124.8,112.21) node [anchor=north west][inner sep=0.75pt]  [font=\scriptsize]  {$a,a'$};
\draw (101.4,135.61) node [anchor=north west][inner sep=0.75pt]  [font=\scriptsize]  {$j,j'$};
\draw (150.4,136.61) node [anchor=north west][inner sep=0.75pt]  [font=\scriptsize]  {$k,k'$};
\draw (175.57,129.7) node [anchor=north west][inner sep=0.75pt]    {$=A_{jk}^{a} (A{_{j'k'}^{a'}})^{*}$};
\draw (306.8,112.21) node [anchor=north west][inner sep=0.75pt]  [font=\scriptsize]  {$a,a'$};
\draw (283.4,135.61) node [anchor=north west][inner sep=0.75pt]  [font=\scriptsize]  {$j,j'$};
\draw (332.4,136.61) node [anchor=north west][inner sep=0.75pt]  [font=\scriptsize]  {$k,k'$};
\draw (357.57,129.7) node [anchor=north west][inner sep=0.75pt]    {$=B_{jk}^{a} (B{_{j'k'}^{a'}})^{*} $};
\end{tikzpicture}
\end{equation}
We impose two-site shift invariance for notational convenience. However, the exact solvability of our model follows directly from the structure of unitaries $U$ and relies neither on the form of matrix-product initial states nor on the shift invariance.

To motivate the introduction of the IM, Fig.~\ref{fig:setting}(a) shows the diagrammatic representation of a spatiotemporal correlation function, $\bra{\Psi_\text{in}}\hat{O}_{x'}(t')\hat{O}_x(t)\ket{\Psi_\text{in}}$, where time-dependent operators are defined in the Heisenberg picture. In this illustrated example, $(x,t)=(1,1)$ and $(x',t')=(0,3)$, with total number of time steps $T=3$. One can isolate the middle subsystem on $x=0$ and $1$ sites, sandwiched between multitime tensors from both sides. These tensors are defined as the left and right IMs [Fig.~\ref{fig:setting}(b)], respectively. Notice that the role of IM is not limited to two-point correlations but extends to multitime correlations of operators with common support in a spatially local region.

The IM framework also naturally applies to quantum-impurity dynamics. As illustrated in Fig.~\ref{fig:setting}(c), an impurity site is coupled to a semi-infinite qudit chain and evolves jointly under quantum-circuit dynamics. The impurity can be controlled and manipulated through quantum operations—such as unitaries or channels—interspersed between the joint evolution at each time step, with final measurements of impurity observables $\hat{O}$.
On the other hand, the impurity together with local operations serves as a flexible probe of nonequilibrium dynamics \cite{Bylander2011Noise, Paladino2014Noise, Pollock2018non, Odonovan2025Diagnosing}, with the IM providing an efficient characterization.
For clarity of presentation, we mainly focus on the setting of impurity dynamics depicted in Fig.~\ref{fig:setting}(c), while most of our results for the left IM apply equally to the right one.

The (left) IM can be decomposed into a triangular tensor $\mathcal{M}$ with two boundaries of open legs, acting on the (left) initial state:
\begin{equation}
    \ket{\text{IM}} = \mathcal{M}(\ket{\Psi_\text{in}^L}\otimes\ket{\Psi_\text{in}^L}^*),
\end{equation}
depicted as follows:
\begin{equation}\label{eq:leftIM}
\tikzset{every picture/.style={line width=0.75pt}} 
\begin{tikzpicture}[x=0.75pt,y=0.75pt,yscale=-1,xscale=1]

\draw  [fill={rgb, 255:red, 155; green, 155; blue, 155 }  ,fill opacity=0.82 ] (211,42) .. controls (211,24.33) and (225.33,10) .. (243,10) -- (349,10) .. controls (366.67,10) and (381,24.33) .. (381,42) -- (381,138) .. controls (381,155.67) and (366.67,170) .. (349,170) -- (243,170) .. controls (225.33,170) and (211,155.67) .. (211,138) -- cycle ;
\draw [fill={rgb, 255:red, 144; green, 19; blue, 254 }  ,fill opacity=1 ][line width=0.75]    (148.39,145.78) -- (165.6,158.16) ;
\draw [fill={rgb, 255:red, 144; green, 19; blue, 254 }  ,fill opacity=1 ][line width=0.75]    (149.43,97) -- (162.77,107.87) ;
\draw [fill={rgb, 255:red, 144; green, 19; blue, 254 }  ,fill opacity=1 ][line width=0.75]    (165.43,72.87) -- (149.54,86.81) ;
\draw [fill={rgb, 255:red, 144; green, 19; blue, 254 }  ,fill opacity=1 ][line width=0.75]    (149.56,47.48) -- (165.77,60.87) ;
\draw [fill={rgb, 255:red, 144; green, 19; blue, 254 }  ,fill opacity=1 ][line width=0.75]    (166.43,25.87) -- (149.43,39) ;
\draw [fill={rgb, 255:red, 144; green, 19; blue, 254 }  ,fill opacity=1 ][line width=0.75]    (163.43,120.47) -- (149.77,134.8) ;
\draw  [color={rgb, 255:red, 0; green, 0; blue, 0 }  ,draw opacity=1 ][fill={rgb, 255:red, 35; green, 245; blue, 240 }  ,fill opacity=0.4 ] (102.43,22) -- (149.43,22) -- (149.43,200) -- (102.43,200) -- cycle ;
\draw [line width=2.25]    (149,182) -- (168,182) ;
\draw  [fill={rgb, 255:red, 144; green, 19; blue, 254 }  ,fill opacity=1 ][line width=1.5]  (336.33,155.06) .. controls (336.33,153.15) and (337.88,151.6) .. (339.79,151.6) -- (347.87,151.6) .. controls (349.78,151.6) and (351.33,153.15) .. (351.33,155.06) -- (351.33,163.14) .. controls (351.33,165.05) and (349.78,166.6) .. (347.87,166.6) -- (339.79,166.6) .. controls (337.88,166.6) and (336.33,165.05) .. (336.33,163.14) -- cycle ;
\draw  [fill={rgb, 255:red, 144; green, 19; blue, 254 }  ,fill opacity=1 ][line width=0.75]  (343.83,155.1) -- (347.83,155.1) -- (347.83,159.1) ;
\draw [fill={rgb, 255:red, 144; green, 19; blue, 254 }  ,fill opacity=1 ][line width=0.75]    (351.13,166.22) -- (361.33,176.6) ;
\draw [fill={rgb, 255:red, 144; green, 19; blue, 254 }  ,fill opacity=1 ][line width=0.75]    (326.33,141.6) -- (336.83,152.1) ;
\draw [fill={rgb, 255:red, 144; green, 19; blue, 254 }  ,fill opacity=1 ][line width=0.75]    (362,141.6) -- (351.11,152.54) ;
\draw [fill={rgb, 255:red, 144; green, 19; blue, 254 }  ,fill opacity=1 ][line width=0.75]    (337,165.27) -- (326.33,176.6) ;
\draw  [fill={rgb, 255:red, 144; green, 19; blue, 254 }  ,fill opacity=1 ][line width=1.5]  (287.33,155.06) .. controls (287.33,153.15) and (288.88,151.6) .. (290.79,151.6) -- (298.87,151.6) .. controls (300.78,151.6) and (302.33,153.15) .. (302.33,155.06) -- (302.33,163.14) .. controls (302.33,165.05) and (300.78,166.6) .. (298.87,166.6) -- (290.79,166.6) .. controls (288.88,166.6) and (287.33,165.05) .. (287.33,163.14) -- cycle ;
\draw  [fill={rgb, 255:red, 144; green, 19; blue, 254 }  ,fill opacity=1 ][line width=0.75]  (294.83,155.1) -- (298.83,155.1) -- (298.83,159.1) ;
\draw [fill={rgb, 255:red, 144; green, 19; blue, 254 }  ,fill opacity=1 ][line width=0.75]    (302.13,166.22) -- (312.33,176.6) ;
\draw [fill={rgb, 255:red, 144; green, 19; blue, 254 }  ,fill opacity=1 ][line width=0.75]    (277.33,141.6) -- (287.83,152.1) ;
\draw [fill={rgb, 255:red, 144; green, 19; blue, 254 }  ,fill opacity=1 ][line width=0.75]    (313,141.6) -- (302.11,152.54) ;
\draw [fill={rgb, 255:red, 144; green, 19; blue, 254 }  ,fill opacity=1 ][line width=0.75]    (288,165.27) -- (277.33,176.6) ;

\draw  [fill={rgb, 255:red, 144; green, 19; blue, 254 }  ,fill opacity=1 ][line width=1.5]  (312.33,131.06) .. controls (312.33,129.15) and (313.88,127.6) .. (315.79,127.6) -- (323.87,127.6) .. controls (325.78,127.6) and (327.33,129.15) .. (327.33,131.06) -- (327.33,139.14) .. controls (327.33,141.05) and (325.78,142.6) .. (323.87,142.6) -- (315.79,142.6) .. controls (313.88,142.6) and (312.33,141.05) .. (312.33,139.14) -- cycle ;
\draw  [fill={rgb, 255:red, 144; green, 19; blue, 254 }  ,fill opacity=1 ][line width=0.75]  (319.83,131.1) -- (323.83,131.1) -- (323.83,135.1) ;
\draw [fill={rgb, 255:red, 144; green, 19; blue, 254 }  ,fill opacity=1 ][line width=0.75]    (327.13,142.22) -- (337.33,152.6) ;
\draw [fill={rgb, 255:red, 144; green, 19; blue, 254 }  ,fill opacity=1 ][line width=0.75]    (302.33,117.6) -- (312.83,128.1) ;
\draw [fill={rgb, 255:red, 144; green, 19; blue, 254 }  ,fill opacity=1 ][line width=0.75]    (338,117.6) -- (327.11,128.54) ;
\draw [fill={rgb, 255:red, 144; green, 19; blue, 254 }  ,fill opacity=1 ][line width=0.75]    (313,141.27) -- (302.33,152.6) ;

\draw  [fill={rgb, 255:red, 144; green, 19; blue, 254 }  ,fill opacity=1 ][line width=1.5]  (337.33,107.06) .. controls (337.33,105.15) and (338.88,103.6) .. (340.79,103.6) -- (348.87,103.6) .. controls (350.78,103.6) and (352.33,105.15) .. (352.33,107.06) -- (352.33,115.14) .. controls (352.33,117.05) and (350.78,118.6) .. (348.87,118.6) -- (340.79,118.6) .. controls (338.88,118.6) and (337.33,117.05) .. (337.33,115.14) -- cycle ;
\draw  [fill={rgb, 255:red, 144; green, 19; blue, 254 }  ,fill opacity=1 ][line width=0.75]  (344.83,107.1) -- (348.83,107.1) -- (348.83,111.1) ;
\draw [fill={rgb, 255:red, 144; green, 19; blue, 254 }  ,fill opacity=1 ][line width=0.75]    (352.13,118.22) -- (362.33,128.6) ;
\draw [fill={rgb, 255:red, 144; green, 19; blue, 254 }  ,fill opacity=1 ][line width=0.75]    (327.33,93.6) -- (337.83,104.1) ;
\draw [fill={rgb, 255:red, 144; green, 19; blue, 254 }  ,fill opacity=1 ][line width=0.75]    (363,93.6) -- (352.11,104.54) ;
\draw [fill={rgb, 255:red, 144; green, 19; blue, 254 }  ,fill opacity=1 ][line width=0.75]    (338,117.27) -- (327.33,128.6) ;

\draw  [fill={rgb, 255:red, 144; green, 19; blue, 254 }  ,fill opacity=1 ][line width=1.5]  (288.33,107.06) .. controls (288.33,105.15) and (289.88,103.6) .. (291.79,103.6) -- (299.87,103.6) .. controls (301.78,103.6) and (303.33,105.15) .. (303.33,107.06) -- (303.33,115.14) .. controls (303.33,117.05) and (301.78,118.6) .. (299.87,118.6) -- (291.79,118.6) .. controls (289.88,118.6) and (288.33,117.05) .. (288.33,115.14) -- cycle ;
\draw  [fill={rgb, 255:red, 144; green, 19; blue, 254 }  ,fill opacity=1 ][line width=0.75]  (295.83,107.1) -- (299.83,107.1) -- (299.83,111.1) ;
\draw [fill={rgb, 255:red, 144; green, 19; blue, 254 }  ,fill opacity=1 ][line width=0.75]    (303.13,118.22) -- (313.33,128.6) ;
\draw [fill={rgb, 255:red, 144; green, 19; blue, 254 }  ,fill opacity=1 ][line width=0.75]    (278.33,93.6) -- (288.83,104.1) ;
\draw [fill={rgb, 255:red, 144; green, 19; blue, 254 }  ,fill opacity=1 ][line width=0.75]    (314,93.6) -- (303.11,104.54) ;
\draw [fill={rgb, 255:red, 144; green, 19; blue, 254 }  ,fill opacity=1 ][line width=0.75]    (289,117.27) -- (278.33,128.6) ;

\draw  [fill={rgb, 255:red, 144; green, 19; blue, 254 }  ,fill opacity=1 ][line width=1.5]  (313.33,83.06) .. controls (313.33,81.15) and (314.88,79.6) .. (316.79,79.6) -- (324.87,79.6) .. controls (326.78,79.6) and (328.33,81.15) .. (328.33,83.06) -- (328.33,91.14) .. controls (328.33,93.05) and (326.78,94.6) .. (324.87,94.6) -- (316.79,94.6) .. controls (314.88,94.6) and (313.33,93.05) .. (313.33,91.14) -- cycle ;
\draw  [fill={rgb, 255:red, 144; green, 19; blue, 254 }  ,fill opacity=1 ][line width=0.75]  (320.83,83.1) -- (324.83,83.1) -- (324.83,87.1) ;
\draw [fill={rgb, 255:red, 144; green, 19; blue, 254 }  ,fill opacity=1 ][line width=0.75]    (328.13,94.22) -- (338.33,104.6) ;
\draw [fill={rgb, 255:red, 144; green, 19; blue, 254 }  ,fill opacity=1 ][line width=0.75]    (303.33,69.6) -- (313.83,80.1) ;
\draw [fill={rgb, 255:red, 144; green, 19; blue, 254 }  ,fill opacity=1 ][line width=0.75]    (339,69.6) -- (328.11,80.54) ;
\draw [fill={rgb, 255:red, 144; green, 19; blue, 254 }  ,fill opacity=1 ][line width=0.75]    (314,93.27) -- (303.33,104.6) ;

\draw  [fill={rgb, 255:red, 144; green, 19; blue, 254 }  ,fill opacity=1 ][line width=1.5]  (238.33,155.46) .. controls (238.33,153.55) and (239.88,152) .. (241.79,152) -- (249.87,152) .. controls (251.78,152) and (253.33,153.55) .. (253.33,155.46) -- (253.33,163.54) .. controls (253.33,165.45) and (251.78,167) .. (249.87,167) -- (241.79,167) .. controls (239.88,167) and (238.33,165.45) .. (238.33,163.54) -- cycle ;
\draw  [fill={rgb, 255:red, 144; green, 19; blue, 254 }  ,fill opacity=1 ][line width=0.75]  (245.83,155.5) -- (249.83,155.5) -- (249.83,159.5) ;
\draw [fill={rgb, 255:red, 144; green, 19; blue, 254 }  ,fill opacity=1 ][line width=0.75]    (253.13,166.62) -- (263.33,177) ;
\draw [fill={rgb, 255:red, 144; green, 19; blue, 254 }  ,fill opacity=1 ][line width=0.75]    (228.33,142) -- (238.83,152.5) ;
\draw [fill={rgb, 255:red, 144; green, 19; blue, 254 }  ,fill opacity=1 ][line width=0.75]    (264,142) -- (253.11,152.94) ;
\draw [fill={rgb, 255:red, 144; green, 19; blue, 254 }  ,fill opacity=1 ][line width=0.75]    (239,165.67) -- (228.33,177) ;

\draw  [fill={rgb, 255:red, 144; green, 19; blue, 254 }  ,fill opacity=1 ][line width=1.5]  (263.33,131.46) .. controls (263.33,129.55) and (264.88,128) .. (266.79,128) -- (274.87,128) .. controls (276.78,128) and (278.33,129.55) .. (278.33,131.46) -- (278.33,139.54) .. controls (278.33,141.45) and (276.78,143) .. (274.87,143) -- (266.79,143) .. controls (264.88,143) and (263.33,141.45) .. (263.33,139.54) -- cycle ;
\draw  [fill={rgb, 255:red, 144; green, 19; blue, 254 }  ,fill opacity=1 ][line width=0.75]  (270.83,131.5) -- (274.83,131.5) -- (274.83,135.5) ;
\draw [fill={rgb, 255:red, 144; green, 19; blue, 254 }  ,fill opacity=1 ][line width=0.75]    (278.13,142.62) -- (288.33,153) ;
\draw [fill={rgb, 255:red, 144; green, 19; blue, 254 }  ,fill opacity=1 ][line width=0.75]    (253.33,118) -- (263.83,128.5) ;
\draw [fill={rgb, 255:red, 144; green, 19; blue, 254 }  ,fill opacity=1 ][line width=0.75]    (289,118) -- (278.11,128.94) ;
\draw [fill={rgb, 255:red, 144; green, 19; blue, 254 }  ,fill opacity=1 ][line width=0.75]    (264,141.67) -- (253.33,153) ;

\draw [fill={rgb, 255:red, 144; green, 19; blue, 254 }  ,fill opacity=1 ][line width=0.75]    (279.63,94.62) -- (289.83,105) ;
\draw  [fill={rgb, 255:red, 255; green, 255; blue, 255 }  ,fill opacity=1 ] (274.53,91.8) .. controls (274.53,90.25) and (275.79,89) .. (277.33,89) .. controls (278.88,89) and (280.13,90.25) .. (280.13,91.8) .. controls (280.13,93.35) and (278.88,94.6) .. (277.33,94.6) .. controls (275.79,94.6) and (274.53,93.35) .. (274.53,91.8) -- cycle ;
\draw  [fill={rgb, 255:red, 255; green, 255; blue, 255 }  ,fill opacity=1 ] (300.53,69.6) .. controls (300.53,68.05) and (301.79,66.8) .. (303.33,66.8) .. controls (304.88,66.8) and (306.13,68.05) .. (306.13,69.6) .. controls (306.13,71.15) and (304.88,72.4) .. (303.33,72.4) .. controls (301.79,72.4) and (300.53,71.15) .. (300.53,69.6) -- cycle ;
\draw  [fill={rgb, 255:red, 144; green, 19; blue, 254 }  ,fill opacity=1 ][line width=1.5]  (338,59.23) .. controls (338,57.32) and (339.55,55.77) .. (341.46,55.77) -- (349.54,55.77) .. controls (351.45,55.77) and (353,57.32) .. (353,59.23) -- (353,67.31) .. controls (353,69.22) and (351.45,70.77) .. (349.54,70.77) -- (341.46,70.77) .. controls (339.55,70.77) and (338,69.22) .. (338,67.31) -- cycle ;
\draw  [fill={rgb, 255:red, 144; green, 19; blue, 254 }  ,fill opacity=1 ][line width=0.75]  (345.5,59.27) -- (349.5,59.27) -- (349.5,63.27) ;
\draw [fill={rgb, 255:red, 144; green, 19; blue, 254 }  ,fill opacity=1 ][line width=0.75]    (352.79,70.38) -- (363,80.77) ;
\draw [fill={rgb, 255:red, 144; green, 19; blue, 254 }  ,fill opacity=1 ][line width=0.75]    (328,45.77) -- (338.5,56.27) ;
\draw [fill={rgb, 255:red, 144; green, 19; blue, 254 }  ,fill opacity=1 ][line width=0.75]    (363.67,45.77) -- (352.78,56.71) ;
\draw  [fill={rgb, 255:red, 144; green, 19; blue, 254 }  ,fill opacity=1 ][line width=1.5]  (361.33,131.23) .. controls (361.33,129.32) and (362.88,127.77) .. (364.79,127.77) -- (372.87,127.77) .. controls (374.78,127.77) and (376.33,129.32) .. (376.33,131.23) -- (376.33,139.31) .. controls (376.33,141.22) and (374.78,142.77) .. (372.87,142.77) -- (364.79,142.77) .. controls (362.88,142.77) and (361.33,141.22) .. (361.33,139.31) -- cycle ;
\draw  [fill={rgb, 255:red, 144; green, 19; blue, 254 }  ,fill opacity=1 ][line width=0.75]  (368.83,131.27) -- (372.83,131.27) -- (372.83,135.27) ;
\draw [fill={rgb, 255:red, 144; green, 19; blue, 254 }  ,fill opacity=1 ][line width=0.75]    (376.13,142.38) -- (386.33,152.77) ;
\draw [fill={rgb, 255:red, 144; green, 19; blue, 254 }  ,fill opacity=1 ][line width=0.75]    (387,117.77) -- (376.11,128.71) ;
\draw  [fill={rgb, 255:red, 144; green, 19; blue, 254 }  ,fill opacity=1 ][line width=1.5]  (362.33,83.23) .. controls (362.33,81.32) and (363.88,79.77) .. (365.79,79.77) -- (373.87,79.77) .. controls (375.78,79.77) and (377.33,81.32) .. (377.33,83.23) -- (377.33,91.31) .. controls (377.33,93.22) and (375.78,94.77) .. (373.87,94.77) -- (365.79,94.77) .. controls (363.88,94.77) and (362.33,93.22) .. (362.33,91.31) -- cycle ;
\draw  [fill={rgb, 255:red, 144; green, 19; blue, 254 }  ,fill opacity=1 ][line width=0.75]  (369.83,83.27) -- (373.83,83.27) -- (373.83,87.27) ;
\draw [fill={rgb, 255:red, 144; green, 19; blue, 254 }  ,fill opacity=1 ][line width=0.75]    (377.13,94.38) -- (387.33,104.77) ;
\draw [fill={rgb, 255:red, 144; green, 19; blue, 254 }  ,fill opacity=1 ][line width=0.75]    (388,69.77) -- (377.11,80.71) ;
\draw  [fill={rgb, 255:red, 144; green, 19; blue, 254 }  ,fill opacity=1 ][line width=1.5]  (362.33,36.23) .. controls (362.33,34.32) and (363.88,32.77) .. (365.79,32.77) -- (373.87,32.77) .. controls (375.78,32.77) and (377.33,34.32) .. (377.33,36.23) -- (377.33,44.31) .. controls (377.33,46.22) and (375.78,47.77) .. (373.87,47.77) -- (365.79,47.77) .. controls (363.88,47.77) and (362.33,46.22) .. (362.33,44.31) -- cycle ;
\draw  [fill={rgb, 255:red, 144; green, 19; blue, 254 }  ,fill opacity=1 ][line width=0.75]  (369.83,36.27) -- (373.83,36.27) -- (373.83,40.27) ;
\draw [fill={rgb, 255:red, 144; green, 19; blue, 254 }  ,fill opacity=1 ][line width=0.75]    (377.13,47.38) -- (387.33,57.77) ;
\draw [fill={rgb, 255:red, 144; green, 19; blue, 254 }  ,fill opacity=1 ][line width=0.75]    (352.33,22.77) -- (362.83,33.27) ;
\draw [fill={rgb, 255:red, 144; green, 19; blue, 254 }  ,fill opacity=1 ][line width=0.75]    (388,22.77) -- (377.11,33.71) ;
\draw  [fill={rgb, 255:red, 255; green, 255; blue, 255 }  ,fill opacity=1 ] (349.53,21.97) .. controls (349.53,20.42) and (350.79,19.17) .. (352.33,19.17) .. controls (353.88,19.17) and (355.13,20.42) .. (355.13,21.97) .. controls (355.13,23.51) and (353.88,24.77) .. (352.33,24.77) .. controls (350.79,24.77) and (349.53,23.51) .. (349.53,21.97) -- cycle ;
\draw  [fill={rgb, 255:red, 255; green, 255; blue, 255 }  ,fill opacity=1 ] (326,45.77) .. controls (326,44.22) and (327.25,42.97) .. (328.8,42.97) .. controls (330.35,42.97) and (331.6,44.22) .. (331.6,45.77) .. controls (331.6,47.31) and (330.35,48.57) .. (328.8,48.57) .. controls (327.25,48.57) and (326,47.31) .. (326,45.77) -- cycle ;
\draw  [fill={rgb, 255:red, 255; green, 255; blue, 255 }  ,fill opacity=1 ] (250.53,118) .. controls (250.53,116.45) and (251.79,115.2) .. (253.33,115.2) .. controls (254.88,115.2) and (256.13,116.45) .. (256.13,118) .. controls (256.13,119.55) and (254.88,120.8) .. (253.33,120.8) .. controls (251.79,120.8) and (250.53,119.55) .. (250.53,118) -- cycle ;
\draw  [fill={rgb, 255:red, 255; green, 255; blue, 255 }  ,fill opacity=1 ] (225.53,142) .. controls (225.53,140.45) and (226.79,139.2) .. (228.33,139.2) .. controls (229.88,139.2) and (231.13,140.45) .. (231.13,142) .. controls (231.13,143.55) and (229.88,144.8) .. (228.33,144.8) .. controls (226.79,144.8) and (225.53,143.55) .. (225.53,142) -- cycle ;
\draw [line width=2.25]    (219.43,183.63) -- (391,183.63) ;
\draw  [fill={rgb, 255:red, 208; green, 2; blue, 27 }  ,fill opacity=1 ] (277.43,175.01) .. controls (277.43,173.75) and (278.45,172.73) .. (279.71,172.73) -- (286.55,172.73) .. controls (287.81,172.73) and (288.83,173.75) .. (288.83,175.01) -- (288.83,189.72) .. controls (288.83,190.98) and (287.81,192) .. (286.55,192) -- (279.71,192) .. controls (278.45,192) and (277.43,190.98) .. (277.43,189.72) -- cycle ;
\draw  [fill={rgb, 255:red, 208; green, 2; blue, 27 }  ,fill opacity=1 ][line width=0.75]  (282.27,174.96) -- (286.27,174.96) -- (286.27,178.96) ;

\draw  [fill={rgb, 255:red, 2; green, 100; blue, 200 }  ,fill opacity=1 ] (351.9,175.01) .. controls (351.9,173.75) and (352.92,172.73) .. (354.18,172.73) -- (361.02,172.73) .. controls (362.28,172.73) and (363.3,173.75) .. (363.3,175.01) -- (363.3,189.72) .. controls (363.3,190.98) and (362.28,192) .. (361.02,192) -- (354.18,192) .. controls (352.92,192) and (351.9,190.98) .. (351.9,189.72) -- cycle ;
\draw  [fill={rgb, 255:red, 2; green, 100; blue, 200 }  ,fill opacity=1 ][line width=0.75]  (356.73,174.96) -- (360.73,174.96) -- (360.73,178.96) ;

\draw  [fill={rgb, 255:red, 208; green, 2; blue, 27 }  ,fill opacity=1 ] (325.3,175.01) .. controls (325.3,173.75) and (326.32,172.73) .. (327.58,172.73) -- (334.42,172.73) .. controls (335.68,172.73) and (336.7,173.75) .. (336.7,175.01) -- (336.7,189.72) .. controls (336.7,190.98) and (335.68,192) .. (334.42,192) -- (327.58,192) .. controls (326.32,192) and (325.3,190.98) .. (325.3,189.72) -- cycle ;
\draw  [fill={rgb, 255:red, 208; green, 2; blue, 27 }  ,fill opacity=1 ][line width=0.75]  (330.13,174.96) -- (334.13,174.96) -- (334.13,178.96) ;

\draw  [fill={rgb, 255:red, 2; green, 100; blue, 200 }  ,fill opacity=1 ] (302.3,175.01) .. controls (302.3,173.75) and (303.32,172.73) .. (304.58,172.73) -- (311.42,172.73) .. controls (312.68,172.73) and (313.7,173.75) .. (313.7,175.01) -- (313.7,189.72) .. controls (313.7,190.98) and (312.68,192) .. (311.42,192) -- (304.58,192) .. controls (303.32,192) and (302.3,190.98) .. (302.3,189.72) -- cycle ;
\draw  [fill={rgb, 255:red, 2; green, 100; blue, 200 }  ,fill opacity=1 ][line width=0.75]  (307.13,174.96) -- (311.13,174.96) -- (311.13,178.96) ;

\draw  [fill={rgb, 255:red, 208; green, 2; blue, 27 }  ,fill opacity=1 ] (227,175.01) .. controls (227,173.75) and (228.02,172.73) .. (229.28,172.73) -- (236.12,172.73) .. controls (237.38,172.73) and (238.4,173.75) .. (238.4,175.01) -- (238.4,189.72) .. controls (238.4,190.98) and (237.38,192) .. (236.12,192) -- (229.28,192) .. controls (228.02,192) and (227,190.98) .. (227,189.72) -- cycle ;
\draw  [fill={rgb, 255:red, 208; green, 2; blue, 27 }  ,fill opacity=1 ][line width=0.75]  (231.83,174.96) -- (235.83,174.96) -- (235.83,178.96) ;

\draw  [fill={rgb, 255:red, 2; green, 100; blue, 200 }  ,fill opacity=1 ] (255,175.01) .. controls (255,173.75) and (256.02,172.73) .. (257.28,172.73) -- (264.12,172.73) .. controls (265.38,172.73) and (266.4,173.75) .. (266.4,175.01) -- (266.4,189.72) .. controls (266.4,190.98) and (265.38,192) .. (264.12,192) -- (257.28,192) .. controls (256.02,192) and (255,190.98) .. (255,189.72) -- cycle ;
\draw  [fill={rgb, 255:red, 2; green, 100; blue, 200 }  ,fill opacity=1 ][line width=0.75]  (259.83,174.96) -- (263.83,174.96) -- (263.83,178.96) ;

\draw  [fill={rgb, 255:red, 0; green, 0; blue, 0 }  ,fill opacity=1 ][line width=2.25]  (213.63,183.83) .. controls (213.63,182.28) and (214.89,181.03) .. (216.43,181.03) .. controls (217.98,181.03) and (219.23,182.28) .. (219.23,183.83) .. controls (219.23,185.37) and (217.98,186.63) .. (216.43,186.63) .. controls (214.89,186.63) and (213.63,185.37) .. (213.63,183.83) -- cycle ;

\draw (113.43,95) node [anchor=north west][inner sep=0.75pt]   [align=left] {left \\IM};
\draw (173,98.4) node [anchor=north west][inner sep=0.75pt]  [font=\large]  {$=$};
\draw (221,37) node [anchor=north west][inner sep=0.75pt]   [align=left] {\begin{minipage}[lt]{49.76pt}\setlength\topsep{0pt}
\begin{center}
spacetime \\mapping\\$\mathcal{M}$
\end{center}

\end{minipage}};

\end{tikzpicture}
\end{equation}
where the solid bullet represents the left steady state $\rho_D$ in the initial MPS auxiliary Hilbert space:
\begin{equation}\label{eq:steady_I}
    \sum_{a,b=0}^{q-1}(A^{a}B^{b})^\dagger \rho_D A^{a}B^{b} = \rho_D
\end{equation}
The triangular shape is obtained by performing the unitarity condition of local gates [Eq.~\eqref{eq:time_unitary}], which reflects the presence of an exact light cone  \cite{Lerose2021Influence, Perez2022Light, Lerose2023Overcoming}, such that those tensors lying outside the light cone do not affect the impurity site.
As the tensor $\mathcal{M}$ maps a spatial state to a temporal state, we refer to it as the \textit{spacetime mapping} operator.

We now present one of the central results of this work: the model defined by the gates from Eq.~\eqref{eq:swap_control} admits analytical MPS representations of the IM. 
More specifically, we derive an MPO representation of $\mathcal{M}$, and classify the corresponding computational complexity according to the scaling of the MPO bond dimension $\chi$ with the evolution time $T$.
For generic MPS initial states with bond dimension $D^2$ (square here arising due to the folded picture), the bond dimension scaling of the IM is given by the product $\chi_{\textrm{IM}}= D^2 \times \chi$. As we restricted ourselves to initial states with a finite bond dimension, both $\chi_{\textrm{IM}}$ and $\chi$ have the same scaling with time,  and, therefore, we use these two measures of complexity interchangeably in the following.

\begin{figure}[h]
\hspace*{-0.5\textwidth}
\includegraphics[width=1\linewidth]{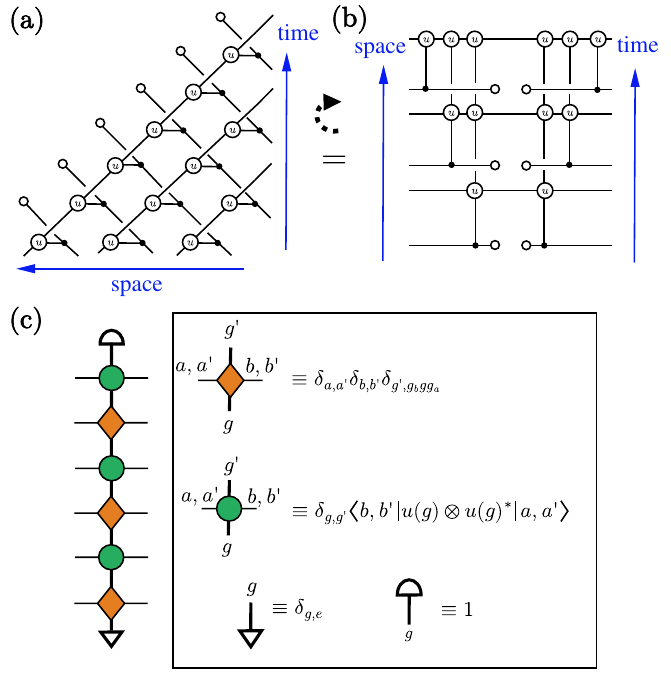} 
\caption{(a) The spacetime mapping in Eq.~\eqref{eq:leftIM} represented in terms of controlled and {\scriptsize${\mathrm{SWAP}}$} gates, according to Eq.~\eqref{eq:swap_control_tensor}.
(b) Rotating and deforming the triangular tensor to a linear structure. Open legs originally at the bottom are relocated to the left boundary, as indicated by the arrows.
(c) The MPO representation and the definition of local tensors. The inner bonds carry the linear space of the group algebra $\mathbb{C}[G]$, with $G=\mathrm{PU}(q)$.
}
\label{fig:MPO_tensors}
\end{figure}

To proceed with this derivation, we recast the tensor network $\mathcal{M}$ in terms of local building blocks: controlled and {\scriptsize${\mathrm{SWAP}}$} gates, as depicted in Fig.~\ref{fig:MPO_tensors}(a). To facilitate an MPO representation, we rotate and deform this tensor network into a linear structure depicted in Fig.~\ref{fig:MPO_tensors}(b). A straightforward construction would glue all vertical legs into a single inner bond, resulting in an exponentially growing bond dimension $\chi(T) \sim q^{2T}$. Instead, we note that the spacetime mapping acts on the physical Hilbert space through the sequence of unitaries $u_{b_t}\cdots u_{b_1}u_{a_1}\cdots u_{a_t}$, shown along the horizontal legs in Fig.~\ref{fig:MPO_tensors}(b), which are controlled by the history of indices traced from bottom to top. This product of unitaries could be organized into a single operator $u(g)$, with a group element $g$ traveling along the vertical legs.

This leads to an observation that the auxiliary linear space associated with the inner bonds can be identified with \textit{the group algebra of $G$ over $\mathbb{C}$}, where $G=\mathrm{PU}(q)$ is the projective unitary group of dimension $q$. 
A natural basis consists of group elements $g\in G$, such that any vector $x$ in this space can be written as a formal linear combination: $x = \sum_{g\in G}x_g g$, with complex coefficients $x_g\in\mathbb{C}$. 
In this sense, the local tensors in Fig.~\ref{fig:MPO_tensors}(c) are defined analogously to conventional tensors, but with an infinite-dimensional bond space.
Concretely, the group element in the diamond tensor is the projective image of the single-site unitary $u_a$ from Eq.~\eqref{eq:swap_control}, and the update rule is $g_bgg_a$. In the circle tensor, the operator $u(g)\otimes u(g)^*$ implements the regular adjoint representation of $g$ on physical indices.
The lower boundary vector in Fig.~\ref{fig:MPO_tensors}(c) corresponds to the unit element $e\in G$, meaning that the inner bond is initialized from the identity. Finally, the upper boundary represents the augmentation ideal of the group algebra, which acts as a uniform summation over all basis vectors. This construction leads to an MPO description with local blocks summarized in Fig.~\ref{fig:MPO_tensors}(c), and the detailed derivation presented in Appendix~\ref{app:group}.

So far, we have focused on the analytical MPO construction of the left IM. Constructions for the right IM can be obtained analogously, as we present in Appendix~\ref{app:right}. We also extend the exact solution to a generalized class of unitaries in Appendix~\ref{app:add_permutation}.

Since the group algebra of $\mathrm{PU}(q)$ has an infinite dimension if considered as a linear space, it is not immediately clear whether the MPO representation in Fig.~\ref{fig:MPO_tensors}(c) offers practical advantages over that in Fig.~\ref{fig:MPO_tensors}(b). As we show later in Sec.~\ref{sec:Hiererchy_bond}, the group-valued bond makes it possible to rigorously characterize the scaling behavior of the effective bond dimension and to identify complexity classes.

\section{Hierarchy of temporal entanglement entropy}\label{sec:Hiererchy_bond}

\subsection{Scaling behavior of the necessary bond dimension}\label{sec:growth}

As discussed above, the bond dimension $\chi$ of an exact MPS representation of the IM generally depends on the total number of time steps $T$, and the asymptotic growth of $\chi(T)$ provides an upper bound on the computational complexity of accurately simulating the IM.
In the literature, another commonly used diagnostic is the temporal entanglement entropy, defined as the bipartite von Neumann entropy of the IM maximized over all contiguous bipartitions. 
The exact representation therefore provides a
rigorous upper bound:
\begin{equation}\label{eq:TEE_bond}
    S(T)\lesssim \ln{[D^2\times\chi(T)]},
\end{equation}
In principle, the scaling behaviors of these two quantities can differ. 
However, for the typical initial states, we find that $S(T)$ follows qualitatively the same scaling suggested by this bound.
In this context, both saturation and polynomial growth of $\chi(T)$ imply classical simulatability of the IM, generally accompanied by the area-law and logarithmic growth of TEE, respectively.

We now count the bond dimension from the exact representation in Fig.~\ref{fig:MPO_tensors}(c). The crucial observation is that the dynamics in the auxiliary linear space can be regarded as a discrete random walk on the group manifold, updated according to the rule $g_\to g_b g g_a$.
Since the initial auxiliary state is fixed as $g(t=0)=e$, only a finite subset of group elements can be reached at finite $T$.
We denote the cumulative set of elements that can be reached within the time step $T$ as $H(T)$,
and the effective bond dimension is given by its size:
\begin{equation}\label{eq:bond_growth}
    \chi(T) = |H(T)|.
\end{equation}

Remarkably, it has been well established that the asymptotic growth rate of $|H(T)|$ -- known as the \textit{growth function} \cite{Grigorchuk1988Cancellative, Okninski1995Generalised, Okninski1996Growth, Okninski1995GrowthII} -- is fully determined by the algebraic structure of the infinite-time reachable set of group elements:
\begin{equation}
    H\equiv\lim_{T\to+\infty}H(T)\subset G.
\end{equation}
$H$ is the (semi)subgroup of $G$ generated by $\{g_a\}_{a=0}^{q-1}$, which is directly related to the set of single-site unitaries chosen in Eq.~\eqref{eq:swap_control}.
With rigorous mathematical formulations deferred to Appendix~\ref{app:growth}, we now present the classification of asymptotic behavior of $| H(T)|$ into three classes, together with the corresponding algebraic structure of $H$.
\begin{description}
    \item[Class I] \textit{Saturation}---If $H$ is a finite subgroup, the bond dimension saturates to $|H|$ after a sufficiently large but finite number of time steps.
    \item[Class II] \textit{Polynomial growth}---If $H$ is infinite but virtually nilpotent \cite{Wolf1968Growth,Milnor1968Growth,Gromov1981Groups}, the bond dimension grows polynomially with $T$. This includes all Abelian and certain non-Abelian cases that are, in a sense, ``Abelian up to a finite part.'' 
    \item[Class III] \textit{Exponential growth}---If $H$ contains a free non-Abelian subgroup, i.e., at least two free elements can be constructed from generators by multiplications, such that all distinct words represent distinct elements, then the bond dimension grows exponentially with $T$. 
\end{description}
According to Eqs.~\eqref{eq:TEE_bond} and \eqref{eq:bond_growth}, these three
classes imply, respectively, area-law, logarithmically growing, and linearly growing bounds on the TEE.
Although rigorous lower bounds are not available, our numerical results for representative models and initial states suggest that these upper bounds capture the corresponding qualitative TEE scaling.

Note that Ref.~\cite{Balasubramanian2024Glassy} also establishes a correspondence between concepts from geometric group theory and the complexity of many-body dynamics. However, the physical and mathematical notions considered there differ from those in this work.

We now comment briefly on each class. 
Class I corresponds to a subclass of quantum circuits introduced in Ref.~\cite{Wang2025Hopf}, which exhibits strong solvability, also manifest in finite local operator entanglement and polynomial recurrence time. Prototypical examples of class II include Abelian $H$ with commutative generators. In this case, the growth function counts the number of integer partitions of $T$ across generators, explicitly leading to polynomial scaling. As an example, dual-unitary circuits introduced in Ref.~\cite{Claeys2021Ergodic} in the form of a
{\scriptsize${\mathrm{SWAP}}$} gate followed by a two-site phase gate fall into this class. More generally, $H$ can take the form of a \textit{semidirect product} $H = K \ltimes N$:
\begin{equation}\label{eq:semiproduct}
    H =\{h=kn|k\in K, n\in N\}
\end{equation}
where $K$ is a finite subgroup and $N$ is an infinite Abelian subgroup. 
All examples showing polynomial growth presented in this work fall into the semidirect-product framework.
Finally, class III is characterized by strong noncommutativity, which arises generally when the controlled unitaries $u_a$ are chosen at random \cite{Epstein1971Almost}. 

In the remainder of this section, we focus on the qubit case $q=2$, where the corresponding group manifold $\mathrm{PU}(2)\cong \mathrm{SO}(3)$ allows for a simple geometric interpretation.
For each class, we present explicit examples of quantum circuits that illustrate the corresponding growth of TEE, thereby building intuition for the abstract mathematical formulation through concrete circuit dynamics. 
All these two-qubit gates can be rewritten as single-qubit rotations dressing an $XXZ$-type two-qubit entangling gate (see Ref.~\cite{Bertini2019Exact} and also Appendix~\ref{app:realization}), which makes them naturally compatible with current programmable quantum platforms~\cite{Mi2021Information, Rosenberg2024Dynamics, Nguyen2024Programmable, Fischer2026Dynamical}.
Extensions to higher $q$ follow analogously. 

\subsection{Abelian groups in classes I and II}\label{sec:Abelian}

We begin with the following controlled unitaries:
\begin{equation}\label{eq:Abelian}
    u_0 = e^{-iK\pi\sigma^z}, u_1 = e^{iK\pi\sigma^z},
\end{equation} 
which we refer to as \textit{model A}.
Here, $\sigma^{x,y,z}$ represent conventional Pauli matrices.
This realization is also known as the dual-unitary $XXZ$ circuit, since the two-qubit gate can be equivalently written as (up to an overall phase)
\begin{equation}
    U_{1,2}=e^{-i(J\sigma^x_1\sigma^x_2+J\sigma^y_1\sigma^y_2+J_z\sigma^z_1\sigma^z_2)},
\end{equation}
with $J=\pi/4$, $J_z=(K+1/4)\pi$ corresponding to the dual-unitary point we studied. The broader class of (1+1) D $XXZ$ circuits with arbitrary parameters is integrable in the sense that extensively many local conservation laws can be constructed rigorously \cite{Vanicat2018Integrable}. 
The dual-unitary point has been studied from various perspectives \cite{Dowling2023Scrambling,Rampp2024Haydenpreskill,Lopez2024Exact, Dowling2025Magic,Alves2025Probes}.
In particular, Ref.~\cite{Giudice2022Temporal} constructed an exact MPS representation of the IM with product initial states using a different approach and demonstrated logarithmic TEE growth for generic values of $K$. Here, we reproduce this behavior from our construction [Fig.~\ref{fig:MPO_tensors}(c)], which applies to generic MPS initial states.

To count the growth function, first we identify the image of $e^{\pm iK\pi\sigma^z}$ in $\mathrm{SO}(3)$: rotations around the $z$ axis by angle $\pm 2\pi K$, denoted by Abelian generators $a$ and $a^{-1}$. For a rational value $K=n/m$ with coprime integers $m$ and $n$, there is $a^m=e$, and, therefore, $H$ is a finite cyclic group with saturation growth function. When $K$ is irrational, any integer power of the generator $a^n$ can never come back to the unit $e$, and $H$ becomes an infinite Abelian group. According to the classification theory described above, the corresponding growth function should scale polynomially with $T$. We can explicitly list all the relevant group elements at time $T$:
\begin{equation}
    H(T) = \{a^{-2T}, a^{-2T+2}, \dots, a^{2T-2}, a^{2T}\},
\end{equation}
with a growth function $|H(T)|=2T+1$. The results agree with those in Ref.~\cite{Giudice2022Temporal}.

\subsection{Non-Abelian groups in classes I and II}\label{sec:nonAbelian}

Reference~\cite{Giudice2022Temporal} conjectures that the logarithmic growth of TEE in the dual-unitary $XXZ$ circuit is closely related to integrability and the presence of quasiparticles. 
Our group-theoretic approach instead shows that such behavior can arise more generally from class II group structure.
To demonstrate this, we consider the following controlled unitaries referred to as \textit{model B}:
\begin{equation}\label{eq:dihedral}
    u_0 = e^{-iK\pi\sigma^z}, u_1 = \sigma^x,
\end{equation}
which presumably define nonintegrable dynamics for generic $K$.
We denote by $r$ and $s$ the images of $u_0$ and $u_1$ in $\mathrm{SO}(3)$, representing, respectively, the rotation around $z$ axis by angle $2\pi K$ and the reflection over the $x$ axis. These non-Abelian generators satisfy relations $s^2=e,srs=r^{-1}$.

For rational $K=n/m$, both $r$ and $s$ have finite order, and $H$ is a finite dihedral group (class I). When $K$ is irrational, $H$ becomes the infinite dihedral group, which is non-Abelian but virtually nilpotent. A more direct way to justify the class II structure is to list the relevant group elements:
\begin{align}
    H(T) &= \{r^{-2T}, r^{-2T+2}, \dots, r^{2T-2}, r^{2T}\}\nonumber\\
    &\cup\{sr^{-2T+1}, sr^{-2T+3}, \dots, sr^{2T-3}, sr^{2T-1}\},
\end{align}
with a linear growth function $|H(T)|=4T+1$.

\begin{figure}[ht]
\hspace{-0.25\textwidth}
\includegraphics[width=0.5\linewidth]{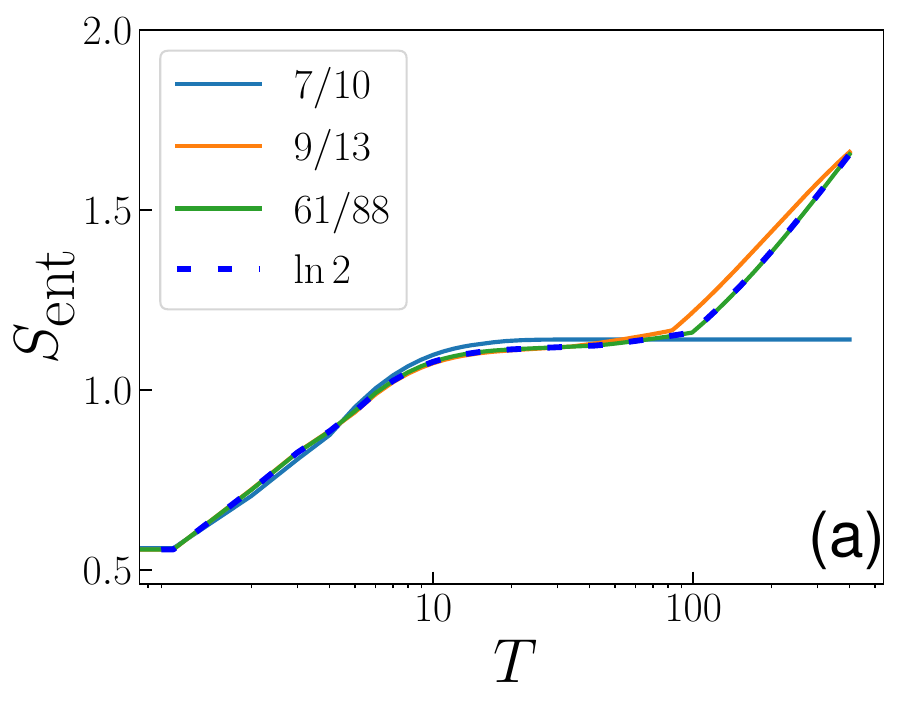} 
\includegraphics[width=0.5\linewidth]{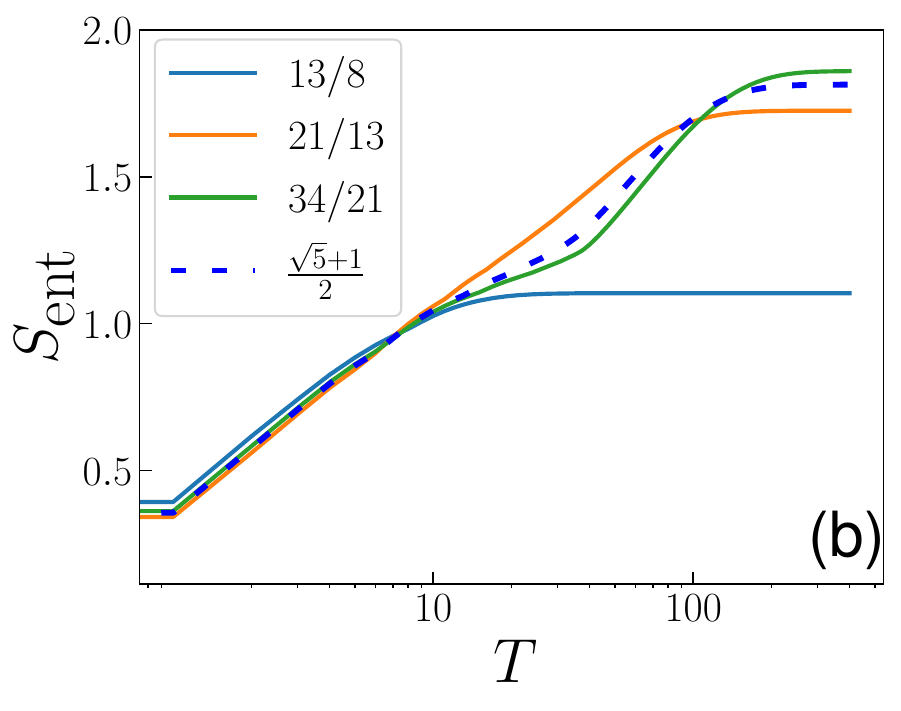} 
\caption{Temporal entanglement entropy for model B as a function of time for different values of $K$ from Eq.~\eqref{eq:dihedral}. Each subplot includes an irrational value of $K$ shown in dashed lines, along with a sequence of rational approximations shown in solid lines. The initial state is set to the product of $(\ket{0}+\ket{1})/\sqrt{2}$.
(a) Values of $K$: $\ln 2$, $7/10$, $9/13$, and $61/88$.
(b) Values of $K$: $(\sqrt{5}+1)/2$, $13/8$, $21/13$, and $34/21$.
}
\label{fig:dihedral}
\end{figure}

We verify this classification by numerically calculating the IM using the light-cone growth algorithm (LCGA), developed in Ref.~\cite{Lerose2023Overcoming}.
We display the growth of TEE for various values of $K$ in Fig.~\ref{fig:dihedral}, where the time axis is shown on a logarithmic scale. 
We consider the initial state to be a product state:
\begin{equation}
    \ket{\Psi_\text{in}}=\bigotimes_x\ket{+}_x,\qquad \ket{+}= \frac{1}{\sqrt{2}}(\ket{0}+\ket{1}),
\end{equation}
although the scaling behavior generalizes to arbitrary MPS initial states. Each subplot includes an irrational value of $K$ along with a sequence of rational approximations. 
Both reveal a consistent pattern of TEE growth: irrational $K$ values lead to logarithmic TEE growth interspersed with plateaus, where each plateau corresponds to the saturation value of TEE of a rational approximation to $K$. In particular, in Fig.~\ref{fig:dihedral}(b), the TEE of $K=(\sqrt{5}+1)/2$ seems to saturate at the plateau of $K=34/21$, but is expected to resume growing at longer time.
This structure is reminiscent of the behavior of the model A reported in Ref.~\cite{Giudice2022Temporal}. 

These observations suggest a universal pattern that irrational values of $K$ produce logarithmic TEE growth interspersed with plateaus and, thus, offer a new perspective on understanding sublinear TEE growth beyond integrability.

\subsection{Class III: Generic controlled unitaries}\label{sec:TEE_generic_group}

In the previous two realizations, we discussed examples exhibiting area-law and logarithmic growth of TEE based on exact MPO constructions. 
These cases rely on carefully engineered controlled unitaries to add certain structures to the algebra.
In contrast, for a generic pair of unitaries $u_0$ and $u_1$ without nontrivial algebraic relations, the reachable set of group elements $H(T)$ expands at the maximal rate, resulting in linear TEE growth.

\begin{figure}[ht]
\hspace{-0.25\textwidth}
\includegraphics[width=0.5\linewidth]{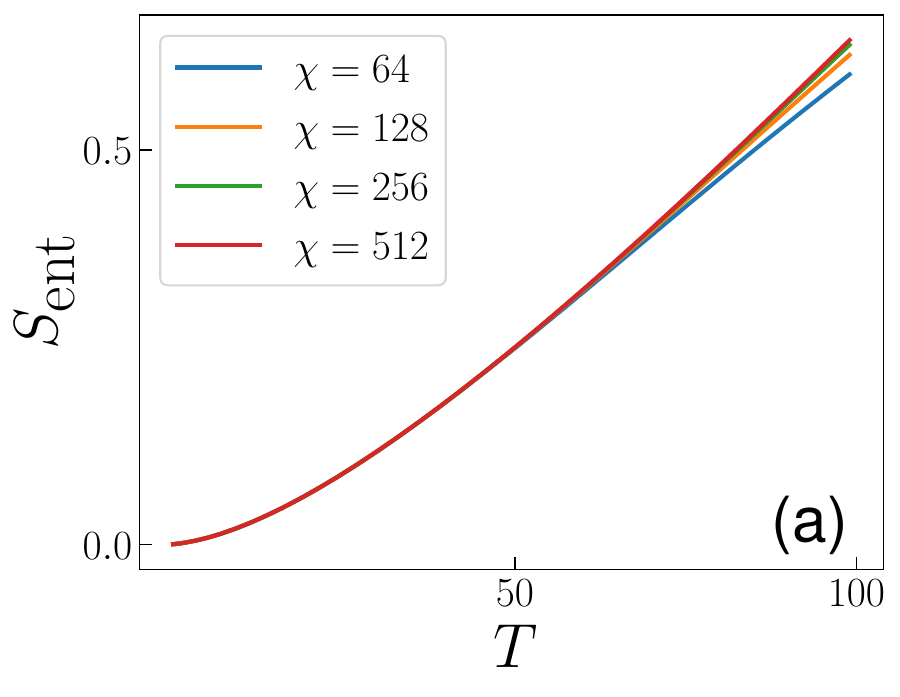} 
\includegraphics[width=0.5\linewidth]{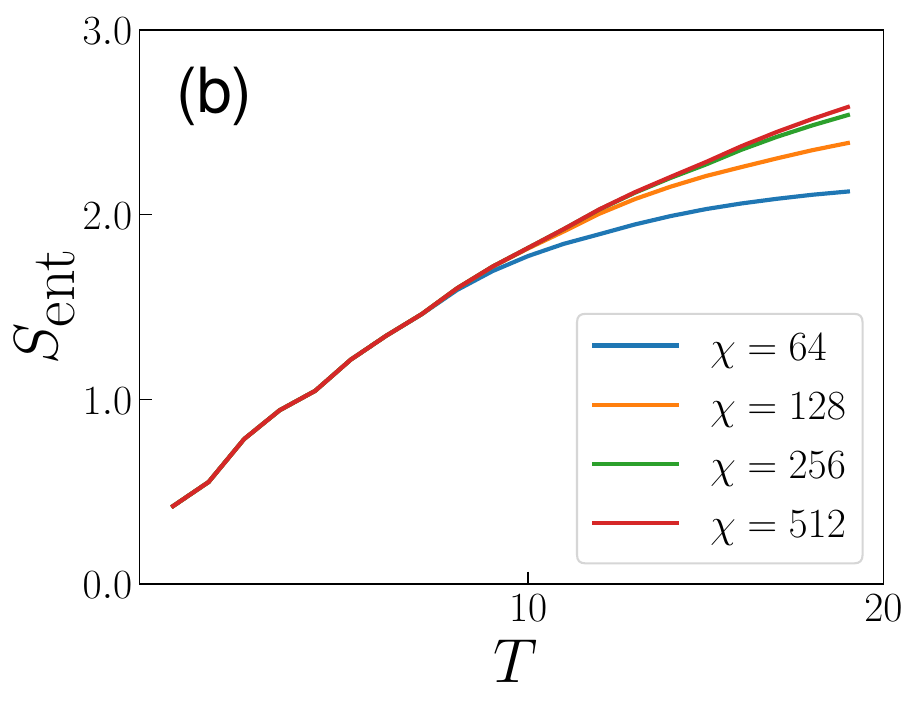} 
\caption{Temporal entanglement entropy for model C as a function of time for different values of $\theta$ in Eq.~\eqref{eq:ModelIII}, with varied truncated bond dimensions $\chi$. 
(a) $\theta=\pi/2+0.05$, close to an area-law parameter point $\pi/2$.
(b) $\theta=\pi/3$, far from area-law points.
}
\label{fig:generic}
\end{figure}

As an example, we consider the following \textit{model C}:
\begin{equation}\label{eq:ModelIII}
    u_0 = e^{-i\theta\sigma^z}, u_1 = e^{-i\theta\sigma^x}.
\end{equation}
Except at special parameter choices ($\theta =\pm \pi/2,\pm 3\pi/2, \pm\pi/4, \pm 3\pi/4$), $H$ falls in class III. We numerically verify the linear growth of TEE using LCGA with truncated bond dimensions. 
In Fig.~\ref{fig:generic}(a), 
we consider a small detuning away from the area-law point at $\theta = \pi/2+\delta$, and view a linear growth with a very small slope (note the $y$-axis scale), which allows the truncation with $\chi=128$ to accurately capture the IM up to time $T\lesssim 100$.
For $\theta=\pi/3$ far from the area-law regime [Fig.~\ref{fig:generic}(b)], the TEE grows with a much larger slope, and the accessible time window is drastically reduced for a given truncation, signaling the breakdown of MPS representations for long time steps. 
Note that, although $H$ itself is not free due to $g_0^3 = g_1^3 = e$, it contains a free subgroup generated by $g_0g_1$ and $g_1g_0$.

To conclude this section, let us note that despite the observed linear growth of TEE, the computational complexity of the two-point dynamical correlation functions can be lower, as outlined in Ref.~\cite{Figueroa2019Almost} in the case of a Haar-random environment, and also in Ref.~\cite{Vilkoviskiy2025TEE}. 
In agreement with these results, we introduce a truncation scheme of IM with polynomially growing bond dimension that captures the two-point correlation functions but presumably fails to correctly reproduce correlations with an extensive number of operator insertions. We rigorously prove certified error for calculations of local observables. 
While we refer the reader to Appendix~\ref{app:polynomial_truncation} for a detailed explanation, here, we provide an intuitive explanation of the scheme.
The key idea is to approximate the manifold of compact group $\mathrm{PU}(q)$ by a $\delta$ covering with a finite set of group elements $\{g_i\}_{i=1}^\mathcal{N}$, such that every group element lies within a ball of radius $\delta$ centered around one of the $g_i$. The set size  $\mathcal{N}$ corresponds to the bond dimension of the approximated MPO and scales proportionally to the volume of the group manifold as 
\begin{equation}
    \mathcal{N} \sim (1/\delta)^{q^2-1}.
\end{equation} 
At each time step, this approximation induces an error of the order of $\delta$ in the auxiliary linear space. The error accumulates linearly in time, leading to a total error bounded above by $T \times \delta$. Since there are $T$ distinct bonds, each carrying a group element with bounded error, the total error on observables scales as $\epsilon \sim T^2 \times \delta$. This results in the relation
\begin{equation}
\mathcal{N} \sim (T^2/\epsilon)^{q^2-1},
\end{equation}
which establishes a bound on the computational complexity of evaluating two-point correlation functions.

\subsection{Level spacing statistics}\label{sec:LSS}

In this section, we explore the connections between the complexity classes of the IM MPS representation and the  (non)integrability of underlying dynamics.
The notion of integrability in quantum many-body systems is rather diverse and somewhat ambiguous. Here, we focus on one of the standard indicators: the level-spacing statistics (LSS) of the effective Floquet Hamiltonian.
In this context, a model is regarded as chaotic if the distribution of its level spacing ratios follows that of random matrix theory, namely the Wigner-Dyson distribution \cite{Wigner1967Random, Dyson1970Correlations, Bohigas1984Characterization, Izrailev1986Limiting, Haake1987Classical}. In contrast, violation of the Wigner-Dyson distribution implies unresolved symmetries and integrability \cite{Berry1977Level}.

To perform exact diagonalization and compute LSS, we consider open boundary conditions of system size $L$, and the Floquet operator takes the form:
\begin{align}\label{eq:unitary_OBC}
    &\mathbb{U}=\mathbb{U}_{\text{odd}}\mathbb{U}_{\text{even}}=e^{-iH_F},\nonumber\\
    \mathbb{U}_{\text{odd}}=\otimes_{x\in {\text{odd},x=1}}^{L-3} &~U_{x,x+1}, \mathbb{U}_{\text{even}}=\otimes_{x\in {\text{even},x=0}}^{L-2} ~U_{x,x+1},
\end{align}
where $H_F$ is the effective Floquet Hamiltonian. Let $\{\theta_n\}_{n=1}^{q^L}$ denote the eigenvalues of $H_F$,
ordered in the increasing order in the branch $(-\pi,\pi]$. The level-spacing ratios are defined as
\begin{equation}
    r_n=\frac{\min(s_n,s_{n+1})}{\max(s_n,s_{n+1})},\quad s_n = \theta_{n+1}-\theta_n.
\end{equation}
We then compute the distribution of $r$ and compare it with predictions from random matrix theory. It should be noted that we consider level-spacing ratios rather than level spacings themselves, since the former require no spectral unfolding \cite{Oganesyan2007Localization, Atas2012Distribution} to reproduce random matrix theory, and, thus, avoid ambiguities associated with the branch choice in the Floquet-operator spectra \cite{Alessio2014Long}. 

\begin{figure}[h]
\hspace{-0.25\textwidth}
\includegraphics[width=0.5\linewidth]{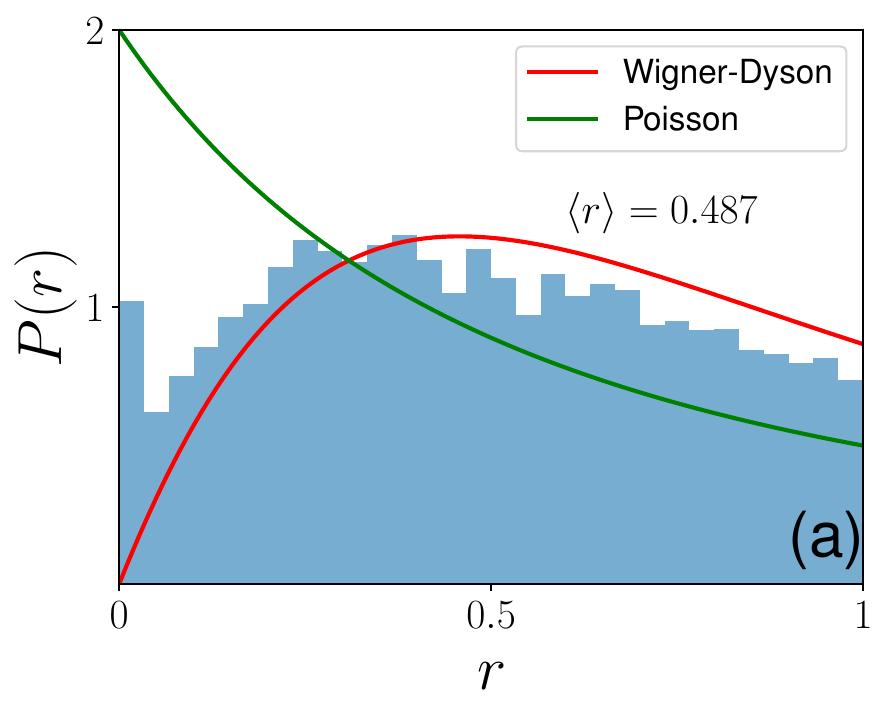}
\includegraphics[width=0.5\linewidth]{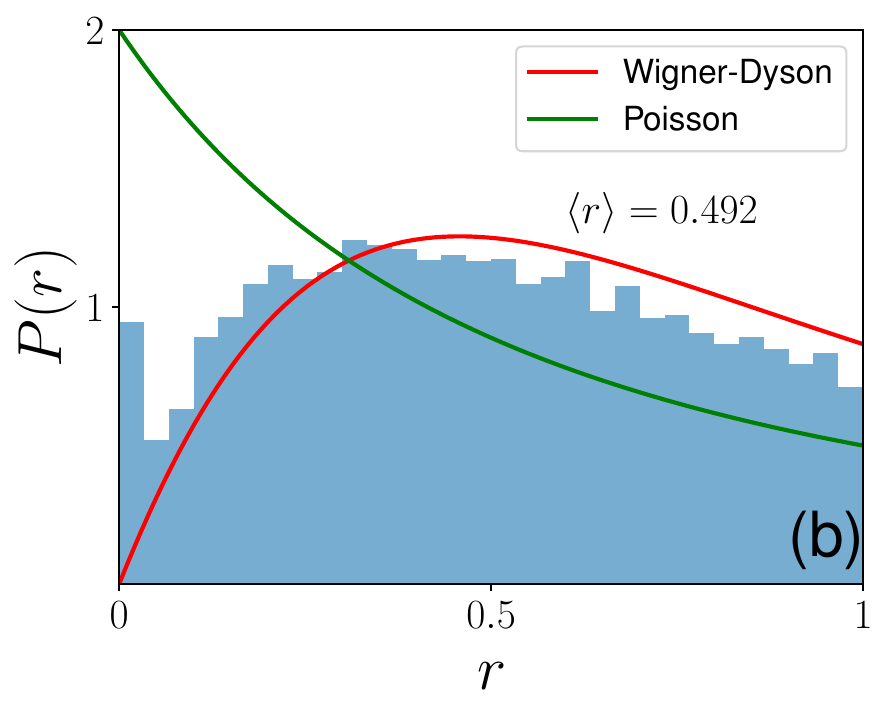} \\
\hspace{-0.25\textwidth}
\includegraphics[width=0.5\linewidth]{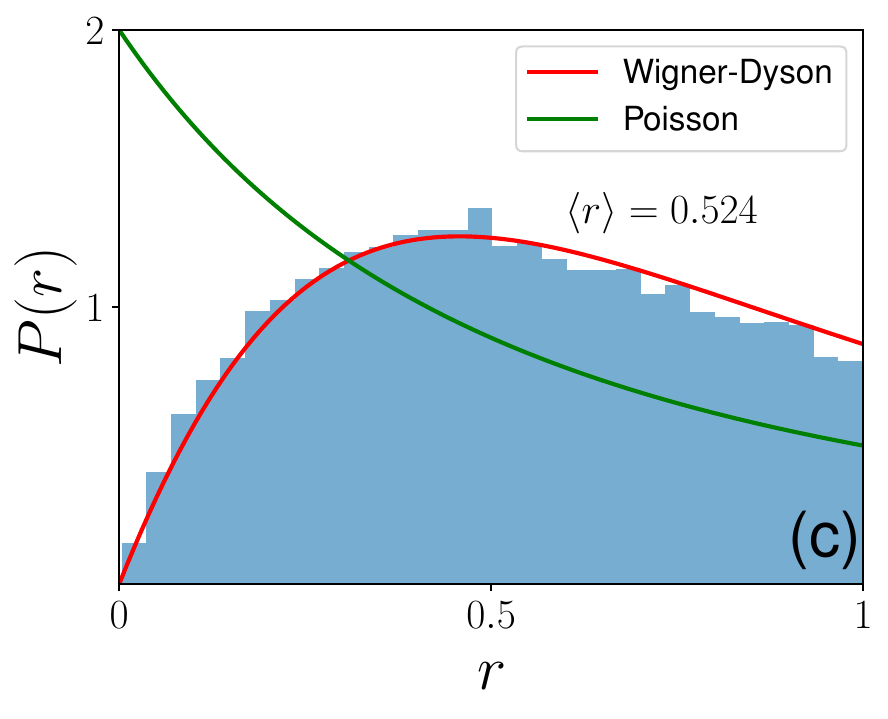} 
\includegraphics[width=0.5\linewidth]{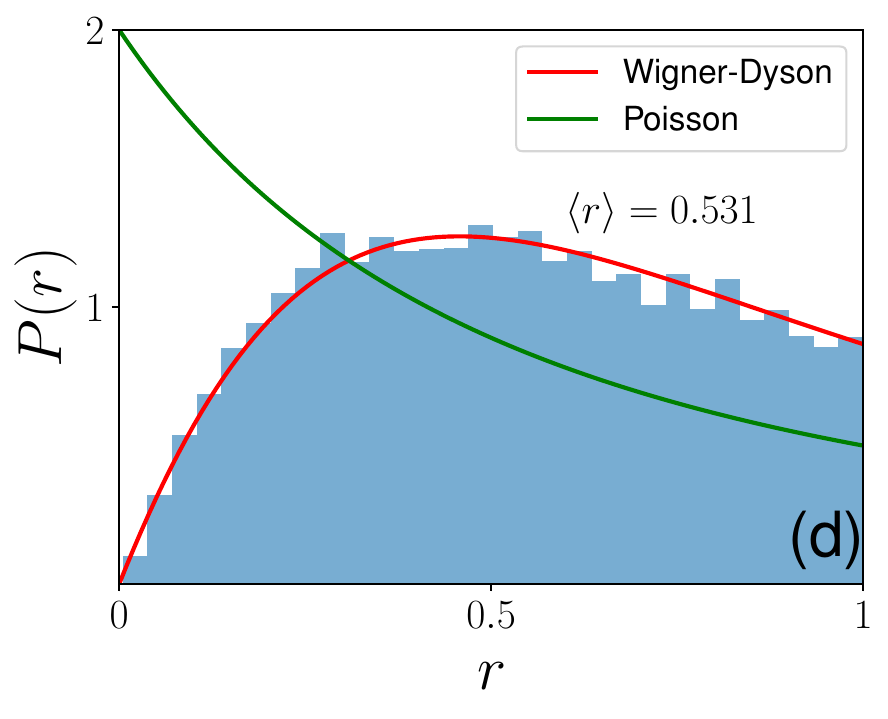} 
\caption{Histograms of level-spacing ratios for the Floquet Hamiltonian defined in Eq.~\eqref{eq:unitary_OBC}, compared with the Wigner-Dyson distribution of circular orthogonal ensembles and the Poisson distribution (solid lines). The mean values of ratios are also shown. System size $L=14$.
(a,b) Model B defined by Eq.~\eqref{eq:dihedral}, deformed by $v=e^{-0.01i\sigma^y}$ as in Eq.~\eqref{eq:deformation}. $K:\ln 2, (\sqrt{5}+1)/2$.
(c,d) Model C defined by Eq.~\eqref{eq:ModelIII}, with $\theta=\pi/2+0.05$ and $\pi/3$.
}
\label{fig:LSS}
\end{figure}

Next, we discuss the LSS of the examples introduced above.
First, model A [Eq.~\eqref{eq:Abelian}] is not only integrable, but, in fact, superintegrable, featuring a complete set of single-site conserved charges propagating in both directions \cite{Bertini2020OperatorII, Gombor2022Superintegrable}. For the LSS, we refer to Ref.~\cite{Alves2025Probes}, which demonstrates that the Floquet spectrum is highly degenerate and provides an explicit formula for energy levels.

In contrast, model B [Eq.~\eqref{eq:dihedral}] exhibits a much richer structure. It features nontrivial multisite conserved charges \cite{Pozsgay2025Soliton} which lead to high degeneracies in the Floquet spectrum. 
To lift these degeneracies, we consider an equivalent class of models by applying a transformation
\begin{equation}\label{eq:deformation}
    u_0\to vu_0 v^\dagger, u_1\to vu_1 v^\dagger,
\end{equation}
where $v$ is an arbitrary single-site unitary. 
This transformation leaves the algebraic structure of the growing set of group elements $H(T)$ unchanged, thereby preserving the bond dimension, while altering the global Floquet operator $\mathbb{U}$. We observe numerically that the LSS of the $v$-deformed Floquet operator $\mathbb{U}$ exhibits universal behavior for generic choices of $v$. We, therefore, regard it as an intrinsic property of the entire equivalent class of models with the same value of $K$.

Figures~\ref{fig:LSS}(a) and \ref{fig:LSS}(b) display the statistics of level-spacing ratios for model B with $K=\ln 2$ and $K = (\sqrt{5}+1)/2$, and the corresponding TEE growth is shown in Fig.s~\ref{fig:dihedral}(a) and \ref{fig:dihedral}(b), respectively. Both the distributions and the mean values exhibit intermediate behavior between chaoticity and integrability. 
Moreover, with increasing system size (not shown here), we observe a suppression of the subpeak around $r=0$ and a corresponding increase in the mean ratio.
A similar scenario can be realized in the Rosenzweig-Porter model, where Hilbert space factorizes into an integrable and a nonintegrable subspace, with the integrable sector becoming negligible in the thermodynamic limit \cite{Berry1984Semiclassical,Prosen1993Energy,Gorin1997Phase,Yan2025Spacing} .
Identifying such an integrable subspace in this model is of independent interest and warrants further investigation.

Finally, we turn to model C [see Eq.~\eqref{eq:ModelIII}]. Before proceeding, we emphasize that the associated Floquet operator $\mathbb{U}$ respects time-reversal symmetry, as can be seen from
\begin{equation}
    \mathbb{U}^T = \mathbb{U}_{\text{even}}^T\mathbb{U}_{\text{odd}}^T=P\mathbb{U}_{\text{even}}\mathbb{U}_{\text{odd}}P=(P\mathbb{U}_{\text{even}})\mathbb{U}(\mathbb{U}_{\text{even}}^\dagger P),
\end{equation}
where $P$ denotes space reflection. It follows that the relevant random matrix ensemble for comparison is the circular orthogonal ensemble rather than the circular unitary ensemble \cite{Wigner1967Random}.

The LSS of $\theta=\pi/2+0.05$ and $\theta=\pi/3$ are shown in Figs.~\ref{fig:LSS}(c) and \ref{fig:LSS}(d).
The distribution $P(r)$ closely follows the time-reversal symmetric Wigner-Dyson distribution, with the mean level-spacing ratio $\langle r\rangle\simeq 0.53$, which confirms the chaoticity. 
Based on these observations, we conjecture that all realizations belonging to class III, characterized by linearly growing TEE, exhibit chaotic LSS following the Wigner-Dyson distribution.

\subsection{Discussion}\label{sec:III_discussion}

Our construction yields a broad family of exactly solvable IMs and provides a controlled realization of distinct regimes of temporal complexity.

In classes I and II, the IM admits an exact MPS representation with, respectively, constant and polynomially growing bond dimension, and the models include both integrable and nonintegrable systems. Area-law TEE has previously been discovered in several solvable settings, including generalizations of dual-unitary circuits with compatible initial states~\cite{Bertini2023Exact, Yu2024Hierarchical, Wang2024Exact, Foligno2023Quantum, Liu2025Solvable, Foligno2025Non} and rule-54-type models~\cite{Klobas2021Exact, Klobas2021ExactII}. 
Instead, our class I and II solutions obtained here apply to generic MPS initial states.

Class III realizes a qualitatively different regime, with exponentially growing bond dimension but efficiently accessible low-order correlations. In particular, in Sec.~\ref{sec:TEE_generic_group}, we derive an error bound showing directly that two-point correlation functions remain efficiently computable despite the large complexity of the full IM.
This separation is consistent with earlier numerical observations for selected local observables~\cite{Banuls2009Matrix, Hermes2012Tensor, Hastings2015Connecting, Perez2022Light}, with the more general relation between temporal complexity and the complexity of local operators evolved in the Heisenberg picture discussed in Ref.~\cite{Carignano2024Temporal}, as well as with the semianalytic results found in Ref.~\cite{Vilkoviskiy2025TEE}.

Finally, we want to emphasize that the distinction among the three classes is not simply tied to integrability or to conventional signatures of quantum chaos. In Sec.~\ref{sec:LSS}, we find rapid thermalization and random-matrix level statistics in all three classes despite different temporal complexities. Moreover, in Sec.~\ref{sec:nonAbelian}, we find logarithmic growth of TEE in both integrable and nonintegrable examples. Among the cases studied here, the only combination not observed is integrability together with exponential TEE growth.
Thus, integrability, thermalization, and temporal complexity characterize distinct aspects of many-body dynamics beyond the usual integrable-versus-chaotic dichotomy. The scaling of IM bond dimension, therefore, provides an additional diagnostic that can distinguish systems with similar signatures of quantum chaos but markedly different complexity encoded in their multitime correlations.

\section{Hierarchy of non-Markovianity}\label{sec:Hiererchy_nMarkovian}

\subsection{Quantum and classical memory}\label{sec:q_and_cl}

We now turn to another notion of temporal complexity, defined in terms of classical and quantum memory. As introduced in Sec.~\ref{sec:MPO} and depicted in Fig.~\ref{fig:setting} (c), the evolved quantum many-body system can be regarded as a bath acting on the impurity. In general, the open-system dynamics of the impurity exhibits non-Markovianity and multitime memory effects, which manifests in the entangled structure of the IM. 

In order to ground the discussion of the nature of memory, we refer to the framework of \textit{Markovian embedding theory} \cite{Siegle2010Markovian,Kretschmer2016Collision,Campbell2018System,Tamascelli2018Non}. 
The key idea is to reduce the simulation complexity of non-Markovian dynamics by replacing the original non-Markovian bath with a minimal ancilla of much smaller Hilbert space, coupled to one or more fictitious Markovian baths. The requirement is that the combined ancilla–Markovian bath reproduces the original bath’s influence on the impurity. In this way, the joint impurity–ancilla dynamics become Markovian and, hence, more tractable.
In the continuous-time setting, this Markovian evolution is described by a Lindblad master equation; in discrete time, it corresponds to a sequence of completely positive and trace-preserving (CPTP) maps. 

The Markovian embedding approach effectively unravels the complex bath as the ancilla and Markovian dynamical generators acting on it, which are the minimal ingredients to reproduce the multitime influence on the impurity.
Structured IMs arising in several different scenarios \cite{Link2024Open, Wang2024Exact, Sonner2025Semigroup} provide concrete examples of Markovian embedding theory. In this context, the inner bonds of IM can be interpreted as carrying the ancilla indices, while the local tensor acts as the generator of the joint impurity-ancilla dynamics. 

We, therefore, say that an IM encodes \textit{classical memory} if the inner bonds admit a description as a classical stochastic dynamics, even though the underlying bath dynamics might be genuinely quantum. In Sec.~\ref{sec:stochastic}, we show that for a specific class of initial states, the IM of our model is classical and can be simulated by Monte Carlo sampling algorithms. By contrast, more general cases of initial states require a quantum ancilla to realize Markovian embedding, as shown in Sec.~\ref{sec:hybrid}. To characterize such situations, in Sec.~\ref{sec:quantum_memory}, we introduce an operational measure that quantifies the quantum memory and validate the corresponding protocol through explicit numerical realization in our model. 
Moreover, since our model of circuits belongs to the dual-unitary class, the IM becomes perfectly Markovian for solvable initial states \cite{Piroli2020Exact,Foligno2025Non}, as we reproduce from our exact MPO constructions in Appendix~\ref{app:reproduce}.
Summarizing the above, the regimes of no memory, classical memory, and genuine quantum memory form another complementary hierarchy of temporal complexities.

\subsection{Classical stochastic processes}\label{sec:stochastic}

We consider initial states in the form of two-site shift-invariant product states \footnote{This construction can be generalized to a product of arbitrary two-site states $\rho_\text{in}=\bigotimes_x\rho_{2x,2x+1}$}:
\begin{equation}
    \ket{\Psi_\text{in}}=\bigotimes_x\ket{\psi_\text{e}}_{2x}\otimes\ket{\psi_\text{o}}_{2x+1},
\end{equation}
with subscripts e and o for even and odd sites, respectively. We aim to show that, with such initial states, the IM can be interpreted as sequential CPTP quantum channels acting jointly on the auxiliary classical state and the impurity site. To this end, we explicitly write down the local tensor constituting the IM obtained by incorporating the MPO of spacetime mapping [see Fig.~\ref{fig:MPO_tensors}(c)] with the initial product state:

\begin{equation}\label{eq: classical_transfer}
\tikzset{every picture/.style={line width=0.75pt}} 
\begin{tikzpicture}[x=0.75pt,y=0.75pt,yscale=-1,xscale=1]

\draw [fill={rgb, 255:red, 144; green, 19; blue, 254 }  ,fill opacity=1 ][line width=0.75]    (39.03,75.93) -- (71.87,75.91) ;
\draw [line width=1.5]    (62.35,68.31) -- (62.25,113.59) ;
\draw [fill={rgb, 255:red, 144; green, 19; blue, 254 }  ,fill opacity=1 ][line width=0.75]    (38.67,104.59) -- (71.51,104.57) ;
\draw  [color={rgb, 255:red, 0; green, 0; blue, 0 }  ,draw opacity=1 ][fill={rgb, 255:red, 229; green, 126; blue, 33 }  ,fill opacity=1 ][line width=0.75]  (62.09,93.76) -- (70.74,104.58) -- (62.09,115.4) -- (53.44,104.58) -- cycle ;
\draw  [color={rgb, 255:red, 0; green, 0; blue, 0 }  ,draw opacity=1 ][fill={rgb, 255:red, 38; green, 173; blue, 95 }  ,fill opacity=1 ][line width=0.75]  (54.45,75.92) .. controls (54.45,71.5) and (58.03,67.92) .. (62.45,67.92) .. controls (66.87,67.92) and (70.45,71.5) .. (70.45,75.92) .. controls (70.45,80.33) and (66.87,83.91) .. (62.45,83.91) .. controls (58.03,83.91) and (54.45,80.33) .. (54.45,75.92) -- cycle ;
\draw [fill={rgb, 255:red, 144; green, 19; blue, 254 }  ,fill opacity=1 ][line width=0.75]    (70.45,75.92) -- (95.3,62.89) ;
\draw [shift={(88.19,66.62)}, rotate = 152.34] [color={rgb, 255:red, 0; green, 0; blue, 0 }  ][line width=0.75]    (10.93,-4.9) .. controls (6.95,-2.3) and (3.31,-0.67) .. (0,0) .. controls (3.31,0.67) and (6.95,2.3) .. (10.93,4.9)   ;
\draw [fill={rgb, 255:red, 144; green, 19; blue, 254 }  ,fill opacity=1 ][line width=0.75]    (71.51,104.57) -- (93.51,117.81) ;
\draw [shift={(76.51,107.58)}, rotate = 31.04] [color={rgb, 255:red, 0; green, 0; blue, 0 }  ][line width=0.75]    (10.93,-3.29) .. controls (6.95,-1.4) and (3.31,-0.3) .. (0,0) .. controls (3.31,0.3) and (6.95,1.4) .. (10.93,3.29)   ;
\draw [line width=1.5]    (62.09,115.4) -- (61.98,139.41) ;
\draw [shift={(62.07,120.91)}, rotate = 90.27] [fill={rgb, 255:red, 0; green, 0; blue, 0 }  ][line width=0.08]  [draw opacity=0] (8.13,-3.9) -- (0,0) -- (8.13,3.9) -- cycle    ;
\draw [line width=1.5]    (62.56,43.9) -- (62.45,67.92) ;
\draw [shift={(62.54,49.41)}, rotate = 90.27] [fill={rgb, 255:red, 0; green, 0; blue, 0 }  ][line width=0.08]  [draw opacity=0] (8.13,-3.9) -- (0,0) -- (8.13,3.9) -- cycle    ;
\draw   (24.32,75.37) -- (38.69,64.82) -- (38.87,85.66) -- cycle ;
\draw   (24.32,104.37) -- (38.69,93.82) -- (38.87,114.66) -- cycle ;

\draw (83,119.4) node [anchor=north west][inner sep=0.75pt]  [font=\small]  {$a,a'$};
\draw (103.3,72.29) node [anchor=north west][inner sep=0.75pt]  [font=\normalsize]  {$ \begin{array}{l}
=\delta _{a,a'}\sum\limits_{c=0}^{q-1} |\bra{c}\psi_\text{o}\rangle |^{2} \delta _{g',g_{a} gg_{c}}\\ 
\ \ \  \ \bra{b} u( g')\ket{\psi_\text{e}}\bra{\psi_\text{e}} u( g')^{\dagger }\ket{b'},
\end{array}$};
\draw (50,135.4) node [anchor=north west][inner sep=0.75pt]  [font=\small]  {$g$};
\draw (49,32.4) node [anchor=north west][inner sep=0.75pt]  [font=\small]  {$g'$};
\draw (83,48.4) node [anchor=north west][inner sep=0.75pt]  [font=\small]  {$b,b'$};
\draw (30,101.4) node [anchor=north west][inner sep=0.75pt]  [font=\scriptsize]  {$\text{o}$};
\draw (30,71.4) node [anchor=north west][inner sep=0.75pt]  [font=\scriptsize]  {$\text{e}$};
\end{tikzpicture}
\end{equation}
where the triangular tensor with symbol e (o) denotes the single-site initial state on even (odd) sites.
Following the direction of the time arrow, this tensor acts as a mapping on the impurity and the classical state on the $\mathrm{PU}(q)$ group manifold. The CPTP properties can be justified by noticing that this tensor realizes a \textit{measure-and-feedback} protocol.
\begin{enumerate}
    \item Perform projective measurements on the impurity qudit and obtain an outcome $a$;
    \item conditioned on the outcome $a$, update the classical state as $g\to g' = g_a g g_c$ with a given probability $|\bra{c}\psi_\text{o}\rangle|^2$;
    \item feed the qudit state $u( g')\ket{\psi_\text{e}}\bra{\psi_\text{e}} u( g')^{\dagger }$ back into the impurity site.
\end{enumerate}
The protocol demonstrates that the impurity–ancilla dynamics is Markovian in an extended hybrid quantum-classical linear space, thus establishing the Markovian embedding of the classical IM. 
Note that it is consistent with the definition of classical memory in terms of process tensors \cite{Giarmatzi2021Witnessing, Nery2021Simple, Taranto2024Characterising}. In addition, initial states in the form of a product of mixed states fit in this framework as well.

Another perspective on classical IM follows directly from the notion of spacetime mapping in Eq.~\eqref{eq:leftIM}. Recall that the IM is obtained by applying the spacetime mapping to the initial state. In the present case, the initial state is a product state and, thus, contains no quantum correlations. The spacetime mapping defines a ``local operations and classical communication'' (LOCC) protocol: as can be seen from Eq.~\eqref{eq: classical_transfer}, it consists only of measurements on odd sites, classical communication through the update of $g$ propagating forward in time, and conditional local unitary operations on even sites. Such a mapping cannot generate quantum memory in the resulting IM.

A key consequence of classical IM is that it allows the efficient classical simulation of observable evolution via mapping to classical stochastic processes. To show it, we consider the setting in Fig.~\ref{fig:setting}(c), where the impurity is initialized as $\rho_\text{imp}$ and evolves under a single-site CPTP quantum channel $\mathcal{K}$, acting between successive interactions with the bath. The expectation value of the observable $\hat{O}$ at time $T$ is graphically represented as
\begin{equation}\label{eq:classical_diagram}
\tikzset{every picture/.style={line width=0.75pt}} 
\begin{tikzpicture}[x=0.75pt,y=0.75pt,yscale=-1,xscale=1]

\draw [line width=1.5]    (130.41,78.24) -- (130.19,148.73) ;
\draw [line width=1.5]    (130.14,170.42) -- (130.04,216.7) ;
\draw    (131.6,126.74) .. controls (147.19,126.73) and (174.41,126.19) .. (174.7,122.64) ;
\draw [fill={rgb, 255:red, 144; green, 19; blue, 254 }  ,fill opacity=1 ][line width=0.75]    (107.03,38.93) -- (153.87,38.91) ;
\draw [line width=1.5]    (130.35,16.31) -- (130.25,60.59) ;
\draw  [fill={rgb, 255:red, 255; green, 255; blue, 255 }  ,fill opacity=1 ][line width=1.5]  (130.33,220.67) -- (123.48,211.56) -- (137.07,211.43) -- cycle ;
\draw  [line width=1.5]  (123.05,17.19) .. controls (123.05,17.15) and (123.05,17.12) .. (123.05,17.08) .. controls (123.03,12.3) and (126.32,8.41) .. (130.42,8.39) .. controls (134.51,8.38) and (137.84,12.24) .. (137.86,17.02) .. controls (137.86,17.06) and (137.86,17.1) .. (137.86,17.13) -- cycle ;
\draw [fill={rgb, 255:red, 144; green, 19; blue, 254 }  ,fill opacity=1 ][line width=0.75]    (106.67,193.21) -- (153.51,193.19) ;
\draw [fill={rgb, 255:red, 144; green, 19; blue, 254 }  ,fill opacity=1 ][line width=0.75]    (107.03,127.64) -- (135.87,127.62) ;
\draw [fill={rgb, 255:red, 144; green, 19; blue, 254 }  ,fill opacity=1 ][line width=0.75]    (106.67,99.4) -- (153.51,99.38) ;
\draw  [color={rgb, 255:red, 0; green, 0; blue, 0 }  ,draw opacity=1 ][fill={rgb, 255:red, 38; green, 173; blue, 95 }  ,fill opacity=1 ][line width=0.75]  (122.15,127.63) .. controls (122.15,123.21) and (125.74,119.63) .. (130.15,119.63) .. controls (134.57,119.63) and (138.15,123.21) .. (138.15,127.63) .. controls (138.15,132.05) and (134.57,135.63) .. (130.15,135.63) .. controls (125.74,135.63) and (122.15,132.05) .. (122.15,127.63) -- cycle ;
\draw  [color={rgb, 255:red, 0; green, 0; blue, 0 }  ,draw opacity=1 ][fill={rgb, 255:red, 229; green, 126; blue, 33 }  ,fill opacity=1 ][line width=0.75]  (130.09,182.37) -- (138.74,193.2) -- (130.09,204.02) -- (121.44,193.2) -- cycle ;
\draw  [color={rgb, 255:red, 0; green, 0; blue, 0 }  ,draw opacity=1 ][fill={rgb, 255:red, 229; green, 126; blue, 33 }  ,fill opacity=1 ][line width=0.75]  (130.09,88.56) -- (138.74,99.39) -- (130.09,110.21) -- (121.44,99.39) -- cycle ;
\draw  [color={rgb, 255:red, 0; green, 0; blue, 0 }  ,draw opacity=1 ][fill={rgb, 255:red, 38; green, 173; blue, 95 }  ,fill opacity=1 ][line width=0.75]  (122.45,38.92) .. controls (122.45,34.5) and (126.03,30.92) .. (130.45,30.92) .. controls (134.87,30.92) and (138.45,34.5) .. (138.45,38.92) .. controls (138.45,43.33) and (134.87,46.91) .. (130.45,46.91) .. controls (126.03,46.91) and (122.45,43.33) .. (122.45,38.92) -- cycle ;
\draw  [fill={rgb, 255:red, 248; green, 231; blue, 28 }  ,fill opacity=1 ] (162.87,33.91) -- (162.87,42.91) -- (153.87,42.91) -- (153.87,33.91) -- cycle ;
\draw  [color={rgb, 255:red, 0; green, 0; blue, 0 }  ,draw opacity=1 ][fill={rgb, 255:red, 231; green, 102; blue, 102 }  ,fill opacity=1 ][line width=1.5]  (165.71,112.72) .. controls (165.71,107.8) and (169.74,103.8) .. (174.7,103.8) .. controls (179.67,103.8) and (183.7,107.8) .. (183.7,112.72) .. controls (183.7,117.65) and (179.67,121.64) .. (174.7,121.64) .. controls (169.74,121.64) and (165.71,117.65) .. (165.71,112.72) -- cycle ;
\draw  [line width=0.75]  (174.65,110.62) -- (178.76,110.62) -- (178.76,114.69) ;
\draw    (149.23,99.5) .. controls (163.19,99.07) and (174.41,98.19) .. (174.7,104.8) ;
\draw   (92.32,127.37) -- (106.69,116.82) -- (106.87,137.66) -- cycle ;
\draw   (92.32,192.37) -- (106.69,181.82) -- (106.87,202.66) -- cycle ;
\draw   (92.32,99.37) -- (106.69,88.82) -- (106.87,109.66) -- cycle ;
\draw   (92.32,39.37) -- (106.69,28.82) -- (106.87,49.66) -- cycle ;

\draw (160.82,80.6) node [anchor=north west][inner sep=0.75pt]    {$\mathcal{K}$};
\draw (158.82,184.4) node [anchor=north west][inner sep=0.75pt]  [font=\normalsize]  {$\rho _{\text{imp}}$};
\draw (164.82,15.4) node [anchor=north west][inner sep=0.75pt]  [font=\normalsize]  {$\hat{O}$};
\draw (132.1,151.5) node [anchor=north west][inner sep=0.75pt]  [rotate=-90]  {$...$};
\draw (132.1,62.5) node [anchor=north west][inner sep=0.75pt]  [rotate=-90]  {$...$};
\draw (25.74,103.79) node [anchor=north west][inner sep=0.75pt]  [font=\normalsize]  {$\langle \hat{O}( T) \rangle =$};
\draw (192,103.4) node [anchor=north west][inner sep=0.75pt]  [font=\normalsize]  {$=\bra{w(\hat{O})}\mathbb{M}^{T-1}\ket{v( \rho_{\text{imp}})}.$};
\draw (98,188.4) node [anchor=north west][inner sep=0.75pt]  [font=\scriptsize]  {$\text{o}$};
\draw (98,123.4) node [anchor=north west][inner sep=0.75pt]  [font=\scriptsize]  {$\text{e}$};
\draw (98,95.4) node [anchor=north west][inner sep=0.75pt]  [font=\scriptsize]  {$\text{o}$};
\draw (98,35.4) node [anchor=north west][inner sep=0.75pt]  [font=\scriptsize]  {$\text{e}$};
\end{tikzpicture}
\end{equation}
In the last equality, we express the tensor network as $(T-1)$ power of the transfer matrix $\mathbb{M}$ sandwiched by vectors on top and bottom, which are all defined in the linear space of group algebra:
\begin{widetext}
\begin{equation}\label{eq:conditional_prob}
\tikzset{every picture/.style={line width=0.75pt}} 
\begin{tikzpicture}[x=0.75pt,y=0.75pt,yscale=-1,xscale=1]

\draw [line width=1.5]    (143.41,83.24) -- (143.19,153.73) ;
\draw    (144.6,131.74) .. controls (160.19,131.73) and (187.41,131.19) .. (187.7,127.64) ;
\draw [fill={rgb, 255:red, 144; green, 19; blue, 254 }  ,fill opacity=1 ][line width=0.75]    (120.03,132.64) -- (148.87,132.62) ;
\draw [fill={rgb, 255:red, 144; green, 19; blue, 254 }  ,fill opacity=1 ][line width=0.75]    (119.67,104.4) -- (166.51,104.38) ;
\draw  [color={rgb, 255:red, 0; green, 0; blue, 0 }  ,draw opacity=1 ][fill={rgb, 255:red, 38; green, 173; blue, 95 }  ,fill opacity=1 ][line width=0.75]  (135.15,132.63) .. controls (135.15,128.21) and (138.74,124.63) .. (143.15,124.63) .. controls (147.57,124.63) and (151.15,128.21) .. (151.15,132.63) .. controls (151.15,137.05) and (147.57,140.63) .. (143.15,140.63) .. controls (138.74,140.63) and (135.15,137.05) .. (135.15,132.63) -- cycle ;
\draw  [color={rgb, 255:red, 0; green, 0; blue, 0 }  ,draw opacity=1 ][fill={rgb, 255:red, 229; green, 126; blue, 33 }  ,fill opacity=1 ][line width=0.75]  (143.09,93.56) -- (151.74,104.39) -- (143.09,115.21) -- (134.44,104.39) -- cycle ;
\draw  [color={rgb, 255:red, 0; green, 0; blue, 0 }  ,draw opacity=1 ][fill={rgb, 255:red, 231; green, 102; blue, 102 }  ,fill opacity=1 ][line width=1.5]  (178.71,117.72) .. controls (178.71,112.8) and (182.74,108.8) .. (187.7,108.8) .. controls (192.67,108.8) and (196.7,112.8) .. (196.7,117.72) .. controls (196.7,122.65) and (192.67,126.64) .. (187.7,126.64) .. controls (182.74,126.64) and (178.71,122.65) .. (178.71,117.72) -- cycle ;
\draw  [line width=0.75]  (187.65,115.62) -- (191.76,115.62) -- (191.76,119.69) ;
\draw    (162.23,104.5) .. controls (176.19,104.07) and (187.41,103.19) .. (187.7,109.8) ;
\draw   (105.32,132.37) -- (119.69,121.82) -- (119.87,142.66) -- cycle ;
\draw   (105.32,104.37) -- (119.69,93.82) -- (119.87,114.66) -- cycle ;

\draw (173.82,85.6) node [anchor=north west][inner sep=0.75pt]    {$\mathcal{K}$};
\draw (27.74,111.79) node [anchor=north west][inner sep=0.75pt]  [font=\normalsize]  {$\bra{g'}\mathbb{M}\ket{g} =$};
\draw (111,128.4) node [anchor=north west][inner sep=0.75pt]  [font=\scriptsize]  {$\text{e}$};
\draw (111,100.4) node [anchor=north west][inner sep=0.75pt]  [font=\scriptsize]  {$\text{o}$};
\draw (138,156.4) node [anchor=north west][inner sep=0.75pt]  [font=\small]  {$g$};
\draw (138,65.4) node [anchor=north west][inner sep=0.75pt]  [font=\small]  {$g'$};
\draw (205,99.79) node [anchor=north west][inner sep=0.75pt]  [font=\normalsize]  {$=\sum\limits_{a,b=0}^{q-1} \delta _{g',g_{b} gg_{a}}\bra{b}\mathcal{K}\left[ u( g)\ket{\psi_e }\bra{\psi_e } u( g)^{\dagger }\right]\ket{b} |\bra{a}\psi_o\rangle |^{2} \equiv P( g'|g)$,};
\end{tikzpicture}
\end{equation}
\begin{equation}
\tikzset{every picture/.style={line width=0.75pt}} 
\begin{tikzpicture}[x=0.75pt,y=0.75pt,yscale=-1,xscale=1]
\draw [line width=1.5]    (151.14,107.42) -- (151.04,153.7) ;
\draw  [fill={rgb, 255:red, 255; green, 255; blue, 255 }  ,fill opacity=1 ][line width=1.5]  (151.33,157.67) -- (144.48,148.56) -- (158.07,148.43) -- cycle ;
\draw [fill={rgb, 255:red, 144; green, 19; blue, 254 }  ,fill opacity=1 ][line width=0.75]    (127.67,130.21) -- (174.51,130.19) ;
\draw  [color={rgb, 255:red, 0; green, 0; blue, 0 }  ,draw opacity=1 ][fill={rgb, 255:red, 229; green, 126; blue, 33 }  ,fill opacity=1 ][line width=0.75]  (151.09,119.37) -- (159.74,130.2) -- (151.09,141.02) -- (142.44,130.2) -- cycle ;
\draw   (113.32,129.37) -- (127.69,118.82) -- (127.87,139.66) -- cycle ;

\draw (174.82,113.4) node [anchor=north west][inner sep=0.75pt]  [font=\normalsize]  {$\rho _{\text{imp}}$};
\draw (17.74,118.79) node [anchor=north west][inner sep=0.75pt]  [font=\normalsize]  {$\bra{g}v( \rho_\text{imp} )\rangle =$};
\draw (119,125.4) node [anchor=north west][inner sep=0.75pt]  [font=\scriptsize]  {$\text{o}$};
\draw (147,89.49) node [anchor=north west][inner sep=0.75pt]  [font=\small]  {$g$};
\draw (212,109.79) node [anchor=north west][inner sep=0.75pt]  [font=\normalsize]  {$=\sum\limits_{a,b=0}^{q-1} \delta _{g,g_{b} g_{a}}\bra{b} \rho_\text{imp} \ket{b} |\bra{a}\psi_\text{o} \rangle |^{2} \equiv P_{0}(g|e) ,$};
\end{tikzpicture}
\end{equation}
\begin{equation}
\tikzset{every picture/.style={line width=0.75pt}} 
\begin{tikzpicture}[x=0.75pt,y=0.75pt,yscale=-1,xscale=1]

\draw [fill={rgb, 255:red, 144; green, 19; blue, 254 }  ,fill opacity=1 ][line width=0.75]    (128.03,99.93) -- (174.87,99.91) ;
\draw [line width=1.5]    (151.35,77.31) -- (151.25,121.59) ;
\draw  [line width=1.5]  (144.05,78.19) .. controls (144.05,78.15) and (144.05,78.12) .. (144.05,78.08) .. controls (144.03,73.3) and (147.32,69.41) .. (151.42,69.39) .. controls (155.51,69.38) and (158.84,73.24) .. (158.86,78.02) .. controls (158.86,78.06) and (158.86,78.1) .. (158.86,78.13) -- cycle ;
\draw  [color={rgb, 255:red, 0; green, 0; blue, 0 }  ,draw opacity=1 ][fill={rgb, 255:red, 38; green, 173; blue, 95 }  ,fill opacity=1 ][line width=0.75]  (143.45,99.92) .. controls (143.45,95.5) and (147.03,91.92) .. (151.45,91.92) .. controls (155.87,91.92) and (159.45,95.5) .. (159.45,99.92) .. controls (159.45,104.33) and (155.87,107.91) .. (151.45,107.91) .. controls (147.03,107.91) and (143.45,104.33) .. (143.45,99.92) -- cycle ;
\draw  [fill={rgb, 255:red, 248; green, 231; blue, 28 }  ,fill opacity=1 ] (183.87,94.91) -- (183.87,103.91) -- (174.87,103.91) -- (174.87,94.91) -- cycle ;
\draw   (113.32,100.37) -- (127.69,89.82) -- (127.87,110.66) -- cycle ;

\draw (180.82,76.4) node [anchor=north west][inner sep=0.75pt]  [font=\footnotesize]  {$\hat{O}$};
\draw (18.74,84.79) node [anchor=north west][inner sep=0.75pt]  [font=\normalsize]  {$\bra{w(\hat{O})}g\rangle =$};
\draw (203,82.4) node [anchor=north west][inner sep=0.75pt]  [font=\normalsize]  {$=\text{Tr}\left[ u( g)\ket{\psi_\text{e} }\bra{\psi_\text{e} } u( g)^{\dagger }\hat{O}\right].$};
\draw (119,96.4) node [anchor=north west][inner sep=0.75pt]  [font=\scriptsize]  {$\text{e}$};
\draw (147.25,126.99) node [anchor=north west][inner sep=0.75pt]  [font=\small]  {$g$};
\end{tikzpicture}
\end{equation}
\end{widetext}

Here, $P(g'|g)$ and $P_0(g|e)$ define conditional probability distributions over the group manifold, as justified by the non-negativity and normalization. Hence, we can interpret the tensor network as a random walk on the group manifold starting from the identity $e$:
\begin{widetext}
\begin{align} \label{eq:dynamics_O}
    \langle \hat{O}(T)\rangle& = \sum_{g_T}P(g_T)\Tr{u(g_T)\ket{\psi_\text{e} }\bra{\psi_\text{e} } u(g_T)^\dagger\hat{O}}\nonumber\\
    &=\sum_{g_T,g_{T-1},\dots,g_2,g_1}\Tr{u(g_T)\ket{\psi_\text{e} }\bra{\psi_\text{e} } u(g_T)^\dagger\hat{O}}P(g_T|g_{T-1})\cdots P(g_2|g_1)P_0(g_1|e).
\end{align}
\end{widetext}
This establishes an equivalence between the quantum dynamics and a Markovian classical stochastic process, of which the validity is independent of the realization of controlled unitaries [Eq.~\eqref{eq:swap_control}].

\begin{figure*}[ht]
\hspace{-0.29\textwidth}
\includegraphics[width=0.25\linewidth]{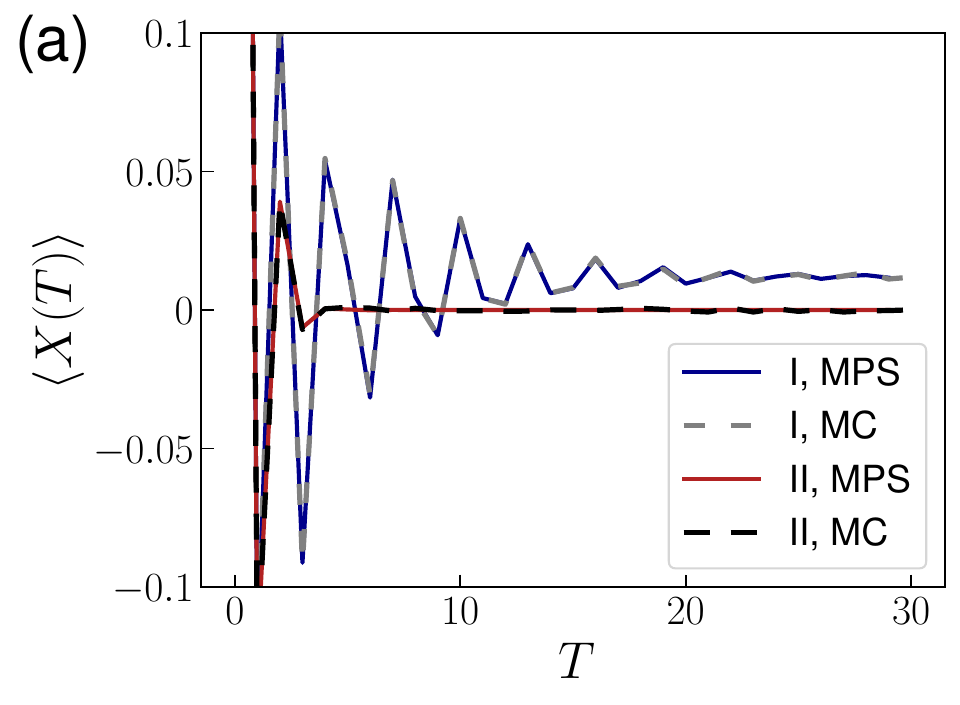}
\includegraphics[width=0.25\linewidth]{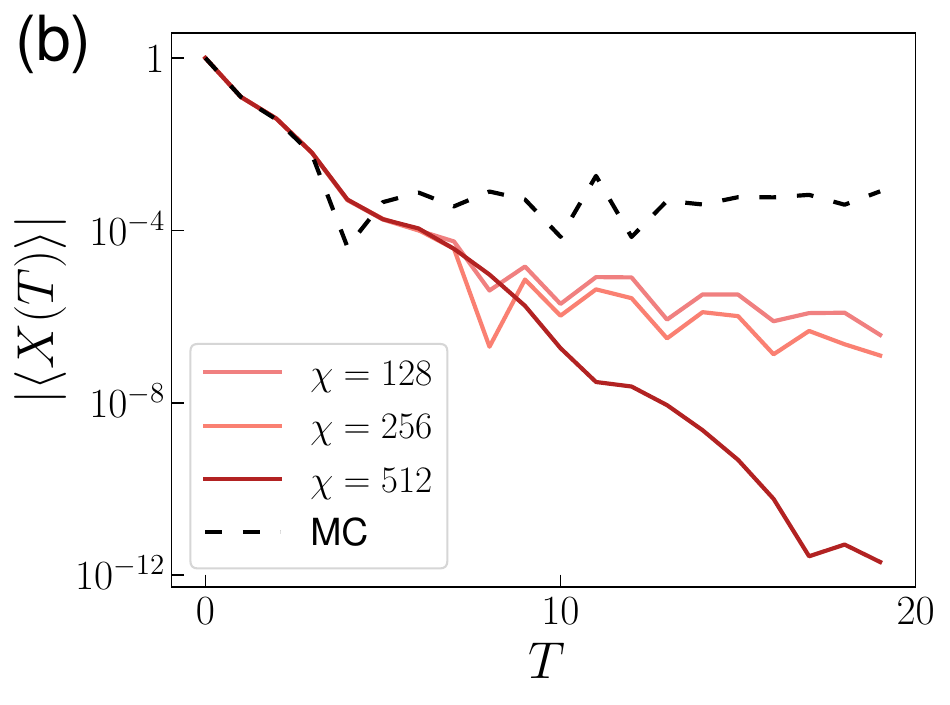} 
\includegraphics[width=0.25\linewidth]{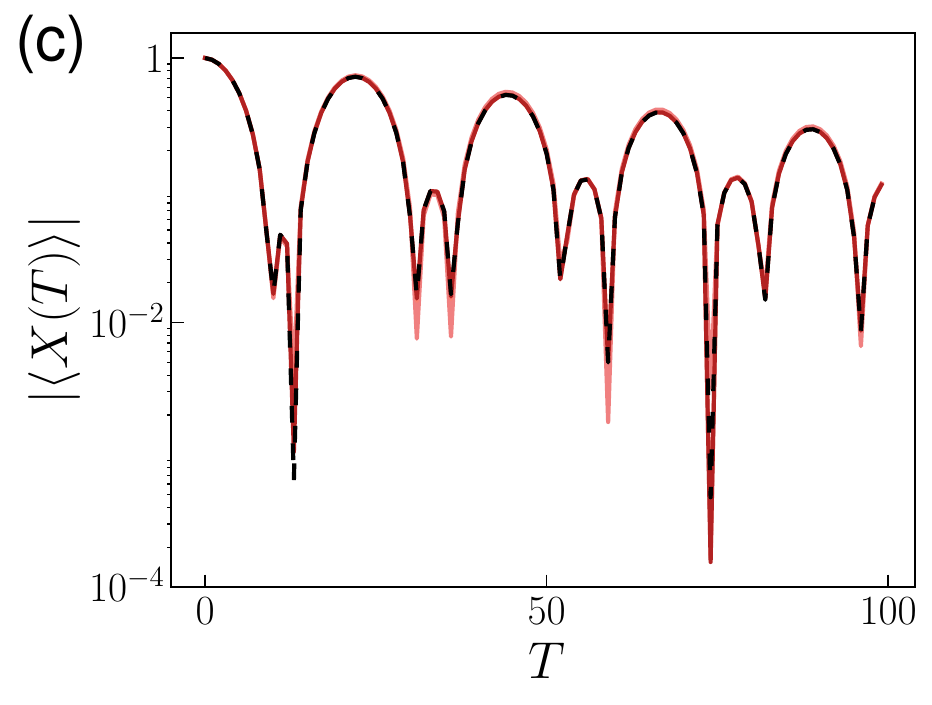} 
\includegraphics[width=0.25\linewidth]{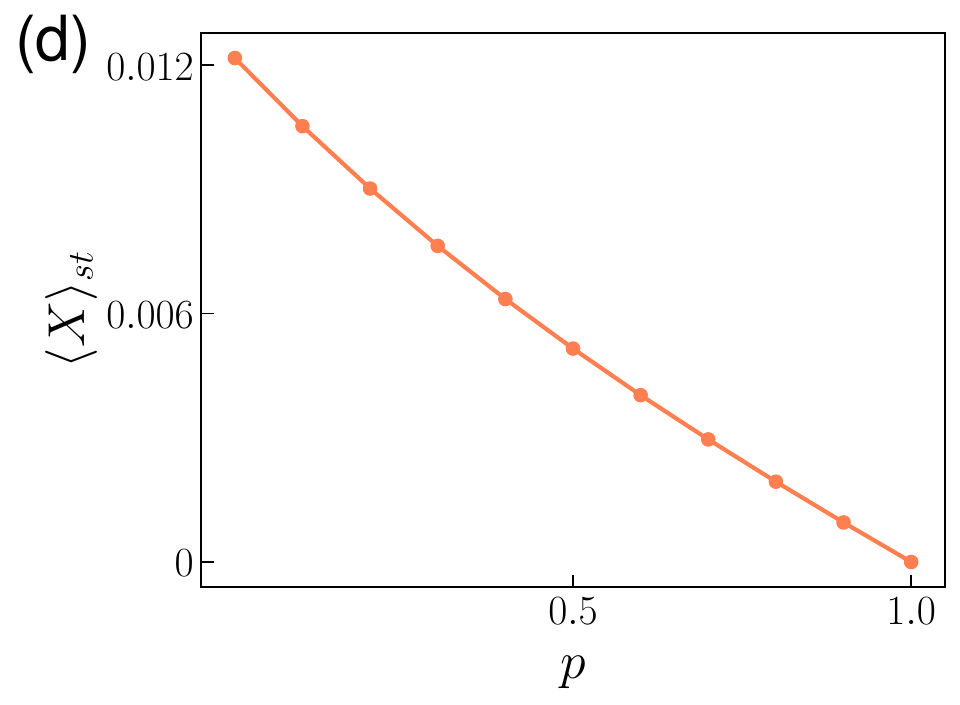} 
\caption{Monte Carlo (MC) simulations of local dynamics of model C defined by Eq.~\eqref{eq:ModelIII}, with $\theta=\pi/3$. Initial states are  $\rho_\text{imp}= \rho_\text{o}= \rho_\text{e}= \ket{+}\bra{+}$. Number of samples $N=10^6$.
(a) Comparison of the Pauli-$X$ expectation values from Monte Carlo and MPS simulations for channels I and II. The truncated MPS bond dimension is $256$.
(b),(c) Log-scale view of exponential decay in $|\langle X(T)\rangle|$ for channel II, $\theta = \pi/3$ compared with $\theta = 0.1$. Bond dimension varies from $128$ to $512$ for both figures.
(d) The stationary expectation value with the mixing rate $p$ between channels I and II.
}
\label{fig:stochastic}
\end{figure*}

It follows that the expectation value depends only on the final-time probability distribution $P(g_T)$. Therefore, late-time observables can be directly computed from the stationary distribution on the group manifold, obtained as the fixed point of the transfer matrix $\mathbb{M}$ or, equivalently, of the conditional probability $P(g'|g)$. 
We obtain analytical solutions for a specific class of single-site quantum channels $\mathcal{K}[\rho_\text{imp}] = \Tr{\rho_\text{imp}}\rho_\text{re}$, which erases the impurity qudit and prepares a new state $\rho_\text{re}$ to replace it (also known as the ``causal break'' in Ref.~\cite{Pollock2018Operational}).  Substituting it into Eq.~\eqref{eq:conditional_prob}, we find
\begin{equation} \label{eq:break_probability}
    P(g'|g) =  \sum_{a,b=0}^{q-1}\delta_{g',g_bgg_a}\bra{b}\rho_\text{re}\ket{b}|\bra{a}\psi_\text{o}\rangle|^2,
\end{equation}
which is homogeneous over the group manifold, i.e., independent of $g$. According to Ref.~\cite{Rosenthal1995Convergence}, the \textit{Haar measure} $d\mu(g)$ is a stationary distribution.
A subtle point is that the Haar measure is defined with respect to the maximal reachable set of group elements, introduced as $H$ in Sec.~\ref{sec:growth}. 
This underscores the role of impurity dynamics as a probe of the underlying many-body evolution.
For clarity, we restrict to class III dynamics, for which the stationary distribution is the Haar measure on the full $\text{PU}(q)$ manifold.
Using Weingarten calculus, the stationary expectation value becomes:
\begin{align}
    \langle \hat{O}(T\to+\infty)\rangle &= \int d\mu(g)\Tr{u(g)\ket{\psi_\text{e} }\bra{\psi_\text{e} } u(g)^\dagger\hat{O}} \nonumber\\
    &= \Tr{\hat{O}}/q,
\end{align}
which corresponds to the thermalization of local observables to the infinite-temperature ensemble.

The framework can be extended to calculations of two-point correlation functions $\langle \hat{O}(T)\hat{O'}\rangle$. Following a similar approach, the quantity is represented as a random walk as in Eq.~\eqref{eq:dynamics_O}, with the initial conditional probability $P_0(g|e)$ replaced by
\begin{equation}
    \sum_{a,b=0}^{q-1} \delta _{g,g_{b} g_{a}}\bra{b}\hat{O'} \rho_\text{imp} \ket{b} |\bra{a}\psi_\text{o} \rangle |^{2}.
\end{equation}
Unlike the one-point case, this quantity is not guaranteed to be real or positive because of the factor $\bra{b}\hat{O'} \rho_\text{imp} \ket{b}$. To overcome this issue, we separate this factor and evaluate the contribution associated with each $b$ independently. For a fixed $b$, we define the conditional probability $P'_0(g|g_b)
    =
    \sum_{a=0}^{q-1}
    \delta_{g,g_b g_a}
    |\bra{a}\psi_{\mathrm{o}}\rangle|^2$.
The two-point correlation function can then be written as
\begin{widetext}
\begin{equation}
    \langle \hat{O}(T)\hat{O'}\rangle =\sum_b\bra{b}\hat{O'} \rho_\text{imp} \ket{b} \sum_{g_T,g_{T-1},\dots,g_2,g_1}\Tr{u(g_T)\ket{\psi_\text{e} }\bra{\psi_\text{e} } u(g_T)^\dagger\hat{O}}P(g_T|g_{T-1})\cdots P(g_2|g_1)P'_0(g_1|g_b)
\end{equation}
\end{widetext}
Thus, the Monte Carlo simulation can be performed separately for each starting point $g_b$, and the resulting estimates are combined at the end with the generally complex weights.
This procedure generalizes straightforwardly to $k$-point time-ordered correlation functions, at an additional cost scaling as $q^{k-1}$.
This exponential dependence on $k$ is natural, since higher-order correlations probe increasingly more information encoded in the IM, whose full representation can require exponentially growing resources.

We verify the stochastic-process approach by simulating it using a Monte Carlo algorithm and comparing it with MPS results obtained via LCGA.
We generate trajectories on the group manifold by sequentially updating the group element $g$ according to the conditional probabilities $P(g'|g)$ and $P_0(g|e)$. The expectation value is then estimated from the final-time distribution by averaging over all trajectories.

We consider unitaries of model C defined by Eq.~\eqref{eq:ModelIII} with $\theta=\pi/3$, initial states $\ket{\psi_\text{e}} = \ket{\psi_\text{o}} =\ket{+}$, and observable $\hat{O}=X$ the Pauli-$X$ operator. 
As the model is in class III, the faithful representation of IM using MPS requires exponential computational resources.
Two limiting cases of $\mathcal{K}$ are simulated: (i) \textit{channel I}, the identity operation, corresponding to a reflecting boundary condition that perfectly sends the output qubit back into the bath, and (ii) \textit{channel II}, the erase-prepare channel with $\rho_\text{re} = \ket{+}\bra{+}$, corresponding to an absorbing boundary condition that erases all information from the bath and induces local thermalization as discussed above.

As shown in Fig.~\ref{fig:stochastic}(a), the two methods agree to high precision for both channels I and II. In particular, channel II leads to the complete decay of the observable, consistent with the convergence of the probability distribution toward the Haar measure.
The decay behavior of $|\langle X(T)\rangle|$ is shown on a finer scale in Fig.~\ref{fig:stochastic}(b). 
The Monte Carlo results develop a plateau near $10^{-4}$, reflecting the expected additive error of $\sim 1/\sqrt{N}$.

To confirm that this deviation is purely statistical and inherently independent of the number of time steps, we alternatively consider $\theta = 0.1$, as shown in Fig.~\ref{fig:stochastic}(c). In this case, both generators $g_{0,1}$ are close to the identity, so the corresponding random walk on the group is expected to relax much more slowly toward the stationary distribution, characterized by a longer mixing time. 
The numerical results—featuring slower decay and pronounced oscillations—support this picture.
Remarkably, since $|\langle X(T)\rangle|$ remains larger than $10^{-4}$ at longer evolution times, the Monte Carlo and MPS results continue to agree perfectly, efficiently capturing multiple dips induced by oscillations throughout the simulation time.

Despite the limitation of additive error, it is worth emphasizing that the Monte Carlo approach differs from MPS simulations in two key aspects:
(i) the computation time scales linearly rather than quadratically with the total number of time steps $T$, and
(ii) it provides a rigorously certified statistical error bound, with the required computational resources scaling quadratically with the desired accuracy, whereas errors in MPS simulations are not \textit{a priori} controlled.

To further investigate the role of the quantum channel, we consider a mixture of channels I and II with rate $0\leq p\leq 1$:
\begin{equation}
\mathcal{K}_\text{mix}[\rho_\text{imp}] = (1-p)\rho_\text{imp} + p\Tr{\rho_\text{imp}}\ket{+}\bra{+},
\end{equation}
where $p=0$ corresponds to purely unitary dynamics of the entire system and $p=1$ complete reset at the boundary.
The stationary expectation value exhibits an approximately linear dependence on $p$ [Fig.~\ref{fig:stochastic}(d)], which warrants future study.

\subsection{Hybrid quantum-classical stochastic process}\label{sec:hybrid}
Next, we consider more general initial states, namely, MPS with bond dimension $D>1$ [Eq.~\eqref{eq:MPSinitial}]. 
In this case, the mapping to a classical stochastic process no longer applies. Nevertheless, the IM still admits a similar description in terms of Markovian embedding.

Specifically, the IM can now be interpreted as a sequence of CPTP maps acting on an enlarged hybrid space consisting of the impurity site, the auxiliary classical state on the PU$(q)$ manifold, and a quantum ancilla associated with the MPS bond. The corresponding local tensor is written as

\begin{equation}
\tikzset{every picture/.style={line width=0.75pt}} 
\begin{tikzpicture}[x=0.75pt,y=0.75pt,yscale=-1,xscale=1]

\draw [fill={rgb, 255:red, 144; green, 19; blue, 254 }  ,fill opacity=1 ][line width=0.75]    (110.51,79.81) -- (138.73,64.95) ;
\draw [shift={(129.93,69.59)}, rotate = 152.23] [color={rgb, 255:red, 0; green, 0; blue, 0 }  ][line width=0.75]    (10.93,-4.9) .. controls (6.95,-2.3) and (3.31,-0.67) .. (0,0) .. controls (3.31,0.67) and (6.95,2.3) .. (10.93,4.9)   ;
\draw [fill={rgb, 255:red, 144; green, 19; blue, 254 }  ,fill opacity=1 ][line width=0.75]    (114.89,118.28) -- (139.87,133.38) ;
\draw [shift={(121.39,122.21)}, rotate = 31.15] [color={rgb, 255:red, 0; green, 0; blue, 0 }  ][line width=0.75]    (10.93,-3.29) .. controls (6.95,-1.4) and (3.31,-0.3) .. (0,0) .. controls (3.31,0.3) and (6.95,1.4) .. (10.93,3.29)   ;
\draw [line width=1.5]    (104.47,128.56) -- (104.34,155.94) ;
\draw [shift={(104.43,135.75)}, rotate = 90.27] [fill={rgb, 255:red, 0; green, 0; blue, 0 }  ][line width=0.08]  [draw opacity=0] (8.13,-3.9) -- (0,0) -- (8.13,3.9) -- cycle    ;
\draw  [color={rgb, 255:red, 0; green, 0; blue, 0 }  ,draw opacity=1 ][fill={rgb, 255:red, 38; green, 173; blue, 95 }  ,fill opacity=1 ][line width=0.75]  (94.78,80.95) .. controls (94.78,75.38) and (99.28,70.85) .. (104.84,70.85) .. controls (110.39,70.85) and (114.89,75.38) .. (114.89,80.95) .. controls (114.89,86.53) and (110.39,91.05) .. (104.84,91.05) .. controls (99.28,91.05) and (94.78,86.53) .. (94.78,80.95) -- cycle ;
\draw  [fill={rgb, 255:red, 229; green, 126; blue, 33 }  ,fill opacity=1 ][line width=0.75]  (104.4,104.62) -- (114.89,118.28) -- (104.4,131.94) -- (93.92,118.28) -- cycle ;
\draw [line width=1.5]    (104.47,44.17) -- (104.34,71.55) ;
\draw [shift={(104.43,51.36)}, rotate = 90.27] [fill={rgb, 255:red, 0; green, 0; blue, 0 }  ][line width=0.08]  [draw opacity=0] (8.13,-3.9) -- (0,0) -- (8.13,3.9) -- cycle    ;
\draw [line width=1.5]    (104.53,89.73) -- (104.4,106.85) ;
\draw  [fill={rgb, 255:red, 208; green, 2; blue, 27 }  ,fill opacity=1 ] (82.23,75.83) .. controls (83.67,75.83) and (84.83,76.99) .. (84.83,78.43) -- (84.82,86.23) .. controls (84.82,87.66) and (83.65,88.83) .. (82.22,88.83) -- (65.54,88.81) .. controls (64.11,88.8) and (62.94,87.64) .. (62.95,86.2) -- (62.96,78.4) .. controls (62.96,76.97) and (64.12,75.8) .. (65.56,75.8) -- cycle ;
\draw  [line width=0.75]  (82.3,81.34) -- (82.29,85.9) -- (77.75,85.89) ;

\draw [fill={rgb, 255:red, 144; green, 19; blue, 254 }  ,fill opacity=1 ][line width=0.75]    (95.5,82.38) -- (85.39,82.42) ;
\draw [line width=2.25]    (73.91,124.89) -- (73.87,152.26) ;
\draw [shift={(73.9,130.28)}, rotate = 90.07] [fill={rgb, 255:red, 0; green, 0; blue, 0 }  ][line width=0.08]  [draw opacity=0] (11.79,-5.66) -- (0,0) -- (11.79,5.66) -- (7.83,0) -- cycle    ;
\draw  [fill={rgb, 255:red, 2; green, 100; blue, 200 }  ,fill opacity=1 ] (82.23,111.75) .. controls (83.67,111.75) and (84.83,112.91) .. (84.83,114.35) -- (84.82,122.15) .. controls (84.82,123.59) and (83.66,124.75) .. (82.22,124.75) -- (65.54,124.73) .. controls (64.11,124.73) and (62.95,123.56) .. (62.95,122.13) -- (62.96,114.33) .. controls (62.96,112.89) and (64.12,111.73) .. (65.56,111.73) -- cycle ;
\draw  [fill={rgb, 255:red, 2; green, 100; blue, 200 }  ,fill opacity=1 ][line width=0.75]  (82.3,117.26) -- (82.29,121.82) -- (77.75,121.82) ;

\draw [fill={rgb, 255:red, 144; green, 19; blue, 254 }  ,fill opacity=1 ][line width=0.75]    (95.5,118.3) -- (85.39,118.34) ;
\draw [line width=2.25]    (73.91,48.48) -- (73.87,75.85) ;
\draw [shift={(73.9,53.87)}, rotate = 90.07] [fill={rgb, 255:red, 0; green, 0; blue, 0 }  ][line width=0.08]  [draw opacity=0] (11.79,-5.66) -- (0,0) -- (11.79,5.66) -- (7.83,0) -- cycle    ;
\draw [line width=2.25]    (73.87,88.4) -- (73.83,112.35) ;

\draw (133.44,136.45) node [anchor=north west][inner sep=0.75pt]  [font=\scriptsize]  {$a,a'$};
\draw (151,80.4) node [anchor=north west][inner sep=0.75pt]  [font=\normalsize]  {$ \begin{array}{l}
=\delta _{a,a'}\sum\limits_{b,c,c'=0}^{q-1} \delta _{g',g_{a} gg_{b}}\\
\ \ \ \ \ \left[ A^{( c)} B^{( b)} \rho _{D} B{^{( b)}}^{\dagger } A{^{( c')}}^{\dagger }\right]\\
\ \ \ \ \ \otimes \left[ u( g')\ket{c}\bra{c'} u( g')^{\dagger }\right],\\
\end{array}$};
\draw (89.69,151.28) node [anchor=north west][inner sep=0.75pt]  [font=\scriptsize]  {$g$};
\draw (89.69,40.66) node [anchor=north west][inner sep=0.75pt]  [font=\scriptsize]  {$g'$};
\draw (53.63,151.35) node [anchor=north west][inner sep=0.75pt]  [font=\scriptsize]  {$\rho _{D}$};
\end{tikzpicture}
\end{equation}
As in the classical case, this map realizes a measure-and-feedback protocol: the impurity site is measured, and, conditioned on the measurement outcome, both the classical state and the quantum ancilla $\rho_D$ are updated.

From the perspective of spacetime mapping, the resulting IM is obtained by applying an LOCC to the initial MPS. More specifically, the IM can be rewritten as a convex mixture of low-bond-dimension MPS, as illustrated below:

\begin{equation}
\tikzset{every picture/.style={line width=0.75pt}} 
\begin{tikzpicture}[x=0.75pt,y=0.75pt,yscale=-1,xscale=1]

\draw [line width=2.25]    (31.92,81.16) -- (31.89,149.11) ;
\draw [line width=2.25]    (31.91,170.14) -- (31.87,213.51) ;
\draw [line width=1.5]    (67.07,78.91) -- (66.86,149.4) ;
\draw [line width=1.5]    (66.81,171.09) -- (66.71,217.37) ;
\draw [fill={rgb, 255:red, 144; green, 19; blue, 254 }  ,fill opacity=1 ][line width=0.75]    (42.7,39.59) -- (90.53,39.57) ;
\draw [line width=1.5]    (67.02,16.98) -- (66.91,61.26) ;
\draw  [fill={rgb, 255:red, 255; green, 255; blue, 255 }  ,fill opacity=1 ][line width=1.5]  (67,221.33) -- (60.15,212.23) -- (73.73,212.1) -- cycle ;
\draw  [line width=1.5]  (59.71,17.86) .. controls (59.71,17.82) and (59.71,17.79) .. (59.71,17.75) .. controls (59.69,12.97) and (62.99,9.08) .. (67.08,9.06) .. controls (71.17,9.04) and (74.51,12.91) .. (74.53,17.69) .. controls (74.53,17.73) and (74.53,17.76) .. (74.53,17.8) -- cycle ;
\draw [fill={rgb, 255:red, 144; green, 19; blue, 254 }  ,fill opacity=1 ][line width=0.75]    (43.34,193.87) -- (90.18,193.85) ;
\draw [fill={rgb, 255:red, 144; green, 19; blue, 254 }  ,fill opacity=1 ][line width=0.75]    (42.7,128.31) -- (72.53,128.29) ;
\draw [fill={rgb, 255:red, 144; green, 19; blue, 254 }  ,fill opacity=1 ][line width=0.75]    (43.34,100.06) -- (90.18,100.04) ;
\draw  [color={rgb, 255:red, 0; green, 0; blue, 0 }  ,draw opacity=1 ][fill={rgb, 255:red, 38; green, 173; blue, 95 }  ,fill opacity=1 ][line width=0.75]  (58.82,128.3) .. controls (58.82,123.88) and (62.4,120.3) .. (66.82,120.3) .. controls (71.24,120.3) and (74.82,123.88) .. (74.82,128.3) .. controls (74.82,132.71) and (71.24,136.3) .. (66.82,136.3) .. controls (62.4,136.3) and (58.82,132.71) .. (58.82,128.3) -- cycle ;
\draw  [color={rgb, 255:red, 0; green, 0; blue, 0 }  ,draw opacity=1 ][fill={rgb, 255:red, 229; green, 126; blue, 33 }  ,fill opacity=1 ][line width=0.75]  (66.76,183.04) -- (75.41,193.86) -- (66.76,204.68) -- (58.11,193.86) -- cycle ;
\draw  [color={rgb, 255:red, 0; green, 0; blue, 0 }  ,draw opacity=1 ][fill={rgb, 255:red, 229; green, 126; blue, 33 }  ,fill opacity=1 ][line width=0.75]  (66.76,89.23) -- (75.41,100.05) -- (66.76,110.88) -- (58.11,100.05) -- cycle ;
\draw  [color={rgb, 255:red, 0; green, 0; blue, 0 }  ,draw opacity=1 ][fill={rgb, 255:red, 38; green, 173; blue, 95 }  ,fill opacity=1 ][line width=0.75]  (59.12,39.58) .. controls (59.12,35.16) and (62.7,31.58) .. (67.12,31.58) .. controls (71.53,31.58) and (75.12,35.16) .. (75.12,39.58) .. controls (75.12,44) and (71.53,47.58) .. (67.12,47.58) .. controls (62.7,47.58) and (59.12,44) .. (59.12,39.58) -- cycle ;
\draw  [fill={rgb, 255:red, 208; green, 2; blue, 27 }  ,fill opacity=1 ] (40.23,122.41) .. controls (41.67,122.41) and (42.83,123.58) .. (42.83,125.01) -- (42.82,132.81) .. controls (42.82,134.25) and (41.65,135.41) .. (40.22,135.41) -- (23.54,135.39) .. controls (22.11,135.39) and (20.94,134.22) .. (20.95,132.79) -- (20.96,124.98) .. controls (20.96,123.55) and (22.12,122.39) .. (23.56,122.39) -- cycle ;
\draw  [line width=0.75]  (40.3,127.92) -- (40.29,132.48) -- (35.75,132.48) ;

\draw [line width=2.25]    (31.88,23.88) -- (31.85,60.25) ;
\draw  [fill={rgb, 255:red, 2; green, 100; blue, 200 }  ,fill opacity=1 ] (40.23,187.33) .. controls (41.67,187.33) and (42.83,188.5) .. (42.83,189.93) -- (42.82,197.73) .. controls (42.82,199.17) and (41.66,200.33) .. (40.22,200.33) -- (23.54,200.31) .. controls (22.11,200.31) and (20.95,199.15) .. (20.95,197.71) -- (20.96,189.91) .. controls (20.96,188.47) and (22.12,187.31) .. (23.56,187.31) -- cycle ;
\draw  [fill={rgb, 255:red, 2; green, 100; blue, 200 }  ,fill opacity=1 ][line width=0.75]  (40.3,192.84) -- (40.29,197.4) -- (35.75,197.4) ;

\draw [fill={rgb, 255:red, 144; green, 19; blue, 254 }  ,fill opacity=1 ][line width=0.75]    (89.93,128.26) -- (74.82,128.3) ;
\draw  [fill={rgb, 255:red, 2; green, 100; blue, 200 }  ,fill opacity=1 ] (40.21,93.33) .. controls (41.65,93.33) and (42.81,94.5) .. (42.81,95.93) -- (42.8,103.73) .. controls (42.8,105.17) and (41.63,106.33) .. (40.2,106.33) -- (23.52,106.31) .. controls (22.09,106.31) and (20.92,105.15) .. (20.93,103.71) -- (20.93,95.91) .. controls (20.94,94.47) and (22.1,93.31) .. (23.54,93.31) -- cycle ;
\draw  [fill={rgb, 255:red, 2; green, 100; blue, 200 }  ,fill opacity=1 ][line width=0.75]  (40.27,98.84) -- (40.27,103.4) -- (35.73,103.4) ;

\draw  [fill={rgb, 255:red, 208; green, 2; blue, 27 }  ,fill opacity=1 ] (40.21,33.07) .. controls (41.65,33.08) and (42.81,34.24) .. (42.81,35.68) -- (42.8,43.48) .. controls (42.8,44.91) and (41.63,46.08) .. (40.2,46.08) -- (23.52,46.06) .. controls (22.09,46.05) and (20.92,44.89) .. (20.92,43.45) -- (20.93,35.65) .. controls (20.94,34.22) and (22.1,33.05) .. (23.54,33.05) -- cycle ;
\draw  [line width=0.75]  (40.27,38.59) -- (40.27,43.15) -- (35.73,43.14) ;

\draw  [fill={rgb, 255:red, 0; green, 0; blue, 0 }  ,fill opacity=1 ] (28.38,23.38) .. controls (28.38,21.45) and (29.95,19.88) .. (31.88,19.88) .. controls (33.82,19.88) and (35.38,21.45) .. (35.38,23.38) .. controls (35.38,25.31) and (33.82,26.88) .. (31.88,26.88) .. controls (29.95,26.88) and (28.38,25.31) .. (28.38,23.38) -- cycle ;
\draw  [fill={rgb, 255:red, 255; green, 255; blue, 255 }  ,fill opacity=1 ] (202.02,16.8) -- (210.02,16.8) -- (210.02,24.8) -- (202.02,24.8) -- cycle ;
\draw  [fill={rgb, 255:red, 255; green, 255; blue, 255 }  ,fill opacity=1 ] (202.29,54.59) -- (210.29,54.59) -- (210.29,62.59) -- (202.29,62.59) -- cycle ;
\draw [line width=2.25]    (170.92,78.98) -- (170.89,154.93) ;
\draw [line width=2.25]    (170.91,175.96) -- (170.87,213.33) ;
\draw [line width=1.5]    (206.07,114.73) -- (205.86,148.22) ;
\draw [fill={rgb, 255:red, 144; green, 19; blue, 254 }  ,fill opacity=1 ][line width=0.75]    (181.7,39.41) -- (229.53,39.39) ;
\draw [line width=1.5]    (206.02,24.8) -- (205.91,55.08) ;
\draw [fill={rgb, 255:red, 144; green, 19; blue, 254 }  ,fill opacity=1 ][line width=0.75]    (182.34,193.69) -- (192.7,193.7) ;
\draw [fill={rgb, 255:red, 144; green, 19; blue, 254 }  ,fill opacity=1 ][line width=0.75]    (181.7,131.13) -- (211.53,131.11) ;
\draw  [color={rgb, 255:red, 0; green, 0; blue, 0 }  ,draw opacity=1 ][fill={rgb, 255:red, 38; green, 173; blue, 95 }  ,fill opacity=1 ][line width=0.75]  (197.82,131.12) .. controls (197.82,126.7) and (201.4,123.12) .. (205.82,123.12) .. controls (210.24,123.12) and (213.82,126.7) .. (213.82,131.12) .. controls (213.82,135.54) and (210.24,139.12) .. (205.82,139.12) .. controls (201.4,139.12) and (197.82,135.54) .. (197.82,131.12) -- cycle ;
\draw  [color={rgb, 255:red, 0; green, 0; blue, 0 }  ,draw opacity=1 ][fill={rgb, 255:red, 38; green, 173; blue, 95 }  ,fill opacity=1 ][line width=0.75]  (198.12,39.4) .. controls (198.12,34.99) and (201.7,31.4) .. (206.12,31.4) .. controls (210.53,31.4) and (214.12,34.99) .. (214.12,39.4) .. controls (214.12,43.82) and (210.53,47.4) .. (206.12,47.4) .. controls (201.7,47.4) and (198.12,43.82) .. (198.12,39.4) -- cycle ;
\draw  [fill={rgb, 255:red, 208; green, 2; blue, 27 }  ,fill opacity=1 ] (179.23,125.23) .. controls (180.67,125.23) and (181.83,126.4) .. (181.83,127.83) -- (181.82,135.63) .. controls (181.82,137.07) and (180.65,138.23) .. (179.22,138.23) -- (162.54,138.21) .. controls (161.11,138.21) and (159.94,137.04) .. (159.95,135.61) -- (159.96,127.81) .. controls (159.96,126.37) and (161.12,125.21) .. (162.56,125.21) -- cycle ;
\draw  [line width=0.75]  (179.3,130.74) -- (179.29,135.3) -- (174.75,135.3) ;

\draw [line width=2.25]    (170.88,23.7) -- (170.85,60.07) ;
\draw  [fill={rgb, 255:red, 2; green, 100; blue, 200 }  ,fill opacity=1 ] (179.23,187.15) .. controls (180.67,187.15) and (181.83,188.32) .. (181.83,189.75) -- (181.82,197.55) .. controls (181.82,198.99) and (180.66,200.15) .. (179.22,200.15) -- (162.54,200.13) .. controls (161.11,200.13) and (159.95,198.97) .. (159.95,197.53) -- (159.96,189.73) .. controls (159.96,188.3) and (161.12,187.13) .. (162.56,187.13) -- cycle ;
\draw  [fill={rgb, 255:red, 2; green, 100; blue, 200 }  ,fill opacity=1 ][line width=0.75]  (179.3,192.66) -- (179.29,197.23) -- (174.75,197.22) ;

\draw [fill={rgb, 255:red, 144; green, 19; blue, 254 }  ,fill opacity=1 ][line width=0.75]    (228.93,131.08) -- (213.82,131.12) ;
\draw  [fill={rgb, 255:red, 2; green, 100; blue, 200 }  ,fill opacity=1 ] (179.21,84.15) .. controls (180.65,84.15) and (181.81,85.32) .. (181.81,86.75) -- (181.8,94.55) .. controls (181.8,95.99) and (180.63,97.15) .. (179.2,97.15) -- (162.52,97.13) .. controls (161.09,97.13) and (159.92,95.97) .. (159.93,94.53) -- (159.93,86.73) .. controls (159.94,85.3) and (161.1,84.13) .. (162.54,84.13) -- cycle ;
\draw  [fill={rgb, 255:red, 2; green, 100; blue, 200 }  ,fill opacity=1 ][line width=0.75]  (179.27,89.66) -- (179.27,94.23) -- (174.73,94.22) ;

\draw  [fill={rgb, 255:red, 208; green, 2; blue, 27 }  ,fill opacity=1 ] (179.21,32.9) .. controls (180.65,32.9) and (181.81,34.06) .. (181.81,35.5) -- (181.8,43.3) .. controls (181.8,44.74) and (180.63,45.9) .. (179.2,45.9) -- (162.52,45.88) .. controls (161.09,45.87) and (159.92,44.71) .. (159.92,43.27) -- (159.93,35.47) .. controls (159.94,34.04) and (161.1,32.87) .. (162.54,32.88) -- cycle ;
\draw  [line width=0.75]  (179.27,38.41) -- (179.27,42.97) -- (174.73,42.96) ;

\draw  [fill={rgb, 255:red, 0; green, 0; blue, 0 }  ,fill opacity=1 ] (167.38,23.2) .. controls (167.38,21.27) and (168.95,19.7) .. (170.88,19.7) .. controls (172.82,19.7) and (174.38,21.27) .. (174.38,23.2) .. controls (174.38,25.13) and (172.82,26.7) .. (170.88,26.7) .. controls (168.95,26.7) and (167.38,25.13) .. (167.38,23.2) -- cycle ;
\draw   (206.7,193.54) -- (192.82,202.7) -- (192.73,184.5) -- cycle ;
\draw   (211.33,193.58) -- (225.8,183.16) -- (225.8,204) -- cycle ;
\draw [fill={rgb, 255:red, 144; green, 19; blue, 254 }  ,fill opacity=1 ][line width=0.75]    (225.34,193.69) -- (235.7,194.04) ;
\draw [fill={rgb, 255:red, 144; green, 19; blue, 254 }  ,fill opacity=1 ][line width=0.75]    (182.34,91.63) -- (192.7,91.7) ;
\draw   (206.7,91.48) -- (192.82,100.64) -- (192.73,82.43) -- cycle ;
\draw   (211.33,91.52) -- (225.8,81.1) -- (225.8,101.93) -- cycle ;
\draw [fill={rgb, 255:red, 144; green, 19; blue, 254 }  ,fill opacity=1 ][line width=0.75]    (225.34,91.63) -- (235.7,91.97) ;
\draw  [fill={rgb, 255:red, 255; green, 255; blue, 255 }  ,fill opacity=1 ] (201.86,146.22) -- (209.86,146.22) -- (209.86,154.22) -- (201.86,154.22) -- cycle ;
\draw  [fill={rgb, 255:red, 255; green, 255; blue, 255 }  ,fill opacity=1 ] (202.07,106.73) -- (210.07,106.73) -- (210.07,114.73) -- (202.07,114.73) -- cycle ;

\draw (68.77,152.17) node [anchor=north west][inner sep=0.75pt]  [rotate=-90]  {$...$};
\draw (68.77,63.17) node [anchor=north west][inner sep=0.75pt]  [rotate=-90]  {$...$};
\draw (33.77,152.17) node [anchor=north west][inner sep=0.75pt]  [rotate=-90]  {$...$};
\draw (33.77,64.17) node [anchor=north west][inner sep=0.75pt]  [rotate=-90]  {$...$};
\draw (93,98.73) node [anchor=north west][inner sep=0.75pt]  [font=\normalsize]  {$ \begin{array}{l}
=\sum\limits_{\{a_{t} ,b_{t}\}}\\
\ \ \ \ \ \\
\end{array}$};
\draw (207.77,158.99) node [anchor=north west][inner sep=0.75pt]  [rotate=-90]  {$...$};
\draw (207.77,66.99) node [anchor=north west][inner sep=0.75pt]  [rotate=-90]  {$...$};
\draw (172.77,159.32) node [anchor=north west][inner sep=0.75pt]  [rotate=-90]  {$...$};
\draw (172.77,61.99) node [anchor=north west][inner sep=0.75pt]  [rotate=-90]  {$...$};
\draw (192.28,188.7) node [anchor=north west][inner sep=0.75pt]  [font=\tiny]  {$b_{1}$};
\draw (213.33,189.6) node [anchor=north west][inner sep=0.75pt]  [font=\tiny]  {$a_{1}$};
\draw (192.28,87.29) node [anchor=north west][inner sep=0.75pt]  [font=\tiny]  {$b_{t}$};
\draw (213.33,87.43) node [anchor=north west][inner sep=0.75pt]  [font=\tiny]  {$a_{t}$};
\draw (178.33,108.49) node [anchor=north west][inner sep=0.75pt]  [font=\scriptsize]  {$g_{t-1}$};
\draw (179,144.49) node [anchor=north west][inner sep=0.75pt]  [font=\scriptsize]  {$g_{t-1}$};
\draw (186,15.82) node [anchor=north west][inner sep=0.75pt]  [font=\scriptsize]  {$g_{T}$};
\draw (186,51.4) node [anchor=north west][inner sep=0.75pt]  [font=\scriptsize]  {$g_{T}$};
\end{tikzpicture}
\end{equation}
This follows by decomposing the diamond tensor in Fig.~\ref{fig:MPO_tensors}(c) into its elements on the basis $a_t$ and $b_t$. Each such component specifies a trajectory of group elements generated iteratively according to the update rule $g_t = g_{a_t}g_{t-1}g_{b_t}$. The hollow squares represent projections onto the corresponding group elements, which serve as basis states of the group algebra. As a result, each component MPS in the summation is obtained from the initial MPS by local measurements, followed by on-site unitaries (action of circle tensors) conditioned on the measurement outcomes. 

This decomposition immediately implies that the entanglement of formation of IM is upper bounded by the bipartite entanglement of the initial MPS~\cite{Bennett1996EOF}. In this sense, the entanglement in the initial MPS upper bounds the genuinely quantum correlations in the IM.

Indeed, the expectation value of any observable or correlation function can be expressed as an average over the individual MPS components. The value of each component remains inexpensive to evaluate, since the MPS bond dimension is no larger than that of the initial one. Sampling the component MPS requires only computing their normalization weights, which is equivalent to sampling the bit-string distribution $b_1b_2\cdots b_T$ from the initial MPS.
For fixed bond dimension $D$, each component can be sampled sequentially with a cost linear in $T$~\cite{Popp2005Localizable, White2009Minimally, Stoudenmire2010Minimally}. Similar to the purely classical case, the observable expectation value can, therefore, be estimated by Monte Carlo sampling, with the number of samples required to achieve an additive error $\epsilon$ scaling as $O(\epsilon^{-2})$, up to a prefactor determined by the variance of the observable.

One may further regard the bipartite entanglement of the initial MPS as a candidate for genuine quantum memory. However, it lacks direct operational meaning, and identifying genuine quantum memory remains challenging, especially in multitime settings~\cite{Milz2020When,Giarmatzi2021Witnessing,Milz2021Genuine,Backer2024Local,Gangwar2025Squashed}. In the next part, we therefore introduce a new operational measure of quantum memory that is naturally compatible with tensor-network techniques, and demonstrate its utility analytically for our model with a specific class of initial MPS.

\subsection{Measure of quantum memory}\label{sec:quantum_memory}

Our approach connects the notion of quantum memory to the information-theoretic task of quantum teleportation. We introduce a new concept of \textit{teleportable entanglement}, defined as the maximal amount of quantum entanglement that can be teleported forward in time by leveraging only the memory encoded in the IM. It naturally depends on the number of time steps, and a subexponential decay implies the presence of long-time quantum memory. Importantly, since it involves operations only on the system degrees of freedom, it does not require direct access to the bath, making it particularly natural for experiments in which the bath is typically inaccessible.

The teleportable entanglement is defined as follows. Consider Alice ($A$) and Bob ($B$) who share a two-qudit entangled state. The qudit $A$ is injected into the qudit chain as an impurity and undergoes the joint dynamics, as illustrated in Fig.~\ref{fig:setting}(c).
The quantum operations on the impurity, performed by experimentalists, are then chosen to be projective measurements followed by local-in-time reinitialization, which prevent any direct transfer of entanglement carried by the qudit $A$ itself.
After multiple time steps $T$, this procedure leads to a sequence of measurement outcomes $M = \{m_1,m_2,\dots,m_{T-1}\}$, and an unnormalized postmeasurement state of the two qudits:
\begin{equation}
\tikzset{every picture/.style={line width=0.75pt}} 
\begin{tikzpicture}[x=0.75pt,y=0.75pt,yscale=-1,xscale=1]

\draw [fill={rgb, 255:red, 144; green, 19; blue, 254 }  ,fill opacity=1 ][line width=0.75]    (136.57,110) -- (160,110) ;
\draw [fill={rgb, 255:red, 144; green, 19; blue, 254 }  ,fill opacity=1 ][line width=0.75]    (160,83) -- (136.57,83) ;
\draw [fill={rgb, 255:red, 144; green, 19; blue, 254 }  ,fill opacity=1 ][line width=0.75]    (137,53) -- (160,53) ;
\draw [fill={rgb, 255:red, 144; green, 19; blue, 254 }  ,fill opacity=1 ][line width=0.75]    (152,22.87) -- (137,36) ;
\draw [fill={rgb, 255:red, 144; green, 19; blue, 254 }  ,fill opacity=1 ][line width=0.75]    (160,140) -- (137.33,139.8) ;
\draw  [color={rgb, 255:red, 0; green, 0; blue, 0 }  ,draw opacity=1 ][fill={rgb, 255:red, 35; green, 245; blue, 240 }  ,fill opacity=0.4 ] (90,22) -- (137,22) -- (137,200) -- (90,200) -- cycle ;
\draw [line width=2.25]    (136.57,187) -- (155.57,187) ;
\draw  [fill={rgb, 255:red, 32; green, 136; blue, 32 }  ,fill opacity=1 ] (147.56,181.85) .. controls (147.58,179.73) and (149.32,178.01) .. (151.44,178.03) -- (186.24,178.3) .. controls (188.36,178.31) and (190.08,180.05) .. (190.06,182.18) -- (189.97,193.74) .. controls (189.95,195.87) and (188.22,197.58) .. (186.09,197.56) -- (151.3,197.3) .. controls (149.17,197.28) and (147.46,195.54) .. (147.47,193.41) -- cycle ;
\draw  [fill={rgb, 255:red, 32; green, 136; blue, 32 }  ,fill opacity=1 ][line width=0.75]  (165.59,180.37) -- (180.5,180.48) -- (180.47,184.48) ;

\draw [fill={rgb, 255:red, 144; green, 19; blue, 254 }  ,fill opacity=1 ][line width=0.75]    (137,159) -- (155.57,178) ;
\draw    (200,170) -- (177,178) ;
\draw   (160,133.25) -- (180,133.25) -- (180,147.15) -- (160,147.15) -- cycle ;
\draw  [draw opacity=0] (161.5,145.54) .. controls (162.44,141.87) and (165.87,139.17) .. (169.92,139.22) .. controls (174.01,139.27) and (177.39,142.09) .. (178.21,145.82) -- (169.82,147.56) -- cycle ; \draw   (161.5,145.54) .. controls (162.44,141.87) and (165.87,139.17) .. (169.92,139.22) .. controls (174.01,139.27) and (177.39,142.09) .. (178.21,145.82) ;  
\draw    (169.71,144.37) -- (171.56,136.07) ;
\draw [shift={(172,134.11)}, rotate = 102.57] [color={rgb, 255:red, 0; green, 0; blue, 0 }  ][line width=0.75]    (4.37,-1.96) .. controls (2.78,-0.92) and (1.32,-0.27) .. (0,0) .. controls (1.32,0.27) and (2.78,0.92) .. (4.37,1.96)   ;

\draw   (160,75) -- (180,75) -- (180,88.89) -- (160,88.89) -- cycle ;
\draw  [draw opacity=0] (161.5,87.28) .. controls (162.44,83.62) and (165.87,80.92) .. (169.92,80.97) .. controls (174.01,81.02) and (177.39,83.84) .. (178.21,87.57) -- (169.82,89.31) -- cycle ; \draw   (161.5,87.28) .. controls (162.44,83.62) and (165.87,80.92) .. (169.92,80.97) .. controls (174.01,81.02) and (177.39,83.84) .. (178.21,87.57) ;  
\draw    (169.71,86.12) -- (171.56,77.81) ;
\draw [shift={(172,75.86)}, rotate = 102.57] [color={rgb, 255:red, 0; green, 0; blue, 0 }  ][line width=0.75]    (4.37,-1.96) .. controls (2.78,-0.92) and (1.32,-0.27) .. (0,0) .. controls (1.32,0.27) and (2.78,0.92) .. (4.37,1.96)   ;

\draw   (174.72,52.52) -- (160.3,63) -- (160.21,42.16) -- cycle ;
\draw   (174.42,110.52) -- (160,121) -- (159.91,100.16) -- cycle ;

\draw (101,90) node [anchor=north west][inner sep=0.75pt]   [align=left] {left \\IM};
\draw (156,10) node [anchor=north west][inner sep=0.75pt]   [align=left] {$A$};
\draw (205,160) node [anchor=north west][inner sep=0.75pt]   [align=left] {$B$};
\draw (162.21,48.66) node [anchor=north west][inner sep=0.75pt]  [font=\footnotesize] [align=left] {r};
\draw (162.21,106.66) node [anchor=north west][inner sep=0.75pt]  [font=\footnotesize] [align=left] {r};
\draw (36,102.4) node [anchor=north west][inner sep=0.75pt]  [font=\normalsize]  {$\tilde{\rho }_{M}^{AB} =$};
\draw (185,133.4) node [anchor=north west][inner sep=0.75pt]  [font=\small]  {$m_{1}$};
\draw (185,73.4) node [anchor=north west][inner sep=0.75pt]  [font=\small]  {$m_{2}$};
\end{tikzpicture}.
\end{equation}
Here, we depict $T=3$ as an example.
Triangular tensors with symbol r denote reinitialized states and can be time independent. In this way, we extract quantum memory encoded in the IM into quantum entanglement of postmeasurement states.

We define the teleportable entanglement as the maximal average entanglement over all measurement outcomes:
\begin{equation}
    \mathbb{E}_\text{TPE}\equiv \max_\mathcal{C}\sum_M p_M E(\rho_M^\text{AB}).
\end{equation}
Here, $p_M = \Tr{\tilde{\rho}_M^{AB}}$ is the probability of outcome $M$ with $\sum_M p_M = 1$, $\rho_M^{AB} = \tilde{\rho}_M^{AB}/p_M$ is the normalized postmeasurement state. The function $E$ is any valid bipartite entanglement measure that vanishes on separable states. 
Since the postmeasurement state is generally mixed, here we use the negativity $E^N$ \cite{Peres1996Separability, Horodecki1996Separability, Vidal2002Computable, Plenio2005Logarithmic}, defined via the partial transpose as
\begin{equation}\label{eq:negativity}
    E^N(\rho_M^{AB}) = [||(\rho_M^{AB})^{\mathrm{T}_{{A}}}||_1-1]/2.
\end{equation}
The maximization is taken over both the class $\mathcal{C}$ of measurement bases and prepared states and the choice of initial two-qudit state. 

Our definition is closely analogous to the notion of localizable entanglement for spatial many-body states~\cite{Verstraete2004Entanglement, Verstraete2004Diverging, Popp2005Localizable}. In that context, localizable entanglement quantifies the maximal amount of entanglement of a quantum state that can be concentrated between two distant qudits A and B by performing local projective measurements on all remaining qudits. The measurements generate an ensemble of postmeasurement states on A and B conditioned on measurement outcomes $M$, together with probabilities: $\{\rho_M^{AB},p_M\}$. The localizable entanglement is then defined as 
\begin{equation}
    \mathbb{E}_\text{LE}\equiv \max_\mathcal{C}\sum_M p_M E(\rho_M^{AB}),
\end{equation}
where the maximization is taken over the class $\mathcal{C}$ of measurement bases.

A key advantage of localizable entanglement is that, by allowing measurements on the remaining degrees of freedom, it can reveal more refined structures of correlations such as long-range string orders in the 1D cluster state ~\cite{Briegel2001Persistent,Verstraete2004Valence} and the ground state of the Affleck-Kennedy-Lieb-Tasaki model \cite{Affleck1988vValence,Verstraete2004Diverging}. Analogously, with measurements and reinitializations at intermediate time steps, 
teleportable entanglement naturally inherits this feature, as we illustrate below with examples. Moreover, the sampling method used to evaluate localizable entanglement in MPS~\cite{Popp2005Localizable} carries over naturally to teleportable entanglement for IM represented as MPS in our settings.

In the case of product-state IM, the postmeasurement state is a product state. For classical IM, correlations between $A$ and $B$ can be mediated only through the inner bond which carries a classical probabilistic distribution, resulting in the absence of quantum entanglement in postmeasurement states. For general initial states, however, computing the teleportable entanglement becomes intractable, as it involves averaging over exponentially many measurement outcomes and optimizing over a high-dimensional continuous parameter space of operations. The latter quickly becomes computationally prohibitive as the evolution time increases.
To overcome these challenges, we consider a class of initial states and controlled unitaries that admit analytical treatment of the teleportable entanglement. Although simplified, such toy models provide valuable insight into the proposed measure of quantum memory.

Specifically, the controlled unitaries are taken as
\begin{equation}\label{eq:semiproduct_unitary}
    u_a = D_a P_a,~ a = 0,1,\dots,q-1,
\end{equation}
where $D_a\in U(1)^{\otimes q}$ are diagonal phase matrices and $P_a\in S_q$ are permutation matrices. From the group-theoretic perspective, the associated subgroup $H$ generated by the projective correspondence takes the form of Eq.~\eqref{eq:semiproduct}, which typically leads to logarithmic growth of TEE. 
To simplify the derivation, we consider a specific class of initial states chosen as a product of pure states $\ket{0}\bra{0}$ on odd sites and an MPS on even sites. The MPS acts on the auxiliary Hilbert space as a chain of controlled unitaries $w^{a}$. Using the notations of Eq.~\eqref{eq:MPSinitial}, the MPS tensors can be written as
\begin{align}
A_{jk}^{a} = \alpha_a w^{a}_{jk},\quad &{w^{a}}^\dagger w^{a} = \mathbbm{1},\quad \sum_{a=0}^{q-1}|\alpha_a|^2 = 1,\nonumber\\
    &B_{jk}^{b} = \delta_{b,0}\delta_{jk}.
\label{eq:MPS_teleportable_entanglement}
\end{align}
The measurements are chosen to be performed in the computational basis $\{\ket{a}\}_{a=0}^{q-1}$, with reinitialized states $\rho_\text{re}=\ket{0}\bra{0}$, while the pure states on odd sites and the reinitialized states could, in principle, be inhomogeneous and correspond to different computational basis elements; we set all of them to $\ket{0}\bra{0}$ for simplicity. Below, we demonstrate that the successive measurements induce the action of unitary transformations on the auxiliary Hilbert space leading to the lossless transmission of the quantum information from $B$ to $A$.

The diagrammatic representation of the postmeasurement state reads:
\begin{equation}
\tikzset{every picture/.style={line width=0.75pt}} 
\begin{tikzpicture}[x=0.75pt,y=0.75pt,yscale=-1,xscale=1]

\draw [line width=1.5]    (276.11,137.72) -- (276,171) ;
\draw [line width=1.5]    (276.4,80.55) -- (276.29,113.83) ;
\draw [fill={rgb, 255:red, 144; green, 19; blue, 254 }  ,fill opacity=1 ][line width=0.75]    (156.19,38) -- (127.77,38.25) ;
\draw  [fill={rgb, 255:red, 32; green, 136; blue, 32 }  ,fill opacity=1 ] (164.12,214.03) .. controls (164.13,211.9) and (165.87,210.19) .. (168,210.2) -- (180.29,210.3) .. controls (182.42,210.31) and (184.13,212.05) .. (184.12,214.18) -- (184.03,225.74) .. controls (184.01,227.87) and (182.27,229.58) .. (180.15,229.56) -- (167.85,229.47) .. controls (165.73,229.45) and (164.01,227.71) .. (164.03,225.59) -- cycle ;
\draw  [fill={rgb, 255:red, 32; green, 136; blue, 32 }  ,fill opacity=1 ][line width=0.75]  (172.61,212.47) -- (179.63,212.52) -- (179.6,216.52) ;

\draw    (184,200) -- (174.06,210) ;
\draw   (153.19,146.25) -- (173.19,146.25) -- (173.19,160.15) -- (153.19,160.15) -- cycle ;
\draw  [draw opacity=0] (154.69,158.54) .. controls (155.63,154.87) and (159.06,152.17) .. (163.12,152.22) .. controls (167.2,152.27) and (170.58,155.09) .. (171.4,158.82) -- (163.01,160.56) -- cycle ; \draw   (154.69,158.54) .. controls (155.63,154.87) and (159.06,152.17) .. (163.12,152.22) .. controls (167.2,152.27) and (170.58,155.09) .. (171.4,158.82) ;  
\draw    (162.91,157.37) -- (164.76,149.07) ;
\draw [shift={(165.19,147.11)}, rotate = 102.57] [color={rgb, 255:red, 0; green, 0; blue, 0 }  ][line width=0.75]    (4.37,-1.96) .. controls (2.78,-0.92) and (1.32,-0.27) .. (0,0) .. controls (1.32,0.27) and (2.78,0.92) .. (4.37,1.96)   ;

\draw   (151.19,89) -- (171.19,89) -- (171.19,102.89) -- (151.19,102.89) -- cycle ;
\draw  [draw opacity=0] (152.69,101.28) .. controls (153.63,97.62) and (157.06,94.92) .. (161.12,94.97) .. controls (165.2,95.02) and (168.58,97.84) .. (169.4,101.57) -- (161.01,103.31) -- cycle ; \draw   (152.69,101.28) .. controls (153.63,97.62) and (157.06,94.92) .. (161.12,94.97) .. controls (165.2,95.02) and (168.58,97.84) .. (169.4,101.57) ;  
\draw    (160.91,100.12) -- (162.76,91.81) ;
\draw [shift={(163.19,89.86)}, rotate = 102.57] [color={rgb, 255:red, 0; green, 0; blue, 0 }  ][line width=0.75]    (4.37,-1.96) .. controls (2.78,-0.92) and (1.32,-0.27) .. (0,0) .. controls (1.32,0.27) and (2.78,0.92) .. (4.37,1.96)   ;

\draw   (164.91,65.52) -- (150.49,76) -- (150.4,55.16) -- cycle ;
\draw   (165.19,124.52) -- (150.77,135) -- (150.68,114.16) -- cycle ;
\draw [fill={rgb, 255:red, 144; green, 19; blue, 254 }  ,fill opacity=1 ][line width=0.75]    (91.36,38.26) -- (134.19,38.24) ;
\draw [line width=1.5]    (127.68,15.65) -- (127.57,199.93) ;
\draw  [line width=1.5]  (127.66,210) -- (120.81,200.89) -- (134.39,200.77) -- cycle ;
\draw  [line width=1.5]  (120.37,16.52) .. controls (120.37,16.49) and (120.37,16.45) .. (120.37,16.42) .. controls (120.35,11.64) and (123.65,7.75) .. (127.74,7.73) .. controls (131.83,7.71) and (135.17,11.57) .. (135.19,16.36) .. controls (135.19,16.39) and (135.19,16.43) .. (135.19,16.47) -- cycle ;
\draw [fill={rgb, 255:red, 144; green, 19; blue, 254 }  ,fill opacity=1 ][line width=0.75]    (104,182.54) -- (150.84,182.52) ;
\draw [fill={rgb, 255:red, 144; green, 19; blue, 254 }  ,fill opacity=1 ][line width=0.75]    (91.06,152.97) -- (152.9,152.95) ;
\draw [fill={rgb, 255:red, 144; green, 19; blue, 254 }  ,fill opacity=1 ][line width=0.75]    (90.36,96.07) -- (151.19,96.05) ;
\draw [fill={rgb, 255:red, 144; green, 19; blue, 254 }  ,fill opacity=1 ][line width=0.75]    (104,66.92) -- (150.84,66.9) ;
\draw [fill={rgb, 255:red, 144; green, 19; blue, 254 }  ,fill opacity=1 ][line width=0.75]    (104,124.73) -- (150.84,124.71) ;
\draw  [color={rgb, 255:red, 0; green, 0; blue, 0 }  ,draw opacity=1 ][fill={rgb, 255:red, 38; green, 173; blue, 95 }  ,fill opacity=1 ][line width=0.75]  (119.48,152.96) .. controls (119.48,148.55) and (123.06,144.96) .. (127.48,144.96) .. controls (131.9,144.96) and (135.48,148.55) .. (135.48,152.96) .. controls (135.48,157.38) and (131.9,160.96) .. (127.48,160.96) .. controls (123.06,160.96) and (119.48,157.38) .. (119.48,152.96) -- cycle ;
\draw  [color={rgb, 255:red, 0; green, 0; blue, 0 }  ,draw opacity=1 ][fill={rgb, 255:red, 229; green, 126; blue, 33 }  ,fill opacity=1 ][line width=0.75]  (127.42,171.71) -- (136.07,182.53) -- (127.42,193.35) -- (118.77,182.53) -- cycle ;
\draw  [color={rgb, 255:red, 0; green, 0; blue, 0 }  ,draw opacity=1 ][fill={rgb, 255:red, 229; green, 126; blue, 33 }  ,fill opacity=1 ][line width=0.75]  (127.42,56.09) -- (136.07,66.91) -- (127.42,77.73) -- (118.77,66.91) -- cycle ;
\draw  [color={rgb, 255:red, 0; green, 0; blue, 0 }  ,draw opacity=1 ][fill={rgb, 255:red, 229; green, 126; blue, 33 }  ,fill opacity=1 ][line width=0.75]  (127.42,113.9) -- (136.07,124.72) -- (127.42,135.54) -- (118.77,124.72) -- cycle ;
\draw  [color={rgb, 255:red, 0; green, 0; blue, 0 }  ,draw opacity=1 ][fill={rgb, 255:red, 38; green, 173; blue, 95 }  ,fill opacity=1 ][line width=0.75]  (119.77,38.25) .. controls (119.77,33.83) and (123.36,30.25) .. (127.77,30.25) .. controls (132.19,30.25) and (135.77,33.83) .. (135.77,38.25) .. controls (135.77,42.67) and (132.19,46.25) .. (127.77,46.25) .. controls (123.36,46.25) and (119.77,42.67) .. (119.77,38.25) -- cycle ;
\draw  [color={rgb, 255:red, 0; green, 0; blue, 0 }  ,draw opacity=1 ][fill={rgb, 255:red, 38; green, 173; blue, 95 }  ,fill opacity=1 ][line width=0.75]  (119.77,96.06) .. controls (119.77,91.64) and (123.36,88.06) .. (127.77,88.06) .. controls (132.19,88.06) and (135.77,91.64) .. (135.77,96.06) .. controls (135.77,100.48) and (132.19,104.06) .. (127.77,104.06) .. controls (123.36,104.06) and (119.77,100.48) .. (119.77,96.06) -- cycle ;
\draw   (164.91,181.52) -- (150.49,192) -- (150.4,171.16) -- cycle ;
\draw   (90,66.58) -- (104.46,56.16) -- (104.46,77) -- cycle ;
\draw   (89,124.58) -- (103.46,114.16) -- (103.46,135) -- cycle ;
\draw   (90,181.58) -- (104.46,171.16) -- (104.46,192) -- cycle ;
\draw [line width=2.25]    (81.42,22.08) -- (81.22,191.08) ;
\draw  [fill={rgb, 255:red, 208; green, 2; blue, 27 }  ,fill opacity=1 ] (89.49,32.93) .. controls (90.75,32.94) and (91.76,33.97) .. (91.75,35.23) -- (91.7,42.07) .. controls (91.69,43.33) and (90.66,44.34) .. (89.4,44.33) -- (74.69,44.22) .. controls (73.44,44.21) and (72.42,43.18) .. (72.43,41.93) -- (72.48,35.09) .. controls (72.49,33.83) and (73.52,32.81) .. (74.78,32.82) -- cycle ;
\draw  [line width=0.75]  (89.5,37.77) -- (89.47,41.77) -- (85.47,41.74) ;

\draw  [fill={rgb, 255:red, 208; green, 2; blue, 27 }  ,fill opacity=1 ] (89.09,146.93) .. controls (90.35,146.94) and (91.36,147.97) .. (91.35,149.23) -- (91.3,156.07) .. controls (91.29,157.33) and (90.26,158.34) .. (89,158.33) -- (74.3,158.22) .. controls (73.04,158.21) and (72.03,157.19) .. (72.03,155.93) -- (72.09,149.09) .. controls (72.1,147.83) and (73.12,146.81) .. (74.38,146.82) -- cycle ;
\draw  [line width=0.75]  (89.1,151.77) -- (89.07,155.77) -- (85.07,155.74) ;

\draw  [fill={rgb, 255:red, 208; green, 2; blue, 27 }  ,fill opacity=1 ] (89.07,89.99) .. controls (90.33,90) and (91.34,91.03) .. (91.33,92.29) -- (91.28,99.13) .. controls (91.27,100.39) and (90.25,101.4) .. (88.99,101.39) -- (74.28,101.28) .. controls (73.02,101.27) and (72.01,100.25) .. (72.02,98.99) -- (72.07,92.15) .. controls (72.08,90.89) and (73.11,89.88) .. (74.37,89.88) -- cycle ;
\draw  [line width=0.75]  (89.09,94.83) -- (89.06,98.83) -- (85.06,98.8) ;

\draw  [fill={rgb, 255:red, 0; green, 0; blue, 0 }  ,fill opacity=1 ] (78,21.58) .. controls (78,19.65) and (79.57,18.08) .. (81.5,18.08) .. controls (83.43,18.08) and (85,19.65) .. (85,21.58) .. controls (85,23.51) and (83.43,25.08) .. (81.5,25.08) .. controls (79.57,25.08) and (78,23.51) .. (78,21.58) -- cycle ;
\draw [line width=2.25]    (81.22,191.08) .. controls (83.09,215.27) and (81.09,221.27) .. (164,220) ;
\draw [fill={rgb, 255:red, 144; green, 19; blue, 254 }  ,fill opacity=1 ][line width=0.75]    (305,38.3) -- (276.58,38.55) ;
\draw  [fill={rgb, 255:red, 32; green, 136; blue, 32 }  ,fill opacity=1 ] (284.12,204.43) .. controls (284.13,202.31) and (285.87,200.59) .. (288,200.61) -- (300.29,200.7) .. controls (302.42,200.72) and (304.13,202.46) .. (304.12,204.59) -- (304.03,216.15) .. controls (304.01,218.27) and (302.27,219.99) .. (300.15,219.97) -- (287.85,219.88) .. controls (285.73,219.86) and (284.01,218.12) .. (284.03,215.99) -- cycle ;
\draw  [fill={rgb, 255:red, 32; green, 136; blue, 32 }  ,fill opacity=1 ][line width=0.75]  (292.61,202.87) -- (299.63,202.93) -- (299.6,206.93) ;

\draw    (304,190) -- (292.61,200.87) ;
\draw   (302,146.56) -- (322,146.56) -- (322,160.45) -- (302,160.45) -- cycle ;
\draw  [draw opacity=0] (303.5,158.84) .. controls (304.44,155.17) and (307.87,152.47) .. (311.92,152.52) .. controls (316.01,152.57) and (319.39,155.39) .. (320.21,159.12) -- (311.82,160.86) -- cycle ; \draw   (303.5,158.84) .. controls (304.44,155.17) and (307.87,152.47) .. (311.92,152.52) .. controls (316.01,152.57) and (319.39,155.39) .. (320.21,159.12) ;  
\draw    (311.71,157.67) -- (313.56,149.37) ;
\draw [shift={(314,147.42)}, rotate = 102.57] [color={rgb, 255:red, 0; green, 0; blue, 0 }  ][line width=0.75]    (4.37,-1.96) .. controls (2.78,-0.92) and (1.32,-0.27) .. (0,0) .. controls (1.32,0.27) and (2.78,0.92) .. (4.37,1.96)   ;

\draw   (300,89.3) -- (320,89.3) -- (320,103.2) -- (300,103.2) -- cycle ;
\draw  [draw opacity=0] (301.5,101.59) .. controls (302.44,97.92) and (305.87,95.22) .. (309.92,95.27) .. controls (314.01,95.32) and (317.39,98.14) .. (318.21,101.87) -- (309.82,103.61) -- cycle ; \draw   (301.5,101.59) .. controls (302.44,97.92) and (305.87,95.22) .. (309.92,95.27) .. controls (314.01,95.32) and (317.39,98.14) .. (318.21,101.87) ;  
\draw    (309.71,100.42) -- (311.56,92.12) ;
\draw [shift={(312,90.17)}, rotate = 102.57] [color={rgb, 255:red, 0; green, 0; blue, 0 }  ][line width=0.75]    (4.37,-1.96) .. controls (2.78,-0.92) and (1.32,-0.27) .. (0,0) .. controls (1.32,0.27) and (2.78,0.92) .. (4.37,1.96)   ;

\draw [fill={rgb, 255:red, 144; green, 19; blue, 254 }  ,fill opacity=1 ][line width=0.75]    (240.16,38.56) -- (283,38.54) ;
\draw [line width=1.5]    (276.64,21.41) -- (276.53,55.69) ;
\draw [fill={rgb, 255:red, 144; green, 19; blue, 254 }  ,fill opacity=1 ][line width=0.75]    (239.87,153.28) -- (301.71,153.26) ;
\draw [fill={rgb, 255:red, 144; green, 19; blue, 254 }  ,fill opacity=1 ][line width=0.75]    (239.16,96.37) -- (300,96.35) ;
\draw  [color={rgb, 255:red, 0; green, 0; blue, 0 }  ,draw opacity=1 ][fill={rgb, 255:red, 38; green, 173; blue, 95 }  ,fill opacity=1 ][line width=0.75]  (268.29,153.27) .. controls (268.29,148.85) and (271.87,145.27) .. (276.29,145.27) .. controls (280.71,145.27) and (284.29,148.85) .. (284.29,153.27) .. controls (284.29,157.68) and (280.71,161.27) .. (276.29,161.27) .. controls (271.87,161.27) and (268.29,157.68) .. (268.29,153.27) -- cycle ;
\draw  [color={rgb, 255:red, 0; green, 0; blue, 0 }  ,draw opacity=1 ][fill={rgb, 255:red, 38; green, 173; blue, 95 }  ,fill opacity=1 ][line width=0.75]  (268.58,38.55) .. controls (268.58,34.13) and (272.16,30.55) .. (276.58,30.55) .. controls (281,30.55) and (284.58,34.13) .. (284.58,38.55) .. controls (284.58,42.97) and (281,46.55) .. (276.58,46.55) .. controls (272.16,46.55) and (268.58,42.97) .. (268.58,38.55) -- cycle ;
\draw  [color={rgb, 255:red, 0; green, 0; blue, 0 }  ,draw opacity=1 ][fill={rgb, 255:red, 38; green, 173; blue, 95 }  ,fill opacity=1 ][line width=0.75]  (268.58,96.36) .. controls (268.58,91.94) and (272.16,88.36) .. (276.58,88.36) .. controls (281,88.36) and (284.58,91.94) .. (284.58,96.36) .. controls (284.58,100.78) and (281,104.36) .. (276.58,104.36) .. controls (272.16,104.36) and (268.58,100.78) .. (268.58,96.36) -- cycle ;
\draw [line width=2.25]    (244.22,22.38) -- (244.03,177.38) ;
\draw  [fill={rgb, 255:red, 208; green, 2; blue, 27 }  ,fill opacity=1 ] (252.29,33.24) .. controls (253.55,33.25) and (254.57,34.27) .. (254.56,35.53) -- (254.51,42.37) .. controls (254.5,43.63) and (253.47,44.65) .. (252.21,44.64) -- (237.5,44.53) .. controls (236.24,44.52) and (235.23,43.49) .. (235.24,42.23) -- (235.29,35.39) .. controls (235.3,34.13) and (236.33,33.12) .. (237.59,33.13) -- cycle ;
\draw  [line width=0.75]  (252.31,38.07) -- (252.28,42.07) -- (248.28,42.04) ;

\draw  [fill={rgb, 255:red, 208; green, 2; blue, 27 }  ,fill opacity=1 ] (251.9,147.24) .. controls (253.16,147.25) and (254.17,148.27) .. (254.16,149.53) -- (254.11,156.37) .. controls (254.1,157.63) and (253.07,158.65) .. (251.81,158.64) -- (237.11,158.53) .. controls (235.85,158.52) and (234.83,157.49) .. (234.84,156.23) -- (234.89,149.39) .. controls (234.9,148.13) and (235.93,147.12) .. (237.19,147.13) -- cycle ;
\draw  [line width=0.75]  (251.91,152.07) -- (251.88,156.07) -- (247.88,156.04) ;

\draw  [fill={rgb, 255:red, 208; green, 2; blue, 27 }  ,fill opacity=1 ] (251.88,90.3) .. controls (253.14,90.31) and (254.15,91.34) .. (254.14,92.6) -- (254.09,99.43) .. controls (254.08,100.69) and (253.05,101.71) .. (251.79,101.7) -- (237.09,101.59) .. controls (235.83,101.58) and (234.82,100.55) .. (234.83,99.29) -- (234.88,92.45) .. controls (234.89,91.19) and (235.91,90.18) .. (237.17,90.19) -- cycle ;
\draw  [line width=0.75]  (251.89,95.13) -- (251.86,99.13) -- (247.86,99.1) ;

\draw [line width=2.25]    (244.03,176.38) .. controls (244.09,189.27) and (248.09,202.27) .. (284,210) ;
\draw  [fill={rgb, 255:red, 0; green, 0; blue, 0 }  ,fill opacity=1 ] (240.72,20.88) .. controls (240.72,18.95) and (242.29,17.38) .. (244.22,17.38) .. controls (246.16,17.38) and (247.72,18.95) .. (247.72,20.88) .. controls (247.72,22.81) and (246.16,24.38) .. (244.22,24.38) .. controls (242.29,24.38) and (240.72,22.81) .. (240.72,20.88) -- cycle ;
\draw  [fill={rgb, 255:red, 255; green, 255; blue, 255 }  ,fill opacity=1 ] (272,167) -- (280,167) -- (280,175) -- (272,175) -- cycle ;
\draw  [fill={rgb, 255:red, 255; green, 255; blue, 255 }  ,fill opacity=1 ] (272,131) -- (280,131) -- (280,139) -- (272,139) -- cycle ;
\draw  [fill={rgb, 255:red, 255; green, 255; blue, 255 }  ,fill opacity=1 ] (272.29,15.83) -- (280.29,15.83) -- (280.29,23.83) -- (272.29,23.83) -- cycle ;
\draw  [fill={rgb, 255:red, 255; green, 255; blue, 255 }  ,fill opacity=1 ] (272.29,52.83) -- (280.29,52.83) -- (280.29,60.83) -- (272.29,60.83) -- cycle ;
\draw  [fill={rgb, 255:red, 255; green, 255; blue, 255 }  ,fill opacity=1 ] (272.29,73.83) -- (280.29,73.83) -- (280.29,81.83) -- (272.29,81.83) -- cycle ;
\draw  [fill={rgb, 255:red, 255; green, 255; blue, 255 }  ,fill opacity=1 ] (272.29,110.83) -- (280.29,110.83) -- (280.29,118.83) -- (272.29,118.83) -- cycle ;

\draw (160,28) node [anchor=north west][inner sep=0.75pt]   [align=left] {$A$};
\draw (185,182) node [anchor=north west][inner sep=0.75pt]   [align=left] {$B$};
\draw (13,102.4) node [anchor=north west][inner sep=0.75pt]  [font=\normalsize]  {$\tilde{\rho }_{M}^{{AB}} =$};
\draw (163.19,135.4) node [anchor=north west][inner sep=0.75pt]  [font=\small]  {$m_{1}$};
\draw (161.19,76.4) node [anchor=north west][inner sep=0.75pt]  [font=\small]  {$m_{2}$};
\draw (151.19,60.4) node [anchor=north west][inner sep=0.75pt]  [font=\scriptsize]  {$0$};
\draw (151.68,119.56) node [anchor=north west][inner sep=0.75pt]  [font=\scriptsize]  {$0$};
\draw (151.19,176.4) node [anchor=north west][inner sep=0.75pt]  [font=\scriptsize]  {$0$};
\draw (96,61.4) node [anchor=north west][inner sep=0.75pt]  [font=\scriptsize]  {$0$};
\draw (95,119.4) node [anchor=north west][inner sep=0.75pt]  [font=\scriptsize]  {$0$};
\draw (96,176.4) node [anchor=north west][inner sep=0.75pt]  [font=\scriptsize]  {$0$};
\draw (308.81,28.3) node [anchor=north west][inner sep=0.75pt]   [align=left] {$A$};
\draw (311,182) node [anchor=north west][inner sep=0.75pt]   [align=left] {$B$};
\draw (312,135.7) node [anchor=north west][inner sep=0.75pt]  [font=\small]  {$m_{1}$};
\draw (310,76.7) node [anchor=north west][inner sep=0.75pt]  [font=\small]  {$m_{2}$};
\draw (257,164.4) node [anchor=north west][inner sep=0.75pt]  [font=\scriptsize]  {$g_{0}^{2}$};
\draw (257,12.4) node [anchor=north west][inner sep=0.75pt]  [font=\scriptsize]  {$g_{0}^{6}$};
\draw (257,47.4) node [anchor=north west][inner sep=0.75pt]  [font=\scriptsize]  {$g_{0}^{6}$};
\draw (257,73.4) node [anchor=north west][inner sep=0.75pt]  [font=\scriptsize]  {$g_{0}^{4}$};
\draw (257,104.4) node [anchor=north west][inner sep=0.75pt]  [font=\scriptsize]  {$g_{0}^{4}$};
\draw (257,127.4) node [anchor=north west][inner sep=0.75pt]  [font=\scriptsize]  {$g_{0}^{2}$};
\draw (200,103.4) node [anchor=north west][inner sep=0.75pt]  [font=\normalsize]  {$=$};
\end{tikzpicture}
\end{equation}
For a generic choice of tensors in Eq.~\eqref{eq:MPS_teleportable_entanglement}, there is a unique steady state in the auxiliary Hilbert space, which is the identity matrix $\rho_D=\mathbbm{1}/D$, according to Eqs.~\eqref{eq:steady_I} and \eqref{eq:MPS_teleportable_entanglement}.
In the last equality, we use explicit components of the diamond tensor from Fig.~\ref{fig:MPO_tensors}(c).
Notably, our choice of initial state and reinitialized states fixes the left and right indices of diamond tensors to be $0$, which implies that the classical state on the inner bond of spacetime mapping is updated deterministically:
\begin{equation}
    g(t=0)=e, g(t=1)=g_0~g(t=0)~g_0 = g_0^2,\dots.
\end{equation}
Consequently, the spacetime mapping factorizes into a product of single-site tensors.

It follows that the postmeasurement state can be expressed as qudits $A$ and $B$ mediated by a sequence of transfer matrices acting on the auxiliary Hilbert space, each given by
\begin{equation}\label{eq:quantum_transfer}
\tikzset{every picture/.style={line width=0.75pt}} 
\begin{tikzpicture}[x=0.75pt,y=0.75pt,yscale=-1,xscale=1]

\draw [line width=2.25]    (85.9,92.06) -- (85.68,114.06) ;
\draw [shift={(85.87,95.36)}, rotate = 90.58] [fill={rgb, 255:red, 0; green, 0; blue, 0 }  ][line width=0.08]  [draw opacity=0] (8.57,-4.12) -- (0,0) -- (8.57,4.12) -- cycle    ;
\draw [line width=1.5]    (109.4,73.17) -- (109.29,106.45) ;
\draw   (133,81.92) -- (153,81.92) -- (153,95.82) -- (133,95.82) -- cycle ;
\draw  [draw opacity=0] (134.5,94.21) .. controls (135.44,90.54) and (138.87,87.84) .. (142.92,87.89) .. controls (147.01,87.94) and (150.39,90.76) .. (151.21,94.49) -- (142.82,96.23) -- cycle ; \draw   (134.5,94.21) .. controls (135.44,90.54) and (138.87,87.84) .. (142.92,87.89) .. controls (147.01,87.94) and (150.39,90.76) .. (151.21,94.49) ;  
\draw    (142.71,93.04) -- (144.56,84.74) ;
\draw [shift={(145,82.78)}, rotate = 102.57] [color={rgb, 255:red, 0; green, 0; blue, 0 }  ][line width=0.75]    (4.37,-1.96) .. controls (2.78,-0.92) and (1.32,-0.27) .. (0,0) .. controls (1.32,0.27) and (2.78,0.92) .. (4.37,1.96)   ;

\draw [fill={rgb, 255:red, 144; green, 19; blue, 254 }  ,fill opacity=1 ][line width=0.75]    (80.16,88.99) -- (133,88.97) ;
\draw  [color={rgb, 255:red, 0; green, 0; blue, 0 }  ,draw opacity=1 ][fill={rgb, 255:red, 38; green, 173; blue, 95 }  ,fill opacity=1 ][line width=0.75]  (101.58,88.98) .. controls (101.58,84.56) and (105.16,80.98) .. (109.58,80.98) .. controls (114,80.98) and (117.58,84.56) .. (117.58,88.98) .. controls (117.58,93.4) and (114,96.98) .. (109.58,96.98) .. controls (105.16,96.98) and (101.58,93.4) .. (101.58,88.98) -- cycle ;
\draw [line width=2.25]    (85.71,65.56) -- (85.48,88.56) ;
\draw [shift={(85.67,69.36)}, rotate = 90.56] [fill={rgb, 255:red, 0; green, 0; blue, 0 }  ][line width=0.08]  [draw opacity=0] (8.57,-4.12) -- (0,0) -- (8.57,4.12) -- cycle    ;
\draw  [fill={rgb, 255:red, 208; green, 2; blue, 27 }  ,fill opacity=1 ] (92.88,82.92) .. controls (94.14,82.93) and (95.15,83.95) .. (95.14,85.21) -- (95.09,92.05) .. controls (95.08,93.31) and (94.05,94.33) .. (92.79,94.32) -- (78.09,94.21) .. controls (76.83,94.2) and (75.82,93.17) .. (75.83,91.91) -- (75.88,85.07) .. controls (75.89,83.81) and (76.91,82.8) .. (78.17,82.81) -- cycle ;
\draw  [line width=0.75]  (92.89,87.75) -- (92.86,91.75) -- (88.86,91.72) ;

\draw  [fill={rgb, 255:red, 255; green, 255; blue, 255 }  ,fill opacity=1 ] (105.4,66.17) -- (113.4,66.17) -- (113.4,74.17) -- (105.4,74.17) -- cycle ;
\draw  [fill={rgb, 255:red, 255; green, 255; blue, 255 }  ,fill opacity=1 ] (105.29,104.45) -- (113.29,104.45) -- (113.29,112.45) -- (105.29,112.45) -- cycle ;

\draw (155,71.4) node [anchor=north west][inner sep=0.75pt]  [font=\normalsize]  {$ \begin{array}{l}
=\sum\limits_{a,a'=0}^{q-1}\bra{m_{t}} u[ g( t)]\ket{a}\bra{a'} u[ g( t)]^{\dagger }\ket{m_{t}}\\
\ \ \ \ \ \  \alpha _{a} \alpha _{a'}^{*} \ w^{( a)} \otimes w{^{( a')}}^{*}
\end{array}$};
\draw (140,67.4) node [anchor=north west][inner sep=0.75pt]  [font=\small]  {$m_{t}$};
\draw (98,50.4) node [anchor=north west][inner sep=0.75pt]  [font=\scriptsize]  {$g( t)$};
\draw (100,117.4) node [anchor=north west][inner sep=0.75pt]  [font=\scriptsize]  {$g( t)$};

\end{tikzpicture}
\end{equation}

According to Eq.~\eqref{eq:semiproduct_unitary}, $u[g(t)]$ acts on computational basis states $\ket{a}$ and $\ket{m_t}$ by permuting to other basis states and attaching a phase factor. We denote this permutation as $\mathcal{P}_{g(t)}$, such that
\begin{equation}
    u[g(t)]\ket{a} \propto \ket{\mathcal{P}_{g(t)}(a)}.
\end{equation}
Therefore, the transfer matrix from Eq.~\eqref{eq:quantum_transfer} reduces to the unitary transformation $w^{\mathcal{P}_{g(t)}^{-1}(m_t)}\otimes {w^{\mathcal{P}_{g(t)}^{-1}(m_t)}}^*$, and for each measurement outcome, the qudits $A$ and $B$ are connected by a product of unitaries:
\begin{equation}
    \mathcal{U}_M = w^{\mathcal{P}_{g(T-1)}^{-1}(m_{T-1})}\cdots w^{\mathcal{P}_{g(2)}^{-1}(m_2)} w^{\mathcal{P}_{g(1)}^{-1}(m_1)}.
\end{equation}
Illustratively, as teleported through a sequence of unitaries, the entanglement between qudits $A$ and $B$ does not decay in time.
The resulting postmeasurement states $\rho_M^\text{AB}$ thus typically possess finite quantum entanglement, implying that the average entanglement remains finite even for a large number of time steps. This confirms the existence of long-time quantum memory. 

To validate our protocol, we consider an initial MPS with bond dimension equal to the local Hilbert space dimension: $D=q$, randomly chosen unitaries $w^{a}$, uniform coefficients $\alpha_a = 1/\sqrt{q}$, and the boundary tensors at qudit $B$ are chosen to be the identity.
Under these conditions, the postmeasurement state conditioned on $M$ takes the form
\begin{equation}
\tikzset{every picture/.style={line width=0.75pt}} 
\begin{tikzpicture}[x=0.75pt,y=0.75pt,yscale=-1,xscale=1]

\draw [line width=1.5]    (145.47,116.12) -- (178.75,116.12) ;
\draw [line width=0.75]    (141.43,141.68) -- (250.42,141.67) ;
\draw [fill={rgb, 255:red, 144; green, 19; blue, 254 }  ,fill opacity=1 ][line width=0.75]    (162.23,131.25) -- (162,91) ;
\draw  [color={rgb, 255:red, 0; green, 0; blue, 0 }  ,draw opacity=1 ][fill={rgb, 255:red, 38; green, 173; blue, 95 }  ,fill opacity=1 ][line width=0.75]  (162.28,123.94) .. controls (158.31,123.91) and (155.12,120.45) .. (155.15,116.22) .. controls (155.19,111.99) and (158.45,108.59) .. (162.43,108.62) .. controls (166.4,108.66) and (169.59,112.12) .. (169.56,116.35) .. controls (169.52,120.58) and (166.26,123.98) .. (162.28,123.94) -- cycle ;
\draw  [fill={rgb, 255:red, 208; green, 2; blue, 27 }  ,fill opacity=1 ] (157.36,133.71) .. controls (157.38,132.45) and (158.42,131.45) .. (159.68,131.47) -- (166.52,131.59) .. controls (167.78,131.61) and (168.78,132.64) .. (168.76,133.9) -- (168.51,148.61) .. controls (168.49,149.87) and (167.45,150.87) .. (166.19,150.85) -- (159.35,150.74) .. controls (158.1,150.71) and (157.09,149.68) .. (157.11,148.42) -- cycle ;
\draw  [line width=0.75]  (162.19,133.74) -- (166.19,133.81) -- (166.13,137.81) ;

\draw  [fill={rgb, 255:red, 0; green, 0; blue, 0 }  ,fill opacity=1 ] (140.39,144.26) .. controls (138.73,144.24) and (137.4,142.89) .. (137.42,141.23) .. controls (137.43,139.58) and (138.79,138.24) .. (140.44,138.26) .. controls (142.1,138.28) and (143.43,139.63) .. (143.41,141.29) .. controls (143.4,142.94) and (142.04,144.27) .. (140.39,144.26) -- cycle ;
\draw  [fill={rgb, 255:red, 255; green, 255; blue, 255 }  ,fill opacity=1 ] (197,120) -- (250,120) -- (250,160) -- (197,160) -- cycle ;
\draw    (250.42,141.67) .. controls (255.27,140.43) and (266.27,146.43) .. (265,94) ;
\draw  [fill={rgb, 255:red, 255; green, 255; blue, 255 }  ,fill opacity=1 ] (176.75,112.12) -- (184.75,112.12) -- (184.75,120.12) -- (176.75,120.12) -- cycle ;
\draw  [fill={rgb, 255:red, 255; green, 255; blue, 255 }  ,fill opacity=1 ] (140,112) -- (148,112) -- (148,120) -- (140,120) -- cycle ;

\draw (83,123.4) node [anchor=north west][inner sep=0.75pt]  [font=\normalsize]  {$\rho _{M}^{{AB}} \ =\ $};
\draw (173,94.4) node [anchor=north west][inner sep=0.75pt]  [font=\footnotesize]  {$g( T)$};
\draw (212,132.4) node [anchor=north west][inner sep=0.75pt]  [font=\normalsize]  {$\mathcal{U}_{M}$};
\draw (121,94.4) node [anchor=north west][inner sep=0.75pt]  [font=\footnotesize]  {$g( T)$};
\draw (151,71) node [anchor=north west][inner sep=0.75pt]   [align=left] {$A$};
\draw (260,71) node [anchor=north west][inner sep=0.75pt]   [align=left] {$B$};

\end{tikzpicture}
\end{equation}
Notice that our chosen measure of mixed-state entanglement negativity is invariant under local unitaries on $\text{A}$ or $\text{B}$, since the spectrum of the partially transposed state remains unchanged. We can, therefore, effectively cancel out the unitaries $\mathcal{U}_M$ and $u[{g(T)}]$ by setting them to the identity and dealing with the effective two-qudit density matrix
\begin{equation}\label{eq:rho_explicit_negativity}
\rho_\text{eff}^{{AB}}=\frac{1}{q^2}\sum\limits_{a,a'=0}^{q-1}(|a\rangle\langle a^\prime|)_{A} \otimes \big(w^{a}(w^{a^\prime})^\dagger\big)_{B},
\end{equation}

Consequently, all postmeasurement states share the same negativity, independent of the specific measurement outcomes. 
We find that for $q=2$ the negativity vanishes exactly, whereas for $q\geq 3$ and randomly drawn auxiliary unitaries $\{w^{a}\}_{a=0}^{q-1}$ the negativity is typically finite, directly implying the presence of long-time quantum memory. We demonstrate these two scenarios in Appendix~\ref{app:negativity}.

As a remark, our construction of the initial MPS encompasses several well-known states, including the Greenberger-Horne-Zeilinger state with long-range connected correlations~\cite{Raussendorf2001One} and the 1D cluster state ~\cite{Briegel2001Persistent,Verstraete2004Valence} and the ground state of the Affleck-Kennedy-Lieb-Tasaki model \cite{Affleck1988vValence,Verstraete2004Diverging} with long-range string order. These states have been proposed as resources for measurement-based quantum teleportation and computation \cite{Raussendorf2001One,Raussendorf2003Measurement,Verstraete2004Entanglement,Popp2005Localizable,Gross2007Novel,Miyake2010Quantum,Wahl2012Matrix,Stephen2017Computational}.
Our present setting follows the same spirit: these initial states, when incorporated with suitable circuit dynamics, serve as resources for measurement-based long-range quantum teleportation \textit{in time}, where entanglement is distributed not across space but across successive moments of time evolution.

Finally, we compare teleportable entanglement with other measures of quantum memory based on the IM formalism. Alternative diagnostics do exist, but they are not tailored to the multitime setting considered here. In particular, the criteria introduced in Refs.~\cite{Backer2024Local,Backer2025Entropic} require only two-time channel tomography on the system but, however, may miss memory structures that become visible only through sequential interventions at multiple times, as in our protocol. Indeed, when applied to the toy model introduced above [Eqs.~\eqref{eq:semiproduct_unitary} and \eqref{eq:MPS_teleportable_entanglement}], those criteria predict the absence of quantum memory.
Another quantity, introduced in Ref.~\cite{Giarmatzi2021Witnessing}, diagnoses quantum memory through the nonseparability of the temporal state; while conceptually direct, its implementation generally requires operations that are highly nonlocal in time and, thus, lacks a clear operational interpretation.

\section{Conclusions and outlook}\label{sec:conclusion}

In summary, we have investigated a family of solvable models of quantum many-body dynamics from a complexity-theoretic perspective. 
This family of dual-unitary quantum circuits spans a broad spectrum of models ranging from integrable to chaotic.
The crucial ingredient for our analysis is the analytical tensor-network representation of the influence matrix as an efficient characterization of nonequilibrium dynamics.
By employing analytical tools from geometric group theory, we rigorously demonstrated three distinct classes of growth behavior of the temporal entanglement entropy, thereby uncovering a hierarchy of computational complexities.
Moreover, using our insights into the memory structure of underlying dynamics, we identified a subset of our models that have classical influence matrices, in the sense that the multitime correlation functions can be faithfully reproduced by fictitious classical stochastic processes, which enables efficient Monte Carlo simulations for local dynamics. 
For the remaining, genuinely quantum cases, we proposed and demonstrated a diagnostic of quantum memory based on an experimentally relevant quantum teleportation protocol.
Our results, thus, provide a unified landscape illuminating the intricate connections between dynamical characteristics and computational complexity of the influence matrix for these models.

The analytical techniques developed here for controlled unitaries point to several natural avenues for extension. 
These include replacing the group algebra by more general Hopf algebras~\cite{Wang2025Hopf}, extending the construction to nonunitary dynamics generated by completely positive maps, as in noisy or disorder-averaged circuits~\cite{Kos2021Correlations,Kos2023Circuits}, and considering systems with infinite-dimensional local Hilbert spaces such as bosons. Such generalizations may reveal new temporal complexity classes beyond the finite-dimensional unitary setting studied here. More broadly, it will be interesting to explore whether similar algebraic structures can also be identified in higher spatial dimensions~\cite{Jonay2021Triunitary,Milbradt2023Ternary,Park2025Simulating} and in cellular automata~\cite{Prosen2021many,Klobas2021Exact,Bertini2023Exact}.

Beyond temporal complexities, these tools provide a useful language for characterizing various facets of nonintegrability and complexity in quantum circuits.
For example, a recent study of a semiergodic dual-unitary circuit closely related to our class III model reported logarithmic growth of operator entanglement in a clean nonintegrable setting at finite size~\cite{Tan2026Logarithmic}. Such behavior is unexpected, since nonintegrable circuits are more commonly associated with linear growth of operator entanglement. Our group-algebra techniques offer a natural framework for understanding this phenomenon by connecting the relaxation of local observables and the restricted structure of time-evolving operators in the Heisenberg picture to an effective random walk on the group manifold. More broadly, we view these methods not only as exact solution techniques, but also as conceptual tools for identifying and analyzing unexpectedly low-complexity sectors within otherwise complicated quantum dynamics.

In addition, the concept of spacetime mapping—generally applicable to quantum circuits and non-Markovian open quantum systems—further extends the scope for constructing analytically tractable nonequilibrium dynamics.
More broadly, this work points toward a promising direction for unveiling the fundamental structural principles that give rise to sublinear growth of temporal entanglement entropy and classical simulatability in interacting systems. Our results, together with related studies~\cite{Klobas2021Exact, Giudice2022Temporal}, suggest that the presence of ballistic conservation laws may play a pivotal role, although the underlying mechanism remains an open and intriguing question~\cite{Aloisio2023Smapling,Carignano2024Temporal}.

Furthermore, our notions of classical and quantum memory of the influence matrix introduce an alternative dimension for characterizing nonequilibrium dynamics. Although demonstrated in solvable models in this work, these notions and associated measures are applicable to generic systems, underscoring the potential of information-theoretic quantities extracted from influence matrices as powerful diagnostics for temporal phases of nonequilibrium quantum matter. 
Along the line of classical memory, this framework can have direct implications for understanding the emergence of universal (multitime) macroscopic phenomena in quantum many-body systems, such as thermalization, relaxation, and equilibration \cite{Figueroa2019Almost,Figueroa2021Markovianization,Dowling2023Equilibration}.
On the other hand, the study of persistent long-time quantum memory in many-body dynamics may inspire the design of non-Markovian quantum error-correction codes \cite{Tanggara2024Strategic,Kobayashi2024Tensornetwork,Kam2025Detrimental}.

Finally, we expect that the mappings developed here, i.e., from influence matrices to classical or hybrid quantum-classical stochastic processes, and more generally to stochastic mixtures of MPS, suggest a broader framework that may extend beyond the exactly solvable setting. 
Similar decompositions are already known for thermal states: thermal expectation values are calculated through sampling over low-entanglement pure states, giving the decomposition a direct algorithmic meaning~\cite{White2009Minimally, Stoudenmire2010Minimally,Bakshi2026Entanglement}.
From this viewpoint, our exact solutions may be understood as a solvable limit of a broader class of influence-matrix decompositions with analogous algorithmic structure. Recent numerical progress for Hamiltonian dynamics points in the same direction, indicating that sampling over suitably structured tensor-network objects can bypass the usual entanglement barrier for local observables~\cite{Carignano2025Overcoming}.

\begin{acknowledgements}
We are grateful to Zhiyuan Wang, Alessio Lerose, and Balázs Pozsgay for helpful discussions. H.-R.~W. thanks Xiao-Yang Yang and Zhong Wang for collaborations on related projects. H.-R.~W. acknowledges Biao Lian for hospitality at Princeton University and the support from the National Natural Science Foundation of China under Grant No.~12125405, the National Key R$\&$D Program of China (No.~2023YFA1406702), and the Tsinghua Visiting Doctoral Students Foundation. This work was also partially supported by a Brown Investigator Award and by the European Research Council (ERC) under the European Union's Horizon 2020 research and innovation program via Grant Agreement No.~864597 (I.~V. and D.~A.~A.).
\end{acknowledgements}

\section*{Data Availability}
The data that support the findings of this article are openly available \cite{wang2025ZenodoTemporal}.

\appendix
\section{DETAILS ABOUT THE EXACT MPO REPRESENTATION}

\subsection{Derivations} \label{app:group}
In this part, we derive the MPO representation of the spacetime mapping, regarding the gap between Figs.~\ref{fig:MPO_tensors}(b) and \ref{fig:MPO_tensors}(c).
The key observation underlying this simplification is that the exponentially large inner bond in Fig.~\ref{fig:MPO_tensors}(b) can be equivalently replaced by the group algebra $\mathbb{C}[\mathrm{PU}(q)]$. To see this, we focus on the tensor between time steps $t-1$ and $t$:
\begin{equation}\label{eq:transfer_tensor}
\tikzset{every picture/.style={line width=0.75pt}} 
\begin{tikzpicture}[x=0.75pt,y=0.75pt,yscale=-1,xscale=1]

\draw [fill={rgb, 255:red, 144; green, 19; blue, 254 }  ,fill opacity=1 ][line width=0.75]    (202.8,96.34) -- (181.82,96.2) ;
\draw [fill={rgb, 255:red, 144; green, 19; blue, 254 }  ,fill opacity=1 ][line width=0.75]    (252.82,64) -- (216.82,64) ;
\draw [fill={rgb, 255:red, 144; green, 19; blue, 254 }  ,fill opacity=1 ][line width=0.75]    (124.8,64) -- (89.8,64) ;
\draw    (114.82,46.4) -- (114.82,92.6) ;
\draw [fill={rgb, 255:red, 144; green, 19; blue, 254 }  ,fill opacity=1 ][line width=0.75]    (202.82,64) -- (139.8,63.86) ;
\draw [fill={rgb, 255:red, 144; green, 19; blue, 254 }  ,fill opacity=1 ][line width=0.75]    (161.82,96) -- (139.8,96) ;
\draw [fill={rgb, 255:red, 144; green, 19; blue, 254 }  ,fill opacity=1 ][line width=0.75]    (252.8,96.36) -- (216.8,96.2) ;
\draw  [fill={rgb, 255:red, 255; green, 255; blue, 255 }  ,fill opacity=1 ] (159.02,96) .. controls (159.02,94.45) and (160.27,93.2) .. (161.82,93.2) .. controls (163.36,93.2) and (164.62,94.45) .. (164.62,96) .. controls (164.62,97.55) and (163.36,98.8) .. (161.82,98.8) .. controls (160.27,98.8) and (159.02,97.55) .. (159.02,96) -- cycle ;
\draw  [fill={rgb, 255:red, 255; green, 255; blue, 255 }  ,fill opacity=1 ] (179.02,96.2) .. controls (179.02,94.65) and (180.27,93.4) .. (181.82,93.4) .. controls (183.36,93.4) and (184.62,94.65) .. (184.62,96.2) .. controls (184.62,97.75) and (183.36,99) .. (181.82,99) .. controls (180.27,99) and (179.02,97.75) .. (179.02,96.2) -- cycle ;
\draw    (114.82,99) -- (114.82,115) ;
\draw    (226.42,99) -- (226.42,115) ;
\draw  [fill={rgb, 255:red, 0; green, 0; blue, 0 }  ,fill opacity=1 ] (99.3,95.8) .. controls (99.3,94.97) and (99.97,94.3) .. (100.8,94.3) .. controls (101.63,94.3) and (102.3,94.97) .. (102.3,95.8) .. controls (102.3,96.63) and (101.63,97.3) .. (100.8,97.3) .. controls (99.97,97.3) and (99.3,96.63) .. (99.3,95.8) -- cycle ;
\draw  [fill={rgb, 255:red, 0; green, 0; blue, 0 }  ,fill opacity=1 ] (239.82,96.5) .. controls (239.82,95.67) and (240.49,95) .. (241.32,95) .. controls (242.15,95) and (242.82,95.67) .. (242.82,96.5) .. controls (242.82,97.33) and (242.15,98) .. (241.32,98) .. controls (240.49,98) and (239.82,97.33) .. (239.82,96.5) -- cycle ;
\draw    (100.82,47.8) -- (100.82,94.8) ;
\draw    (241.32,47.8) -- (241.32,95) ;
\draw  [fill={rgb, 255:red, 255; green, 255; blue, 255 }  ,fill opacity=1 ] (95.82,63.5) .. controls (95.82,60.74) and (98.06,58.5) .. (100.82,58.5) .. controls (103.58,58.5) and (105.82,60.74) .. (105.82,63.5) .. controls (105.82,66.26) and (103.58,68.5) .. (100.82,68.5) .. controls (98.06,68.5) and (95.82,66.26) .. (95.82,63.5) -- cycle ;

\draw  [fill={rgb, 255:red, 255; green, 255; blue, 255 }  ,fill opacity=1 ] (236.82,64) .. controls (236.82,61.24) and (239.06,59) .. (241.82,59) .. controls (244.58,59) and (246.82,61.24) .. (246.82,64) .. controls (246.82,66.76) and (244.58,69) .. (241.82,69) .. controls (239.06,69) and (236.82,66.76) .. (236.82,64) -- cycle ;

\draw  [fill={rgb, 255:red, 255; green, 255; blue, 255 }  ,fill opacity=1 ] (109.82,64) .. controls (109.82,61.24) and (112.06,59) .. (114.82,59) .. controls (117.58,59) and (119.82,61.24) .. (119.82,64) .. controls (119.82,66.76) and (117.58,69) .. (114.82,69) .. controls (112.06,69) and (109.82,66.76) .. (109.82,64) -- cycle ;

\draw    (149.32,99) -- (149.32,115) ;
\draw    (193.32,99) -- (193.32,115) ;
\draw    (149.32,47.8) -- (149.32,92) ;
\draw  [fill={rgb, 255:red, 255; green, 255; blue, 255 }  ,fill opacity=1 ] (143.82,64) .. controls (143.82,61.24) and (146.06,59) .. (148.82,59) .. controls (151.58,59) and (153.82,61.24) .. (153.82,64) .. controls (153.82,66.76) and (151.58,69) .. (148.82,69) .. controls (146.06,69) and (143.82,66.76) .. (143.82,64) -- cycle ;

\draw    (226.82,47.8) -- (226.82,93) ;
\draw    (193.32,47.8) -- (193.32,93) ;
\draw  [fill={rgb, 255:red, 255; green, 255; blue, 255 }  ,fill opacity=1 ] (188.82,64) .. controls (188.82,61.24) and (191.06,59) .. (193.82,59) .. controls (196.58,59) and (198.82,61.24) .. (198.82,64) .. controls (198.82,66.76) and (196.58,69) .. (193.82,69) .. controls (191.06,69) and (188.82,66.76) .. (188.82,64) -- cycle ;

\draw  [fill={rgb, 255:red, 255; green, 255; blue, 255 }  ,fill opacity=1 ] (221.82,64) .. controls (221.82,61.24) and (224.06,59) .. (226.82,59) .. controls (229.58,59) and (231.82,61.24) .. (231.82,64) .. controls (231.82,66.76) and (229.58,69) .. (226.82,69) .. controls (224.06,69) and (221.82,66.76) .. (221.82,64) -- cycle ;

\draw [fill={rgb, 255:red, 144; green, 19; blue, 254 }  ,fill opacity=1 ][line width=0.75]    (124.8,96.2) -- (89.82,96.2) ;

\draw (222.82,61) node [anchor=north west][inner sep=0.75pt]  [font=\tiny]  {$u$};
\draw (189.82,61) node [anchor=north west][inner sep=0.75pt]  [font=\tiny]  {$u$};
\draw (144.82,61) node [anchor=north west][inner sep=0.75pt]  [font=\tiny]  {$u$};
\draw (110.82,61) node [anchor=north west][inner sep=0.75pt]  [font=\tiny]  {$u$};
\draw (237.82,61) node [anchor=north west][inner sep=0.75pt]  [font=\tiny]  {$u$};
\draw (96.82,60.5) node [anchor=north west][inner sep=0.75pt]  [font=\tiny]  {$u$};
\draw (126,62) node [anchor=north west][inner sep=0.75pt]   [align=left] {...};
\draw (126,93.4) node [anchor=north west][inner sep=0.75pt]   [align=left] {...};
\draw (203.82,93.4) node [anchor=north west][inner sep=0.75pt]   [align=left] {...};
\draw (203.82,62) node [anchor=north west][inner sep=0.75pt]   [align=left] {...};
\draw (113.97,131.94) node [anchor=north west][inner sep=0.75pt]  [font=\Huge,rotate=-270] [align=left] {{\Huge \{}};
\draw (127.5,129.5) node [anchor=north west][inner sep=0.75pt]    {$t$};
\end{tikzpicture}
\end{equation}
Because of the identity constraints indicated by the solid dots, the inner bond at time $t$ encodes the linear space spanned by the history of odd-numbered open legs, denoted as 
\begin{equation}
    (a_{[1:2t-1]},b_{[1:2t-1]})\equiv(a_1, b_1, a_3, b_3, \dots, a_{2t-1}, b_{2t-1}),
\end{equation}
where $a$ and $b$ refer to open legs on the left and right, respectively.
Note that it is unnecessary to distinguish between indices on forward and backward branches, as they are dephased into the same computational basis state, indicated by hollow dots.

The tensor Eq.~\eqref{eq:transfer_tensor} consists of two layers. Upon crossing the first layer, the indices of the inner bond are enlarged and updated from $(a_{[1:2t-3]},b_{[1:2t-3]})$ to $(a_{[1:2t-1]},b_{[1:2t-1]})$. When crossing the second layer, a product of unitaries is generated, acting on the physical legs from left to right as
\begin{equation}
    u_{a_{[1:2t-1]},b_{[1:2t-1]}} = u_{b_{2t-1}}\cdots u_{b_1}u_{a_1}\cdots u_{a_{2t-1}}.
\end{equation}
Notably, the unitaries at two successive time steps are related by unitary multiplications:
\begin{equation}\label{eq:history_unitary}
    u_{a_{[1:2t-1]},b_{[1:2t-1]}} = u_{b_{2t-1}}u_{a_{[1:2t-3]},b_{[1:2t-3]}}u_{a_{2t-1}}.
\end{equation}

It follows that the relevant information affecting the system is not the entire sequence of indices but rather this unitary [up to an overall $\mathrm{U}(1)$ phase]. We, therefore, replace the inner bond with the linear space of all single-site unitaries modulo the phase, which form the group algebra of $G=\mathrm{PU}(q)$.
This observation leads to an equivalent but shift-invariant MPO representation for the spacetime mapping, as summarized in Fig.~\ref{fig:MPO_tensors}(c).
In the diamond tensor, the group element $g$ is then updated to $g_bgg_a$, corresponding to updating $u_{a_{[1:2t-3]},b_{[1:2t-3]}}$ under the rule of Eq.~\eqref{eq:history_unitary}.
In the round tensor, the term $u(g)\otimes u(g)^*$ implements the regular adjoint representation of $g$, effectively realizing the operator  $u_{a_{[1:2t-1]},b_{[1:2t-1]}}$ acting on physical legs. The upper and lower boundary vectors are chosen appropriately.

\subsection{Right influence matrix} \label{app:right}
\begin{figure}[ht]
\hspace*{-0.48\textwidth}
\includegraphics[width=0.95\linewidth]{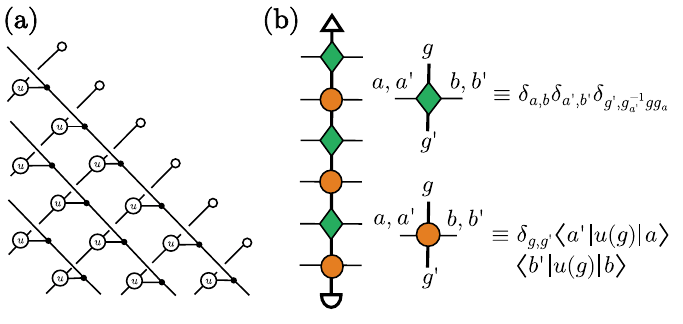} 
\caption{Spacetime mapping for the right influence matrix. (a) Representation in terms of controlled and {\scriptsize${\mathrm{SWAP}}$} gates. (b) The analytical MPO representation and the definition of local tensors.
}
\label{fig:MPO_right}
\end{figure}

For completeness, we now present the analytical MPO representation for the right influence matrix.
Figure~\ref{fig:MPO_right} (a) depicts the associated spacetime mapping in terms of local building blocks.
Following a similar procedure as for the left spacetime mapping in Appendix~\ref{app:group}, we first deform the tensor network into a triangular shape and then replace the inner bond with the group algebra of $\mathrm{PU}(q)$. The resulting shift-invariant MPO form is shown in Fig.~\ref{fig:MPO_right}(b).

\subsection{Generalizations} \label{app:add_permutation}

We generalize the two-qudit unitaries in Eq.~\eqref{eq:swap_control} by introducing an additional permutation acting on the control site \cite{Wang2025Hopf}:
\begin{equation}
    U=S\left[\sum_{a=0}^{q-1}u_a\otimes (P\ket{a})\bra{a}\right],
\end{equation}
where $P\in S_q$ is a permutation matrix acting on the computational basis $\{\ket{a}\}_{a=0}^{q-1}$ in the way $P\ket{a}=\ket{\mathcal{P}(a)}$. We assume that $P$ has finite order $l$, i.e., $P^l = I_q$. Following the same procedure outlined in Sec.~\ref{sec:MPO} and Appendix~\ref{app:group}, we first note that the unitary that affects the system at the time step $t$ takes the form
\begin{align}\label{eq:history_permu}
    u_{a_{[1:2t-1]},b_{[1:2t-1]}} = &u_{b_{2t-1}}\cdots u_{\mathcal{P}^{t-2}(b_3)}u_{\mathcal{P}^{t-1}(b_1)}\nonumber\\
    &u_{\mathcal{P}^{t-1}(a_1)}u_{\mathcal{P}^{t-2}(a_3)}\cdots u_{a_{2t-1}}.
\end{align}
Compared to Eq.~\eqref{eq:history_unitary}, the unitaries at two successive time steps are no longer simply related by unitary multiplications. To address this, we introduce the tensor product of $l$ copies of $\mathrm{PU}(q)$ for the inner bond, denoted as: $(g_0,g_1,\dots,g_{l-1})$. At time step $t$, given the odd-numbered leg indices $a$ and $b$, the state is updated according to the rule:
\begin{equation}
(g_bg_1g_a, g_{\mathcal{P}(b)}g_2g_{\mathcal{P}(a)},\dots,g_{\mathcal{P}^{l-1}(b)}g_0g_{\mathcal{P}^{l-1}(a)}).
\end{equation}
Using the $ l$ periodicity of $\mathcal{P}$, it can be verified that the first entry correctly generates the adjoint action of Eq.~\eqref{eq:history_permu}. The corresponding local tensors that replace those in Fig.~\ref{fig:MPO_tensors}(c) are presented below:
\begin{equation}
\tikzset{every picture/.style={line width=0.75pt}} 
\begin{tikzpicture}[x=0.75pt,y=0.75pt,yscale=-1,xscale=1]

\draw [line width=1.5]    (58.95,28.36) -- (57.91,78.27) ;
\draw [fill={rgb, 255:red, 144; green, 19; blue, 254 }  ,fill opacity=1 ][line width=0.75]    (33.67,54.12) -- (83.67,54.1) ;
\draw  [fill={rgb, 255:red, 229; green, 126; blue, 33 }  ,fill opacity=1 ][line width=0.75]  (58.67,42.13) -- (67.91,54.11) -- (58.67,66.09) -- (49.44,54.11) -- cycle ;
\draw [line width=1.5]    (58.95,135.36) -- (57.91,185.27) ;
\draw [fill={rgb, 255:red, 144; green, 19; blue, 254 }  ,fill opacity=1 ][line width=0.75]    (33.67,161.12) -- (83.67,161.1) ;
\draw  [color={rgb, 255:red, 0; green, 0; blue, 0 }  ,draw opacity=1 ][fill={rgb, 255:red, 38; green, 173; blue, 95 }  ,fill opacity=1 ][line width=0.75]  (49.82,161.11) .. controls (49.82,156.22) and (53.78,152.26) .. (58.67,152.26) .. controls (63.57,152.26) and (67.53,156.22) .. (67.53,161.11) .. controls (67.53,166) and (63.57,169.97) .. (58.67,169.97) .. controls (53.78,169.97) and (49.82,166) .. (49.82,161.11) -- cycle ;

\draw (18,38.4) node [anchor=north west][inner sep=0.75pt]  [font=\normalsize]  {$a,a'$};
\draw (80,39.4) node [anchor=north west][inner sep=0.75pt]  [font=\normalsize]  {$b,b'$};
\draw (29,80.4) node [anchor=north west][inner sep=0.75pt]  [font=\normalsize]  {$g_{0} ,\cdots ,g_{l-1}$};
\draw (100,48.4) node [anchor=north west][inner sep=0.75pt]  [font=\normalsize]  {$\equiv \delta _{a,a'} \ \delta _{b,b'} \ \delta _{g'_{0} ,\ g_{b} g_{1} g_{a}} \cdots \delta _{g'_{l-1} ,\ g_{\mathcal{P}^{l-1}( b)} g_{0} g_{\mathcal{P}^{l-1}( a)}}$};
\draw (22,145.4) node [anchor=north west][inner sep=0.75pt]  [font=\normalsize]  {$a,a'$};
\draw (77,145.4) node [anchor=north west][inner sep=0.75pt]  [font=\normalsize]  {$b,b'$};
\draw (103,151.4) node [anchor=north west][inner sep=0.75pt]  [font=\normalsize]  {$\equiv \bra{b,b'} u( g_{0}) \otimes u( g_{0})^{*}\ket{a,a'} \delta _{g'_{0} ,\ g_{0}} \cdots \delta _{g'_{l-1} ,\ g_{l-1}}$};
\draw (27,11.4) node [anchor=north west][inner sep=0.75pt]  [font=\normalsize]  {$g'_{0} ,\cdots ,g'_{l-1}$};
\draw (55,47.4) node [anchor=north west][inner sep=0.75pt]  [font=\small]  {$l$};
\draw (27,118.4) node [anchor=north west][inner sep=0.75pt]  [font=\normalsize]  {$g'_{0} ,\cdots ,g'_{l-1}$};
\draw (27,188.4) node [anchor=north west][inner sep=0.75pt]  [font=\normalsize]  {$g_{0} ,\cdots ,g_{l-1}$};
\draw (55,154.4) node [anchor=north west][inner sep=0.75pt]  [font=\small]  {$l$};

\end{tikzpicture}.
\end{equation}

\section{GROWTH FUNCTION OF INFINITELY GENERATED GROUPS}\label{app:growth}
In this appendix, we formalize the concept of growth function introduced in Sec.~\ref{sec:growth}. 
To this end, we follow the generation of $H(T)$ for small time steps and identify its generators.
Recall that the inner bond is initialized in the unit element: 
\begin{equation}
    H(T=0)=\{e\}.
\end{equation}
After one time step, the reachable set is generated through group multiplications: 
 \begin{equation}
     H(T=1) = \{g_a g_b\}^{q-1}_{a,b = 0}\cup \{e\}.
 \end{equation}
Accordingly, after two time steps, there is
\begin{equation}
    H(T=2) = \{g_ag_bg_cg_d\}_{a,b,c,d = 0}^{q-1}\cup\{g_a g_b\}^{q-1}_{a,b = 0}\cup \{e\}
\end{equation}
More generally, $H(T)$ consists of all words of even length not exceeding $2T$, in terms of the generating set $S = \{g_a\}_{a=0}^{q-1}$, formally written as
\begin{equation}
    H(T) \equiv \{g|g = g_{a_1}g_{a_2}\cdots g_{a_{n}}, g_a\in S, n\in\text{even}, n\le2T\}.
\end{equation}
For convenience, we define an alternative generating set 
\begin{equation}
    \tilde{S} = \{e\}\cup \{g_ag_b\}^{q-1}_{a, b=0},
\end{equation}
which leads to an equivalent definition of $H(T)$:
\begin{equation}\label{eq:growth_function}
     H(T) \equiv \{h|h = h_{i_1}h_{i_2}\cdots h_{i_{T}}, h_i\in \tilde{S}\}.
\end{equation}
By construction, $H(T)\subseteq H(T+1)$; hence, the growth function as the size of $H(T)$ is a nondecreasing function of $T$.

The infinite-time limit $H \equiv \lim\limits_{T \to \infty} H(T)$ forms a \textit{finitely generated cancellative linear semigroup},  justified as follows:
\begin{itemize}
    \item \textit{Semigroup}---Closure under multiplication, but not necessarily under inverses.
    \item \textit{Cancellative}---For $h_{1,2,3}\in H$, it holds that if $h_1 h_2 = h_1 h_3$, then $h_2 = h_3$; similarly, if $h_2h_1=h_3h_1 $, then $h_2=h_3$.
    \item \textit{Finitely generated}---All elements of $H$ can be generated within finite time of multiplications from $\tilde{S}$.
    \item \textit{Linear}---$H$ admits a faithful finite-dimensional matrix representation (in this context, the adjoint representation of the unitaries).
\end{itemize}
All four properties follow directly from the constructive definition of $H(T)$. The classification of the growth behavior of such semigroups is well established in Refs.~\cite{Grigorchuk1988Cancellative,Okninski1995Generalised,Okninski1996Growth,Okninski1995GrowthII}, corresponding precisely to the three classes described in the main text. 

Furthermore, associated with $H$ there is an enveloping group $H'$, defined as the group generated by $\tilde{S}\cup\tilde{S}^{-1}$. The growth function of the group $H'$ is qualitatively identical to that of the semigroup $H$, under the above four properties~\cite{Grigorchuk1988Cancellative}.
Therefore, for simplicity, we do not distinguish between the notions of a semigroup and its enveloping group in the main text.

\section{EQUIVALENT REPRESENTATION OF THE TWO-QUBIT GATE}\label{app:realization}
For $q=2$, the local gate takes the form
\begin{equation}
    U = S (u_0\otimes\ket{0}\bra{0}+u_1\otimes\ket{1}\bra{1}).
\end{equation}
Here, the {\scriptsize${\mathrm{SWAP}}$} gate can be recast in terms of Pauli matrices as
\begin{equation}
    S = e^{i\pi/4} e^{-i\pi/4(\sigma^x_1\sigma^x_2+\sigma^y_1\sigma^y_2+\sigma^z_1\sigma^z_2)},
\end{equation}
and the controlled unitary can be rewritten as
\begin{equation}
    u_0\otimes\ket{0}\bra{0}+u_1\otimes\ket{1}\bra{1} = (v\otimes I)e^{-iK\pi\sigma^z_1\sigma^z_2}(w\otimes I),
\end{equation}
where $v,w\in\text{SU}(2)$ and $K\in[0,1)$ are chosen such that
\begin{equation}
    u_0 = ve^{-iK\pi\sigma^z}w, ~u_1 = ve^{iK\pi\sigma^z}w.
\end{equation}
Equivalently, they satisfy
\begin{equation}
    ve^{-2iK\pi\sigma^z}v^\dagger = u_0u_1^\dagger, ~w^\dagger e^{-2iK\pi\sigma^z}w = u_1^\dagger u_0.
\end{equation}
Thus, $v$ and $w^\dagger$ diagonalize $u_0u_1^\dagger$ and $u_1^\dagger u_0$ respectively, and the eigenvalues are $e^{\pm 2iK\pi}$. Since $u_0u_1^\dagger$ and $u_1^\dagger u_0$ are related by a similarity transformation, they share the same eigenvalues, and, therefore, the above decomposition always exists.

To conclude, the two-qubit gate can be written in the equivalent form
\begin{equation}\label{eq:repara}
    U = e^{i\pi/4} (I\otimes v)V\left[\frac{\pi}{4},\frac{\pi}{4},\frac{\pi}{4}+K\pi\right](w\otimes I),
\end{equation}
where $V[J_x,J_y,J_z] = e^{-i(J_x\sigma^x_1\sigma^x_2+J_y\sigma^y_1\sigma^y_2+J_z\sigma^z_1\sigma^z_2)}$. More generally, two-qubit dual-unitary gates admit such a parametrization~\cite{Bertini2019Exact}; here, we make the connection to Eq.~\eqref{eq:swap_control} explicit.

Next, we discuss possible realizations of such gates. The required ingredients are already accessible in current superconducting-qubit platforms, where high-fidelity single- and two-qubit operations enable programmable circuit implementations and have been used to probe a broad range of quantum many-body phenomena~\cite{Mi2021Information, Rosenberg2024Dynamics, Nguyen2024Programmable, Fischer2026Dynamical}.

\section{POLYNOMIALLY SCALING TRUNCATION OF THE BOND DIMENSION}\label{app:polynomial_truncation}

As demonstrated in Sec.~\ref{sec:TEE_generic_group}, the TEE typically exhibits linear growth for models in class III. In this section, we present a truncation scheme that approximates the MPS representation of the influence matrix with polynomially growing bond dimension while ensuring a certified upper bound on the error in local observables.
To be precise, we fix a CPTP quantum channel $\mathcal{K}$ acting on the system qudit between successive interactions with the bath, in analogy with the setup discussed in Fig.~\ref{fig:setting}(c) and Sec.~\ref{sec:stochastic}. 


Recall that the virtual bond of the MPO representation of the spacetime quantum channel takes values on the $\mathrm{PU}(q)$ group manifold. Our strategy of truncation is to construct a $\delta$ covering of the group manifold with a finite set of group elements $\{g_i\}_{i=1}^\mathcal{N}$, such that any group element lies within a ball of radius $\delta$ centered at $g_i$. The size of the set $\mathcal{N}$ corresponds to the bond dimension of the approximated MPO.
We define the distance between group elements according to the bi-invariant metric arising from the Killing form:
\begin{equation}
    d^2(g_1,g_2) = \Tr{\log(g_1^{-1}g_2)^{\dagger}\log(g_1^{-1}g_2)}.
\end{equation}
The size of the covering set scales as $\mathcal{N}\sim (1/\delta)^{q^2-1}$. We show below that replacing each group element in the virtual bond by its nearest neighbor in the covering results in an overall error in the expectation value of local observables bounded by $\epsilon = \delta \times T^2$. This leads to a trade-off between the size of the covering set and error, scaling polynomially with time:
\begin{equation}\label{eq:bond_dimension_bound}
    \mathcal{N}\sim C_q(\frac{1}{\delta})^{q^2-1}=C_q(\frac{T^2}{\epsilon})^{q^2-1}.
\end{equation}
Here, the additional prefactor $C_q$ is related to the volume of $\mathrm{PU}(q)$ and does not scale with $T$ or $\epsilon$:
\begin{equation}
C_q = \text{Vol}[\mathrm{PU}(q)]+o(\epsilon).
\end{equation}
We consider the example of $q=2$ featuring in the $\mathrm{PU}(2)$ group, using the Hopf parametrization:
\begin{equation}
g = \begin{pmatrix}
\cos(\theta)e^{i\phi} & \sin(\theta)e^{i\psi}\\
-\sin(\theta)e^{-i\psi} & \cos(\theta)e^{-i\phi}
\end{pmatrix}.
\end{equation}
The invariant distance on the group is given by
\begin{equation}
ds^2 = d\theta^2 + \cos^2(\theta)\,d\phi^2 + \sin^2(\theta)\,d\psi^2.
\end{equation}
We choose the following discretization of the group:
\begin{equation}
\theta = \frac{\delta n_{\theta}}{N}, \quad \phi = \frac{\delta n_{\phi}}{N}, \quad \psi = \frac{\delta n_{\psi}}{N},
\end{equation}
where \(n_{\theta}, n_{\phi}\) and \(n_{\psi}\) are integers ranging from $0$ to \(N_{\theta}, N_{\phi}\) and \(N_{\psi}\), respectively, with
\begin{equation}
N_{\theta} = \frac{\pi}{2\delta}, \quad N_{\phi} = N_{\psi} = \frac{2\pi}{\delta}.
\end{equation}
In this way, we select approximately $2\pi^3/\delta^3$ points on the group manifold, such that the distance between any two points is clearly less than $\delta$. While this discretization does not achieve the optimal constant $C_q$, it reproduces the correct scaling behavior.

\subsection{Bound in the case of classical non-Markovianity}
We begin by analyzing the classical case described by Eqs.~\eqref{eq:classical_diagram} and \eqref{eq:dynamics_O}. Our goal is to bound the error in the observable $\langle \hat{O}(T)\rangle$ when replacing the continuous group manifold by the discretized covering set as $\epsilon < \delta \times T^2$.

Note that according to the delta function in the conditional probability [Eq.~\eqref{eq:conditional_prob}] of the group random walk, group element $g_t$ is updated recursively in time:
\begin{equation}\label{eq:g_i_recursion}
    g_{t+1} = g_{b_{t}} g_t g_{a_t}.
\end{equation}
This equation allows us to rewrite the correlation function as a summation over $q^{2T}$ trajectories, each labeled by a sequence of $(a_t,b_t)$:
\begin{align} \label{eq:O_sum_over_trajectories}
    \langle \hat{O}(T)\rangle = &\sum_{\{(a_t,b_t)\}}\Tr{u(g_T)\ket{\psi_\text{e} }\bra{\psi_\text{e} } u(g_T)^\dagger\hat{O}}\nonumber\\
    &\times P(g_T|g_{T-1})\cdots P(g_2|g_1)P_0(g_1|e).
\end{align}
We now approximate each $g_t$ by the closest neighborhood $\tilde{g}_t$ from the chosen covering set. We define the finite difference of an arbitrary operator as $\Delta_{g_t} A(g_t) = A(g_t)-A(\tilde{g}_t)$. According to Eq.~\eqref{eq:g_i_recursion}, the approximation error accumulates as
\begin{equation}
d(g_{t},\tilde{g}_t) \leq \delta \times t<\delta \times T.
\end{equation}
To estimate the total error in the observable, we use an exact telescoping decomposition of the trajectory probabilities in Eq.~\eqref{eq:O_sum_over_trajectories}. For each trajectory, one has the identity
\begin{align}
&\prod_{t=2}^{T} P(g_t|g_{t-1})
-
\prod_{t=2}^{T} P(\tilde g_t|\tilde g_{t-1})
\nonumber\\
&\qquad =
\sum_{s=2}^{T}
P(\tilde g_T|\tilde g_{T-1})\cdots
P(\tilde g_{s+1}|\tilde g_s)
\,
\Delta_{g_{s-1}}P(g_s|g_{s-1})
\,\nonumber \\ &\qquad \times 
P(g_{s-1}|g_{s-2})\cdots
P(g_2|g_1),
\label{eq:telescoping_P}
\end{align}
where
\begin{equation}
\Delta_{g_{s-1}}P(g_s|g_{s-1})
\equiv
P(g_s|g_{s-1})
-
P(\tilde g_s|\tilde g_{s-1}).
\end{equation}
Using Eq.~\eqref{eq:telescoping_P}, together with the corresponding difference in the final observable, we obtain:
\begin{widetext}
\begin{align} \label{eq:O_estimate}
\Delta\langle \hat{O}(T)\rangle
&\le
\sum_{\{(a_t,b_t)\}}
\Big\|
\Delta_{g_T}
\Tr{
u(g_T)\ket{\psi_\text{e}}\bra{\psi_\text{e}}
u(g_T)^\dagger\hat{O}
}
P(g_T|g_{T-1})\cdots P(g_2|g_1)P_0(g_1|e)
\Big\|
\nonumber\\
&\quad+
\sum_{t=2}^{T}
\sum_{\{(a_t,b_t)\}}
\Big\|
\Tr{
u(\tilde g_T)\ket{\psi_\text{e}}\bra{\psi_\text{e}}
u(\tilde g_T)^\dagger\hat{O}
}
P(\tilde g_T|\tilde g_{T-1})\cdots
P(\tilde g_{t+1}|\tilde g_t)
\nonumber\\
&\qquad\qquad\times
\Delta_{g_{t-1}}P(g_t|g_{t-1})
P(g_{t-1}|g_{t-2})\cdots
P(g_2|g_1)P_0(g_1|e)
\Big\|.
\end{align}
\end{widetext}
We now bound each term in this expression. Assuming $| \hat{O} | \le 1$, we can estimate
\begin{align}\label{eq:uniform_O}
&\Big\|\Delta_{g_T}\Tr{u(g_T)\ket{\psi_\text{e} }\bra{\psi_\text{e}} u(g_T)^\dagger\hat{O}}\Big\| \le \delta \times T, \\ \label{eq:uniform_P}
&\Big\|\Delta_{g_{t-1}}P(g_t|g_{t-1})\Big\|\le q\delta \times T p_{a_t,b_t},
\end{align}
where $p_{a_t,b_t}=\frac{1}{q}|\bra{a}\psi_o\rangle |^{2} $ is a probability distribution independent of $g_t$. Here, we have used the relation between the distance bound and the bound on matrix elements:
\begin{align}
&\|\bra{a}\mathcal{K}[u(g)\ket{\psi_\text{e} }\bra{\psi_\text{e}} u(g)^\dagger]\ket{a}-\bra{a}\mathcal{K}[u(\tilde{g})\ket{\psi_\text{e} }\bra{\psi_\text{e}} u(\tilde{g})^\dagger]\ket{a}\|\nonumber\\
&< d(g,\tilde{g})<\delta \times T.
\end{align}

The first term in Eq.~\eqref{eq:O_estimate} can be bounded by substituting Eq.~\eqref{eq:uniform_O} and using the property that the products of conditional probabilities sum up to one.
In order to bound remaining terms, we use Eq.~\eqref{eq:uniform_P}:
\begin{widetext}
\begin{align}
&\sum\limits_{t=1}^{T}\sum_{\{(a_t,b_t)\}}\Big\|\Tr{u(g_T)\rho_e u(g_T)^\dagger\hat{O}}P(g_T|g_{T-1})\cdots \Delta_{g_{t-1}}P(g_t|g_{t-1})\cdots P(g_2|g_1)P_0(g_1|e)\Big\| \nonumber\\ 
\le & q \delta \times T\sum\limits_{t=1}^{T}\sum_{\{(a_t,b_t)\}}P(g_T|g_{T-1})\cdots p_{a_t,b_t}P(g_{t-1}|g_{t-2})\cdots P(g_2|g_1)P_0(g_1|e)\nonumber\\
= & q \delta \times T^2.
\end{align}
\end{widetext}
This finally leads to
\begin{equation}
\Delta\langle \hat{O}(T)\rangle\le \delta \times T +  q\delta \times T^2.
\end{equation}
By choosing $\delta = \epsilon/(q T^2)$, we can guarantee that the total error remains below a threshold $\epsilon$. We conclude that the approximated MPS representation with bond dimension $\mathcal{N} \sim C_q \left(T^2/\epsilon \right)^{q^2 - 1}$ achieves a certified bound $\epsilon$ of error for local observables.

\subsection{Bound in general cases}

We now consider a generic initial state given by the tensor product of a bath and impurity density matrices: 
\begin{equation}
    \rho_\text{in} = \rho_\text{bath}\otimes \rho_\text{imp},
\end{equation}
where the bath state $\rho_\text{bath}$ can be a matrix-product state or density operator.
The impurity evolves under a CPTP quantum channel $\mathcal{K}(\rho)=\sum\limits_\mu K_\mu \rho K_\mu^{\dagger}$. As demonstrated in Sec.~\ref{sec:quantum_memory}, generic initial states will induce quantum correlations, rendering Eq.~\eqref{eq:dynamics_O} inapplicable. In this case, we introduce the approach that evaluates the expectation value as a sum of overlaps between $\rho_\text{bath}$ and product operators, depicted as follows:
\begin{equation}
\tikzset{every picture/.style={line width=0.75pt}} 
\begin{tikzpicture}[x=0.75pt,y=0.75pt,yscale=-1,xscale=1]

\draw   (240,32.3) -- (280,32.3) -- (280,222.3) -- (240,222.3) -- cycle ;
\draw [fill={rgb, 255:red, 144; green, 19; blue, 254 }  ,fill opacity=1 ][line width=0.75]    (280.25,50.6) -- (330.5,50.59) ;
\draw [fill={rgb, 255:red, 144; green, 19; blue, 254 }  ,fill opacity=1 ][line width=0.75]    (280.21,111.86) -- (300.46,111.85) ;
\draw [fill={rgb, 255:red, 144; green, 19; blue, 254 }  ,fill opacity=1 ][line width=0.75]    (280.04,139.3) -- (330.29,139.29) ;
\draw [fill={rgb, 255:red, 144; green, 19; blue, 254 }  ,fill opacity=1 ][line width=0.75]    (279.75,204.8) -- (302,204.79) ;
\draw [line width=1.5]    (353.73,89.88) -- (353.52,160.37) ;
\draw [line width=1.5]    (353.47,182.06) -- (353.36,228.34) ;
\draw    (354.92,138.38) .. controls (370.51,138.37) and (397.73,137.83) .. (398.03,134.28) ;
\draw [fill={rgb, 255:red, 144; green, 19; blue, 254 }  ,fill opacity=1 ][line width=0.75]    (330.36,50.56) -- (377.19,50.54) ;
\draw [line width=1.5]    (353.68,27.95) -- (353.57,72.23) ;
\draw  [fill={rgb, 255:red, 255; green, 255; blue, 255 }  ,fill opacity=1 ][line width=1.5]  (353.66,232.3) -- (346.81,223.2) -- (360.39,223.07) -- cycle ;
\draw  [line width=1.5]  (346.37,28.83) .. controls (346.37,28.79) and (346.37,28.76) .. (346.37,28.72) .. controls (346.35,23.94) and (349.65,20.05) .. (353.74,20.03) .. controls (357.83,20.01) and (361.17,23.88) .. (361.19,28.66) .. controls (361.19,28.7) and (361.19,28.73) .. (361.19,28.77) -- cycle ;
\draw [fill={rgb, 255:red, 144; green, 19; blue, 254 }  ,fill opacity=1 ][line width=0.75]    (330,204.84) -- (369.84,204.82) ;
\draw [fill={rgb, 255:red, 144; green, 19; blue, 254 }  ,fill opacity=1 ][line width=0.75]    (330.36,139.28) -- (359.19,139.26) ;
\draw [fill={rgb, 255:red, 144; green, 19; blue, 254 }  ,fill opacity=1 ][line width=0.75]    (330,111.03) -- (369.84,111.01) ;
\draw  [color={rgb, 255:red, 0; green, 0; blue, 0 }  ,draw opacity=1 ][fill={rgb, 255:red, 38; green, 173; blue, 95 }  ,fill opacity=1 ][line width=0.75]  (345.48,139.27) .. controls (345.48,134.85) and (349.06,131.27) .. (353.48,131.27) .. controls (357.9,131.27) and (361.48,134.85) .. (361.48,139.27) .. controls (361.48,143.68) and (357.9,147.27) .. (353.48,147.27) .. controls (349.06,147.27) and (345.48,143.68) .. (345.48,139.27) -- cycle ;
\draw  [color={rgb, 255:red, 0; green, 0; blue, 0 }  ,draw opacity=1 ][fill={rgb, 255:red, 229; green, 126; blue, 33 }  ,fill opacity=1 ][line width=0.75]  (353.42,194.01) -- (362.07,204.83) -- (353.42,215.65) -- (344.77,204.83) -- cycle ;
\draw  [color={rgb, 255:red, 0; green, 0; blue, 0 }  ,draw opacity=1 ][fill={rgb, 255:red, 229; green, 126; blue, 33 }  ,fill opacity=1 ][line width=0.75]  (353.42,100.2) -- (362.07,111.02) -- (353.42,121.85) -- (344.77,111.02) -- cycle ;
\draw  [color={rgb, 255:red, 0; green, 0; blue, 0 }  ,draw opacity=1 ][fill={rgb, 255:red, 38; green, 173; blue, 95 }  ,fill opacity=1 ][line width=0.75]  (345.77,50.55) .. controls (345.77,46.13) and (349.36,42.55) .. (353.77,42.55) .. controls (358.19,42.55) and (361.77,46.13) .. (361.77,50.55) .. controls (361.77,54.97) and (358.19,58.55) .. (353.77,58.55) .. controls (349.36,58.55) and (345.77,54.97) .. (345.77,50.55) -- cycle ;
\draw  [fill={rgb, 255:red, 248; green, 231; blue, 28 }  ,fill opacity=1 ] (386.19,45.54) -- (386.19,54.54) -- (377.19,54.54) -- (377.19,45.54) -- cycle ;
\draw  [color={rgb, 255:red, 0; green, 0; blue, 0 }  ,draw opacity=1 ][fill={rgb, 255:red, 231; green, 102; blue, 102 }  ,fill opacity=1 ][line width=1.5]  (391.03,124.36) .. controls (391.03,119.43) and (395.06,115.44) .. (400.03,115.44) .. controls (405,115.44) and (409.02,119.43) .. (409.02,124.36) .. controls (409.02,129.29) and (405,133.28) .. (400.03,133.28) .. controls (395.06,133.28) and (391.03,129.29) .. (391.03,124.36) -- cycle ;
\draw  [line width=0.75]  (400.98,122.25) -- (405.09,122.25) -- (405.09,126.33) ;
\draw    (386.56,111.13) .. controls (400.52,110.7) and (397.73,109.83) .. (398.03,116.44) ;
\draw [fill={rgb, 255:red, 144; green, 19; blue, 254 }  ,fill opacity=1 ][line width=0.75]    (387.84,204.82) -- (398.67,204.8) ;

\draw (308.35,199.61) node [anchor=north west][inner sep=0.75pt]  [font=\normalsize]  {$a_{1}$};
\draw (308,105.52) node [anchor=north west][inner sep=0.75pt]  [font=\normalsize]  {$a_{t}$};
\draw (244,115.7) node [anchor=north west][inner sep=0.75pt]  [font=\normalsize]  {$\rho_\text{bath}$};
\draw (411,104.7) node [anchor=north west][inner sep=0.75pt]    {$\mathcal{K}$};
\draw (401.14,198.04) node [anchor=north west][inner sep=0.75pt]  [font=\normalsize]  {$\rho_\text{imp}$};
\draw (388.14,27.04) node [anchor=north west][inner sep=0.75pt]  [font=\normalsize]  {$\hat{O}$};
\draw (365.43,163.14) node [anchor=north west][inner sep=0.75pt]  [rotate=-90]  {$...$};
\draw (366.43,74.14) node [anchor=north west][inner sep=0.75pt]  [rotate=-90]  {$...$};
\draw (121,117.4) node [anchor=north west][inner sep=0.75pt]  [font=\normalsize]  {$\langle \hat{O}( T) \rangle =\sum\limits_{\{( a_{t} ,b_{t})\}}$};
\draw (307.6,74.3) node [anchor=north west][inner sep=0.75pt]  [rotate=-90]  {$...$};
\draw (307.6,163.3) node [anchor=north west][inner sep=0.75pt]  [rotate=-90]  {$...$};
\draw (372,102.7) node [anchor=north west][inner sep=0.75pt]  [font=\normalsize]  {$b_{t}$};
\draw (372,196.4) node [anchor=north west][inner sep=0.75pt]  [font=\normalsize]  {$b_{1}$};

\end{tikzpicture},
\end{equation}
which is formulated as
\begin{align}
    \langle \hat{O}(T)\rangle = &\text{Tr}[\rho_\text{bath} \sum_{\{(a_t,b_t)\}} \langle b_1|\rho_{\text{imp}}|b_1\rangle|a_1\rangle\langle a_1| \otimes\pi_{a_2,b_2}(g_1)\nonumber\\
    &\otimes\pi_{a_3,b_3}(g_2)\cdots \otimes\pi_{a_{T},b_{T}}(g_{T-1})\otimes M_{\hat{O}}(g_T)],
\end{align}
where $M_{\hat{O}}(g_T)= u(g_T)^\dagger \hat{O} u(g_T)$, and positive operators $\pi_{a_t,b_t}(g_{t-1})$ are defined as:
\begin{equation}
\tikzset{every picture/.style={line width=0.75pt}} 
\begin{tikzpicture}[x=0.75pt,y=0.75pt,yscale=-1,xscale=1]

\draw [fill={rgb, 255:red, 144; green, 19; blue, 254 }  ,fill opacity=1 ][line width=0.75]    (138.04,131.3) -- (153.29,131.29) ;
\draw [line width=1.5]    (176.73,78.88) -- (176.52,152.37) ;
\draw    (177.92,130.38) .. controls (193.51,130.37) and (220.73,129.83) .. (221.03,126.28) ;
\draw [fill={rgb, 255:red, 144; green, 19; blue, 254 }  ,fill opacity=1 ][line width=0.75]    (153.36,131.28) -- (182.19,131.26) ;
\draw [fill={rgb, 255:red, 144; green, 19; blue, 254 }  ,fill opacity=1 ][line width=0.75]    (153,103.03) -- (192.84,103.01) ;
\draw  [color={rgb, 255:red, 0; green, 0; blue, 0 }  ,draw opacity=1 ][fill={rgb, 255:red, 38; green, 173; blue, 95 }  ,fill opacity=1 ][line width=0.75]  (168.48,131.27) .. controls (168.48,126.85) and (172.06,123.27) .. (176.48,123.27) .. controls (180.9,123.27) and (184.48,126.85) .. (184.48,131.27) .. controls (184.48,135.68) and (180.9,139.27) .. (176.48,139.27) .. controls (172.06,139.27) and (168.48,135.68) .. (168.48,131.27) -- cycle ;
\draw  [color={rgb, 255:red, 0; green, 0; blue, 0 }  ,draw opacity=1 ][fill={rgb, 255:red, 229; green, 126; blue, 33 }  ,fill opacity=1 ][line width=0.75]  (176.42,92.2) -- (185.07,103.02) -- (176.42,113.85) -- (167.77,103.02) -- cycle ;
\draw  [color={rgb, 255:red, 0; green, 0; blue, 0 }  ,draw opacity=1 ][fill={rgb, 255:red, 231; green, 102; blue, 102 }  ,fill opacity=1 ][line width=1.5]  (214.03,116.36) .. controls (214.03,111.43) and (218.06,107.44) .. (223.03,107.44) .. controls (228,107.44) and (232.02,111.43) .. (232.02,116.36) .. controls (232.02,121.29) and (228,125.28) .. (223.03,125.28) .. controls (218.06,125.28) and (214.03,121.29) .. (214.03,116.36) -- cycle ;
\draw  [line width=0.75]  (223.98,114.25) -- (228.09,114.25) -- (228.09,118.33) ;
\draw    (209.56,103.13) .. controls (223.52,102.7) and (220.73,101.83) .. (221.03,108.44) ;

\draw (196,94.4) node [anchor=north west][inner sep=0.75pt]  [font=\normalsize]  {$b$};
\draw (139,96.4) node [anchor=north west][inner sep=0.75pt]  [font=\normalsize]  {$a$};
\draw (171,154.4) node [anchor=north west][inner sep=0.75pt]  [font=\normalsize]  {$g$};
\draw (158,59.4) node [anchor=north west][inner sep=0.75pt]  [font=\normalsize]  {$g_{b} gg_{a}$};
\draw (61,101.4) node [anchor=north west][inner sep=0.75pt]  [font=\normalsize]  {$\pi _{a,b}( g) =$};
\draw (107,182.4) node [anchor=north west][inner sep=0.75pt]  [font=\normalsize]  {$=\sum\limits_{\mu }\left[ u( g)^{\dagger } K_{\mu }^{\dagger }\ket{b}\bra{b} K_{\mu } u( g)\right] \otimes \ket{a}\bra{a}$};

\end{tikzpicture}
\end{equation}
Note that the quantum channel $\mathcal{K}$ acts on basis states $\ket{b}$ by the adjoint action.
This decomposition allows us to distribute the overall error across time steps, enabling a localized estimate of the contribution from
$g_t$ respectively. 
Similarly with the previous classical case, we approximate each $g_t$ in the formula by the closest neighborhood $\tilde{g}_t$ from the covering set. The first-order error reads
\begin{widetext}
\begin{align}
&\Delta\langle \hat{O}(T)\rangle =\nonumber\\
& \sum\limits_{t=1}^{T}\Tr{\rho_{\text{bath}}\sum\limits_{\{(a_t,b_t)\}} \langle b_1|\rho_{\text{imp}}|b_1\rangle|a_1\rangle\langle a_1| \otimes\pi_{a_2,b_2}(g_1)\otimes\cdots \otimes\Delta_{g_{t-1}}\pi_{a_{t},b_t}(g_{t-1})\otimes\cdots\otimes\pi_{a_{T},b_{T}}(g_{T-1})\otimes M_O(g_T)}.
\end{align}
\end{widetext}
Again, since the distance between $g_t$ and $\tilde{g}_t$ is bounded, the error in $\pi_{a_{t},b_t}(g_{t-1})$ can also be uniformly bounded as
\begin{equation}\label{eq:uniform_bound}
\|\Delta_{g_{t-1}}\pi_{a_{t},b_t}(g_{t-1})\|\le q \delta \times T \|\tilde{\pi}_{a_{t},b_t}\|,
\end{equation}
where the bound holds for any matrix element in the computational basis as well. The $g$-independent operator $\tilde{\pi}_{a_{t},b_t}$ is defined as
\begin{equation}
\tilde{\pi}_{a,b}=\sum\limits_\mu [\tilde{K}^{\dagger}_\mu\ket{b}\bra{b}\tilde{K}_\mu]\otimes\ket{a}\bra{a},    
\end{equation}
with $(\tilde{K}_\mu)_{a,b}=\delta_{a,\mu}/\sqrt{q}$.The corresponding fictitious quantum channel $\tilde{\mathcal{K}}(\rho)=\sum\limits_\mu \tilde{K}_\mu\rho \tilde{K}_\mu^\dagger$ is nothing but a reset channel that prepares the pure state $\ket{\psi}=\sum\limits_a{\ket{a}}/\sqrt{q}$.

Proceeding in analogy with the classical case, we use Eq.~\eqref{eq:uniform_bound} and further replace $M_{\hat{O}}(g_T)$ with $M_{|\hat{O}|}(g_T)=u(g_T)^\dagger |\hat{O}_{a}||a\rangle\langle a|u(g_T)$. This leads to the estimate
\begin{widetext}
\begin{align}
\|\delta\langle \hat{O}(T)\rangle\|&\le \delta \times T \times \sum\limits_{t=1}^{T}\Tr{\rho_{\text{bath}}\sum\limits_{\{a_t,b_t\}} \langle b_1|\rho_{\text{imp}}|b_1\rangle|a_1\rangle\langle a_1| \otimes\pi_{a_2,b_2}(g_1) \otimes\cdots \otimes \tilde{\pi}_{a_t,b_t}\cdots\otimes\pi_{a_{T},b_{T}}(g_{T-1})\otimes M_{|\hat{O}|}(g_T)}\nonumber\\
&\le\delta \times T^2.
\end{align}
\end{widetext}
In the final inequality, we have used the fact that each term under the trace corresponds to a physical correlation and, therefore, is bounded by one. This reproduces the same trade-off between error $\epsilon$ and the truncated bond dimension $\mathcal{N}$ as in the classical case.

\section{SOLVABLE INITIAL STATES AND PRODUCT-STATE INFLUENCE MATRICES
}\label{app:reproduce}

Reference~\cite{Piroli2020Exact} identified a class of initial MPS, termed ``solvable initial states,'' that leads to a product-state IM and area-law TEE for arbitrary dual-unitary circuits. Since our unitaries [Eq.~\eqref{eq:swap_control}] belong to the dual-unitary class, the same results are expected to hold.
In this appendix, we verify this by starting from the exact MPO construction [Fig.~\ref{fig:MPO_tensors}(c)] without invoking dual unitarity to show the consistency of our results with the literature.

We begin by quoting the condition of solvable initial states:
\begin{equation}\label{eq:solvable_condition}
\tikzset{every picture/.style={line width=0.75pt}} 
\begin{tikzpicture}[x=0.75pt,y=0.75pt,yscale=-1,xscale=1]

\draw [line width=2.25]    (120,146.04) -- (149,146.04) ;
\draw  [fill={rgb, 255:red, 2; green, 100; blue, 200 }  ,fill opacity=1 ] (128.47,138.68) .. controls (128.47,137.42) and (129.49,136.4) .. (130.75,136.4) -- (137.59,136.4) .. controls (138.85,136.4) and (139.87,137.42) .. (139.87,138.68) -- (139.87,153.39) .. controls (139.87,154.65) and (138.85,155.67) .. (137.59,155.67) -- (130.75,155.67) .. controls (129.49,155.67) and (128.47,154.65) .. (128.47,153.39) -- cycle ;
\draw  [fill={rgb, 255:red, 2; green, 100; blue, 200 }  ,fill opacity=1 ][line width=0.75]  (133.3,138.63) -- (137.3,138.63) -- (137.3,142.63) ;

\draw [fill={rgb, 255:red, 144; green, 19; blue, 254 }  ,fill opacity=1 ][line width=0.75]    (134.2,126) -- (134.25,135.9) ;
\draw [line width=2.25]    (148,146.04) -- (177,146.04) ;
\draw  [fill={rgb, 255:red, 208; green, 2; blue, 27 }  ,fill opacity=1 ] (156.47,138.68) .. controls (156.47,137.42) and (157.49,136.4) .. (158.75,136.4) -- (165.59,136.4) .. controls (166.85,136.4) and (167.87,137.42) .. (167.87,138.68) -- (167.87,153.39) .. controls (167.87,154.65) and (166.85,155.67) .. (165.59,155.67) -- (158.75,155.67) .. controls (157.49,155.67) and (156.47,154.65) .. (156.47,153.39) -- cycle ;
\draw  [fill={rgb, 255:red, 208; green, 2; blue, 27 }  ,fill opacity=1 ][line width=0.75]  (161.3,138.63) -- (165.3,138.63) -- (165.3,142.63) ;

\draw [fill={rgb, 255:red, 119; green, 241; blue, 254 }  ,fill opacity=1 ][line width=0.75]    (162.2,125) -- (162.25,135.9) ;
\draw  [fill={rgb, 255:red, 241; green, 238; blue, 28 }  ,fill opacity=1 ] (105.57,145.8) -- (113.4,137.95) -- (121.25,145.79) -- (113.41,153.64) -- cycle ;
\draw  [fill={rgb, 255:red, 255; green, 255; blue, 255 }  ,fill opacity=1 ] (131.4,123.2) .. controls (131.4,121.65) and (132.65,120.4) .. (134.2,120.4) .. controls (135.75,120.4) and (137,121.65) .. (137,123.2) .. controls (137,124.75) and (135.75,126) .. (134.2,126) .. controls (132.65,126) and (131.4,124.75) .. (131.4,123.2) -- cycle ;
\draw [line width=2.25]    (232,145.54) -- (232,145.54) ;
\draw [line width=2.25]    (231,145.54) -- (260,145.54) ;
\draw [fill={rgb, 255:red, 144; green, 19; blue, 254 }  ,fill opacity=1 ][line width=0.75]    (245.1,129.8) -- (245.06,118.9) ;
\draw  [fill={rgb, 255:red, 241; green, 238; blue, 28 }  ,fill opacity=1 ] (229.57,145.3) -- (237.4,137.45) -- (245.25,145.29) -- (237.41,153.14) -- cycle ;
\draw  [fill={rgb, 255:red, 255; green, 255; blue, 255 }  ,fill opacity=1 ] (248.06,132.57) .. controls (248.08,134.11) and (246.84,135.38) .. (245.29,135.4) .. controls (243.75,135.42) and (242.48,134.19) .. (242.46,132.64) .. controls (242.44,131.09) and (243.67,129.82) .. (245.22,129.8) .. controls (246.77,129.78) and (248.04,131.02) .. (248.06,132.57) -- cycle ;

\draw (190,135.16) node [anchor=north west][inner sep=0.75pt]    {$=\frac{1}{q}$};

\end{tikzpicture}.
\end{equation}
The colored diamond tensor represents the left steady state $S_D$ of the quantum channel in the auxiliary Hilbert space: 
\begin{equation}
    \sum_{a,b=0}^{q-1}(B^{b}A^{a})^\dagger S_D B^{b}A^{a} = S_D,
\end{equation}
which should not be confused with the $\rho_D$ in Eq.~\eqref{eq:steady_I}. We also introduce two tensor equations to be used later:
\begin{equation}\label{eq:tensor_relation_1}
\tikzset{every picture/.style={line width=0.75pt}} 
\begin{tikzpicture}[x=0.75pt,y=0.75pt,yscale=-1,xscale=1]

\draw [fill={rgb, 255:red, 144; green, 19; blue, 254 }  ,fill opacity=1 ][line width=0.75]    (195.83,90.51) -- (234.08,90.49) ;
\draw [line width=1.5]    (212.36,70.16) -- (212.26,112.05) ;
\draw  [line width=1.5]  (205.62,70.95) .. controls (205.62,70.92) and (205.62,70.88) .. (205.62,70.85) .. controls (205.6,66.55) and (208.64,63.04) .. (212.42,63.03) .. controls (216.2,63.01) and (219.28,66.49) .. (219.3,70.79) .. controls (219.3,70.83) and (219.3,70.86) .. (219.3,70.9) -- cycle ;
\draw  [color={rgb, 255:red, 0; green, 0; blue, 0 }  ,draw opacity=1 ][fill={rgb, 255:red, 38; green, 173; blue, 95 }  ,fill opacity=1 ][line width=1]  (204.79,89.69) .. controls (204.79,85.71) and (208.22,82.49) .. (212.45,82.49) .. controls (216.68,82.49) and (220.11,85.71) .. (220.11,89.69) .. controls (220.11,93.67) and (216.68,96.89) .. (212.45,96.89) .. controls (208.22,96.89) and (204.79,93.67) .. (204.79,89.69) -- cycle ;
\draw  [fill={rgb, 255:red, 255; green, 255; blue, 255 }  ,fill opacity=1 ] (194.72,87.44) .. controls (196.26,87.44) and (197.51,88.7) .. (197.51,90.25) .. controls (197.5,91.79) and (196.24,93.04) .. (194.7,93.04) .. controls (193.15,93.03) and (191.9,91.78) .. (191.91,90.23) .. controls (191.91,88.68) and (193.17,87.43) .. (194.72,87.44) -- cycle ;
\draw [line width=1.5]    (302.69,95.49) -- (302.6,113.38) ;
\draw  [line width=1.5]  (295.95,96.28) .. controls (295.95,96.25) and (295.95,96.22) .. (295.95,96.18) .. controls (295.93,91.88) and (298.98,88.38) .. (302.76,88.36) .. controls (306.53,88.35) and (309.61,91.82) .. (309.63,96.13) .. controls (309.63,96.16) and (309.63,96.2) .. (309.63,96.23) -- cycle ;
\draw [fill={rgb, 255:red, 144; green, 19; blue, 254 }  ,fill opacity=1 ][line width=0.75]    (287.5,79.51) -- (316.74,79.49) ;
\draw  [fill={rgb, 255:red, 255; green, 255; blue, 255 }  ,fill opacity=1 ] (287.38,76.44) .. controls (288.93,76.44) and (290.18,77.7) .. (290.17,79.25) .. controls (290.17,80.79) and (288.91,82.04) .. (287.37,82.04) .. controls (285.82,82.03) and (284.57,80.78) .. (284.57,79.23) .. controls (284.58,77.68) and (285.84,76.43) .. (287.38,76.44) -- cycle ;

\draw (250.33,89.4) node [anchor=north west][inner sep=0.75pt]    {$=$};

\end{tikzpicture},
\end{equation}
and
\begin{equation}\label{eq:tensor_relation_2}
\tikzset{every picture/.style={line width=0.75pt}} 
\begin{tikzpicture}[x=0.75pt,y=0.75pt,yscale=-1,xscale=1]

\draw [line width=1.5]    (315.69,95.49) -- (315.6,133.38) ;
\draw  [line width=1.5]  (308.95,96.28) .. controls (308.95,96.25) and (308.95,96.22) .. (308.95,96.18) .. controls (308.93,91.88) and (311.98,88.38) .. (315.76,88.36) .. controls (319.53,88.35) and (322.61,91.82) .. (322.63,96.13) .. controls (322.63,96.16) and (322.63,96.2) .. (322.63,96.23) -- cycle ;
\draw [fill={rgb, 255:red, 144; green, 19; blue, 254 }  ,fill opacity=1 ][line width=0.75]    (293.83,114.65) -- (338.08,114.63) ;
\draw  [fill={rgb, 255:red, 229; green, 126; blue, 33 }  ,fill opacity=1 ][line width=1]  (315.46,104.9) -- (323.45,114.64) -- (315.46,124.38) -- (307.47,114.64) -- cycle ;
\draw [line width=1.5]    (400.36,119.82) -- (400.26,132.71) ;
\draw  [line width=1.5]  (393.62,120.61) .. controls (393.62,120.58) and (393.62,120.55) .. (393.62,120.52) .. controls (393.6,116.21) and (396.64,112.71) .. (400.42,112.7) .. controls (404.2,112.68) and (407.28,116.16) .. (407.3,120.46) .. controls (407.3,120.49) and (407.3,120.53) .. (407.3,120.56) -- cycle ;
\draw [fill={rgb, 255:red, 144; green, 19; blue, 254 }  ,fill opacity=1 ][line width=0.75]    (375.83,107.31) -- (388.08,107.3) ;
\draw  [fill={rgb, 255:red, 255; green, 255; blue, 255 }  ,fill opacity=1 ] (390.89,104.5) .. controls (392.44,104.51) and (393.69,105.77) .. (393.68,107.31) .. controls (393.68,108.86) and (392.42,110.11) .. (390.87,110.1) .. controls (389.33,110.1) and (388.08,108.84) .. (388.08,107.3) .. controls (388.09,105.75) and (389.34,104.5) .. (390.89,104.5) -- cycle ;
\draw [fill={rgb, 255:red, 144; green, 19; blue, 254 }  ,fill opacity=1 ][line width=0.75]    (409.83,107.51) -- (426.08,107.49) ;
\draw  [fill={rgb, 255:red, 255; green, 255; blue, 255 }  ,fill opacity=1 ] (409.72,104.44) .. controls (411.26,104.44) and (412.51,105.7) .. (412.51,107.25) .. controls (412.5,108.79) and (411.24,110.04) .. (409.7,110.04) .. controls (408.15,110.03) and (406.9,108.78) .. (406.91,107.23) .. controls (406.91,105.68) and (408.17,104.43) .. (409.72,104.44) -- cycle ;

\draw (349.33,105.07) node [anchor=north west][inner sep=0.75pt]    {$=$};

\end{tikzpicture}.
\end{equation}
Both of them can be verified via explicit tensor contractions using the definitions of the MPO components in Fig.~\ref{fig:MPO_tensors}(c).

We now contract the spacetime quantum channel $\mathcal{M}$ with solvable initial states, proceeding from top to bottom:
\begin{widetext}
\begin{equation}
\tikzset{every picture/.style={line width=0.75pt}} 
\begin{tikzpicture}[x=0.75pt,y=0.75pt,yscale=-1,xscale=1]

\draw [line width=2.25]    (81.61,53.67) -- (81.42,138.67) ;
\draw [fill={rgb, 255:red, 144; green, 19; blue, 254 }  ,fill opacity=1 ][line width=0.75]    (88.5,94.18) -- (131.74,94.16) ;
\draw [line width=1.5]    (110.03,73.82) -- (109.93,138.71) ;
\draw  [line width=1.5]  (103.28,74.61) .. controls (103.28,74.58) and (103.28,74.55) .. (103.28,74.52) .. controls (103.26,70.21) and (106.31,66.71) .. (110.09,66.7) .. controls (113.87,66.68) and (116.95,70.16) .. (116.97,74.46) .. controls (116.97,74.49) and (116.97,74.53) .. (116.97,74.56) -- cycle ;
\draw  [color={rgb, 255:red, 0; green, 0; blue, 0 }  ,draw opacity=1 ][fill={rgb, 255:red, 38; green, 173; blue, 95 }  ,fill opacity=1 ][line width=1]  (102.46,93.36) .. controls (102.46,89.38) and (105.89,86.16) .. (110.12,86.16) .. controls (114.35,86.16) and (117.78,89.38) .. (117.78,93.36) .. controls (117.78,97.33) and (114.35,100.56) .. (110.12,100.56) .. controls (105.89,100.56) and (102.46,97.33) .. (102.46,93.36) -- cycle ;
\draw [fill={rgb, 255:red, 144; green, 19; blue, 254 }  ,fill opacity=1 ][line width=0.75]    (88.17,119.98) -- (132.41,119.96) ;
\draw  [fill={rgb, 255:red, 229; green, 126; blue, 33 }  ,fill opacity=1 ][line width=1]  (109.79,110.23) -- (117.78,119.97) -- (109.79,129.71) -- (101.8,119.97) -- cycle ;
\draw  [fill={rgb, 255:red, 208; green, 2; blue, 27 }  ,fill opacity=1 ] (89.68,88.52) .. controls (90.94,88.53) and (91.96,89.56) .. (91.95,90.82) -- (91.9,97.66) .. controls (91.89,98.92) and (90.86,99.93) .. (89.6,99.92) -- (74.89,99.81) .. controls (73.63,99.8) and (72.62,98.77) .. (72.63,97.51) -- (72.68,90.68) .. controls (72.69,89.42) and (73.72,88.4) .. (74.98,88.41) -- cycle ;
\draw  [line width=0.75]  (89.7,93.36) -- (89.67,97.36) -- (85.67,97.33) ;

\draw  [fill={rgb, 255:red, 2; green, 100; blue, 200 }  ,fill opacity=1 ] (89.48,114.02) .. controls (90.74,114.03) and (91.76,115.06) .. (91.75,116.32) -- (91.7,123.16) .. controls (91.69,124.42) and (90.66,125.43) .. (89.4,125.42) -- (74.69,125.31) .. controls (73.44,125.3) and (72.42,124.28) .. (72.43,123.02) -- (72.48,116.18) .. controls (72.49,114.92) and (73.52,113.9) .. (74.78,113.91) -- cycle ;
\draw  [fill={rgb, 255:red, 2; green, 100; blue, 200 }  ,fill opacity=1 ][line width=0.75]  (89.5,118.86) -- (89.47,122.85) -- (85.47,122.83) ;

\draw  [fill={rgb, 255:red, 241; green, 238; blue, 28 }  ,fill opacity=1 ] (73.77,48.83) -- (81.61,40.98) -- (89.45,48.82) -- (81.61,56.67) -- cycle ;
\draw [fill={rgb, 255:red, 144; green, 19; blue, 254 }  ,fill opacity=1 ][line width=0.75]    (95.57,68.02) -- (86.67,67.88) ;
\draw  [fill={rgb, 255:red, 255; green, 255; blue, 255 }  ,fill opacity=1 ] (97.38,65.11) .. controls (98.93,65.11) and (100.18,66.37) .. (100.17,67.91) .. controls (100.17,69.46) and (98.91,70.71) .. (97.37,70.71) .. controls (95.82,70.7) and (94.57,69.44) .. (94.57,67.9) .. controls (94.58,66.35) and (95.84,65.1) .. (97.38,65.11) -- cycle ;
\draw  [fill={rgb, 255:red, 2; green, 100; blue, 200 }  ,fill opacity=1 ] (89.39,62.67) .. controls (90.65,62.67) and (91.67,63.69) .. (91.67,64.95) -- (91.67,71.79) .. controls (91.67,73.05) and (90.65,74.07) .. (89.39,74.07) -- (74.68,74.07) .. controls (73.42,74.07) and (72.4,73.05) .. (72.4,71.79) -- (72.4,64.95) .. controls (72.4,63.69) and (73.42,62.67) .. (74.68,62.67) -- cycle ;
\draw  [fill={rgb, 255:red, 2; green, 100; blue, 200 }  ,fill opacity=1 ][line width=0.75]  (89.44,67.5) -- (89.44,71.5) -- (85.44,71.5) ;

\draw [line width=2.25]    (183.95,103) -- (183.75,135) ;
\draw [fill={rgb, 255:red, 144; green, 19; blue, 254 }  ,fill opacity=1 ][line width=0.75]    (195.83,90.51) -- (234.08,90.49) ;
\draw [line width=1.5]    (212.36,70.16) -- (212.26,135.05) ;
\draw  [line width=1.5]  (205.62,70.95) .. controls (205.62,70.92) and (205.62,70.88) .. (205.62,70.85) .. controls (205.6,66.55) and (208.64,63.04) .. (212.42,63.03) .. controls (216.2,63.01) and (219.28,66.49) .. (219.3,70.79) .. controls (219.3,70.83) and (219.3,70.86) .. (219.3,70.9) -- cycle ;
\draw  [color={rgb, 255:red, 0; green, 0; blue, 0 }  ,draw opacity=1 ][fill={rgb, 255:red, 38; green, 173; blue, 95 }  ,fill opacity=1 ][line width=1]  (204.79,89.69) .. controls (204.79,85.71) and (208.22,82.49) .. (212.45,82.49) .. controls (216.68,82.49) and (220.11,85.71) .. (220.11,89.69) .. controls (220.11,93.67) and (216.68,96.89) .. (212.45,96.89) .. controls (208.22,96.89) and (204.79,93.67) .. (204.79,89.69) -- cycle ;
\draw [fill={rgb, 255:red, 144; green, 19; blue, 254 }  ,fill opacity=1 ][line width=0.75]    (190.5,116.31) -- (234.75,116.3) ;
\draw  [fill={rgb, 255:red, 229; green, 126; blue, 33 }  ,fill opacity=1 ][line width=1]  (212.12,106.56) -- (220.11,116.3) -- (212.12,126.05) -- (204.13,116.3) -- cycle ;
\draw  [fill={rgb, 255:red, 2; green, 100; blue, 200 }  ,fill opacity=1 ] (191.82,110.35) .. controls (193.08,110.36) and (194.09,111.39) .. (194.08,112.65) -- (194.03,119.49) .. controls (194.02,120.75) and (192.99,121.76) .. (191.73,121.75) -- (177.03,121.65) .. controls (175.77,121.64) and (174.76,120.61) .. (174.76,119.35) -- (174.81,112.51) .. controls (174.82,111.25) and (175.85,110.24) .. (177.11,110.25) -- cycle ;
\draw  [fill={rgb, 255:red, 2; green, 100; blue, 200 }  ,fill opacity=1 ][line width=0.75]  (191.83,115.19) -- (191.8,119.19) -- (187.8,119.16) ;

\draw  [fill={rgb, 255:red, 241; green, 238; blue, 28 }  ,fill opacity=1 ] (176.1,98.16) -- (183.94,90.32) -- (191.78,98.16) -- (183.95,106) -- cycle ;
\draw  [fill={rgb, 255:red, 255; green, 255; blue, 255 }  ,fill opacity=1 ] (194.72,87.44) .. controls (196.26,87.44) and (197.51,88.7) .. (197.51,90.25) .. controls (197.5,91.79) and (196.24,93.04) .. (194.7,93.04) .. controls (193.15,93.03) and (191.9,91.78) .. (191.91,90.23) .. controls (191.91,88.68) and (193.17,87.43) .. (194.72,87.44) -- cycle ;
\draw [line width=2.25]    (287.28,101.33) -- (287.09,133.33) ;
\draw [line width=1.5]    (315.69,95.49) -- (315.6,133.38) ;
\draw  [line width=1.5]  (308.95,96.28) .. controls (308.95,96.25) and (308.95,96.22) .. (308.95,96.18) .. controls (308.93,91.88) and (311.98,88.38) .. (315.76,88.36) .. controls (319.53,88.35) and (322.61,91.82) .. (322.63,96.13) .. controls (322.63,96.16) and (322.63,96.2) .. (322.63,96.23) -- cycle ;
\draw [fill={rgb, 255:red, 144; green, 19; blue, 254 }  ,fill opacity=1 ][line width=0.75]    (293.83,114.65) -- (338.08,114.63) ;
\draw  [fill={rgb, 255:red, 229; green, 126; blue, 33 }  ,fill opacity=1 ][line width=1]  (315.46,104.9) -- (323.45,114.64) -- (315.46,124.38) -- (307.47,114.64) -- cycle ;
\draw  [fill={rgb, 255:red, 2; green, 100; blue, 200 }  ,fill opacity=1 ] (295.15,108.69) .. controls (296.41,108.7) and (297.42,109.73) .. (297.41,110.98) -- (297.36,117.82) .. controls (297.35,119.08) and (296.33,120.1) .. (295.07,120.09) -- (280.36,119.98) .. controls (279.1,119.97) and (278.09,118.94) .. (278.1,117.68) -- (278.15,110.84) .. controls (278.16,109.58) and (279.19,108.57) .. (280.44,108.58) -- cycle ;
\draw  [fill={rgb, 255:red, 2; green, 100; blue, 200 }  ,fill opacity=1 ][line width=0.75]  (295.17,113.52) -- (295.14,117.52) -- (291.14,117.49) ;

\draw  [fill={rgb, 255:red, 241; green, 238; blue, 28 }  ,fill opacity=1 ] (279.43,96.5) -- (287.27,88.65) -- (295.12,96.49) -- (287.28,104.33) -- cycle ;
\draw [line width=2.25]    (396.28,94) -- (396.09,129) ;
\draw [line width=1.5]    (429.69,117.16) -- (429.6,129.05) ;
\draw  [line width=1.5]  (422.95,117.95) .. controls (422.95,117.92) and (422.95,117.88) .. (422.95,117.85) .. controls (422.93,113.55) and (425.98,110.04) .. (429.76,110.03) .. controls (433.53,110.01) and (436.61,113.49) .. (436.63,117.79) .. controls (436.63,117.83) and (436.63,117.86) .. (436.63,117.9) -- cycle ;
\draw [fill={rgb, 255:red, 144; green, 19; blue, 254 }  ,fill opacity=1 ][line width=0.75]    (402.83,107.31) -- (412.08,107.3) ;
\draw  [fill={rgb, 255:red, 2; green, 100; blue, 200 }  ,fill opacity=1 ] (404.15,101.35) .. controls (405.41,101.36) and (406.42,102.39) .. (406.41,103.65) -- (406.36,110.49) .. controls (406.35,111.75) and (405.33,112.76) .. (404.07,112.75) -- (389.36,112.65) .. controls (388.1,112.64) and (387.09,111.61) .. (387.1,110.35) -- (387.15,103.51) .. controls (387.16,102.25) and (388.19,101.24) .. (389.44,101.25) -- cycle ;
\draw  [fill={rgb, 255:red, 2; green, 100; blue, 200 }  ,fill opacity=1 ][line width=0.75]  (404.17,106.19) -- (404.14,110.19) -- (400.14,110.16) ;

\draw  [fill={rgb, 255:red, 241; green, 238; blue, 28 }  ,fill opacity=1 ] (388.43,89.16) -- (396.27,81.32) -- (404.12,89.16) -- (396.28,97) -- cycle ;
\draw [fill={rgb, 255:red, 144; green, 19; blue, 254 }  ,fill opacity=1 ][line width=0.75]    (322.5,79.51) -- (338.74,79.49) ;
\draw  [fill={rgb, 255:red, 255; green, 255; blue, 255 }  ,fill opacity=1 ] (322.38,76.44) .. controls (323.93,76.44) and (325.18,77.7) .. (325.17,79.25) .. controls (325.17,80.79) and (323.91,82.04) .. (322.37,82.04) .. controls (320.82,82.03) and (319.57,80.78) .. (319.57,79.23) .. controls (319.58,77.68) and (320.84,76.43) .. (322.38,76.44) -- cycle ;
\draw  [fill={rgb, 255:red, 255; green, 255; blue, 255 }  ,fill opacity=1 ] (414.89,104.5) .. controls (416.44,104.51) and (417.69,105.77) .. (417.68,107.31) .. controls (417.68,108.86) and (416.42,110.11) .. (414.87,110.1) .. controls (413.33,110.1) and (412.08,108.84) .. (412.08,107.3) .. controls (412.09,105.75) and (413.34,104.5) .. (414.89,104.5) -- cycle ;
\draw [fill={rgb, 255:red, 144; green, 19; blue, 254 }  ,fill opacity=1 ][line width=0.75]    (424.5,80.18) -- (440.74,80.16) ;
\draw  [fill={rgb, 255:red, 255; green, 255; blue, 255 }  ,fill opacity=1 ] (424.38,77.11) .. controls (425.93,77.11) and (427.18,78.37) .. (427.17,79.91) .. controls (427.17,81.46) and (425.91,82.71) .. (424.37,82.71) .. controls (422.82,82.7) and (421.57,81.44) .. (421.57,79.9) .. controls (421.58,78.35) and (422.84,77.1) .. (424.38,77.11) -- cycle ;
\draw [fill={rgb, 255:red, 144; green, 19; blue, 254 }  ,fill opacity=1 ][line width=0.75]    (424.5,99.51) -- (440.74,99.49) ;
\draw  [fill={rgb, 255:red, 255; green, 255; blue, 255 }  ,fill opacity=1 ] (424.38,96.44) .. controls (425.93,96.44) and (427.18,97.7) .. (427.17,99.25) .. controls (427.17,100.79) and (425.91,102.04) .. (424.37,102.04) .. controls (422.82,102.03) and (421.57,100.78) .. (421.57,99.23) .. controls (421.58,97.68) and (422.84,96.43) .. (424.38,96.44) -- cycle ;

\draw (140.67,93.4) node [anchor=north west][inner sep=0.75pt]    {$=\frac{1}{q}$};
\draw (244.33,93.4) node [anchor=north west][inner sep=0.75pt]    {$=\frac{1}{q}$};
\draw (344.33,93.07) node [anchor=north west][inner sep=0.75pt]    {$=\frac{1}{q}$};

\end{tikzpicture}.
\end{equation}
\end{widetext}
The first equality applies the solvability condition Eq.~\eqref{eq:solvable_condition}, and the second and third steps use Eqs.~\eqref{eq:tensor_relation_1} and \eqref{eq:tensor_relation_2}, respectively. This contraction can be performed iteratively. At each step, the nontrivial tensors cancel, leaving only identity operators. Therefore, the influence matrix reduces to a product state, representing a perfectly Markovian bath with infinite temperature that completely thermalizes local qudits. 

Furthermore, Ref.~\cite{Foligno2025Non} extended the concept of solvable initial states to dual-unitaries featuring local conservation laws. Specifically, in the presence of right-moving single-site conserved charges, there are more initial MPS that lead to product-state influence matrices. In our model [Eq.~\eqref{eq:swap_control}], a conserved charge $o$ satisfies
\begin{align}
    & U^\dagger (o\otimes I_q) U = \sum_{a=0}^{q-1} \ket{a}\bra{a}\otimes u_a^\dagger ou_a = I_q\otimes o\nonumber\\
    &\Longleftrightarrow \qquad u_a^\dagger ou_a = o\nonumber\\
    &\Longleftrightarrow  \qquad  [o,u_a] = 0, \forall a.
\end{align}
It follows that $o$ must commute with all the controlled unitaries $\{u_a\}$, as well as any unitary generated by their compositions and inverses. We denote by $N=\langle\{u_a\}_{a=0}^{q-1}\rangle$ the subgroup of $\mathrm{U}(q)$ generated by these unitaries, which should not be confused with $H$ as a subgroup of $\mathrm{PU}(q)$ introduced in Sec.~\ref{sec:growth}. To find the complete set of right-moving charges, notice that $N$ may act reducibly on the Hilbert space of dimension $q$, even though the unitaries $u_a$ themselves are irreducible as elements of $\mathrm{U}(q)$. Let $u\in N$ admit a decomposition into a direct sum of irreducible representations (irreps): $u = \oplus_\lambda u_\lambda$, where each $\lambda$ labels an irrep of $N$ with dimension $d_\lambda$. Then, the linearly independent charges are given by the projectors onto these irreps: $o_\lambda = \Pi_\lambda$, each of rank $d_\lambda$.

We quote the condition for the so-called ``charged solvable initial states'' as follows \cite{Foligno2025Non}: 
\begin{equation}
\tikzset{every picture/.style={line width=0.75pt}} 
\begin{tikzpicture}[x=0.75pt,y=0.75pt,yscale=-1,xscale=1]

\draw [line width=2.25]    (120,143.04) -- (149,143.04) ;
\draw  [fill={rgb, 255:red, 2; green, 100; blue, 200 }  ,fill opacity=1 ] (128.47,135.68) .. controls (128.47,134.42) and (129.49,133.4) .. (130.75,133.4) -- (137.59,133.4) .. controls (138.85,133.4) and (139.87,134.42) .. (139.87,135.68) -- (139.87,150.39) .. controls (139.87,151.65) and (138.85,152.67) .. (137.59,152.67) -- (130.75,152.67) .. controls (129.49,152.67) and (128.47,151.65) .. (128.47,150.39) -- cycle ;
\draw  [fill={rgb, 255:red, 2; green, 100; blue, 200 }  ,fill opacity=1 ][line width=0.75]  (133.3,135.63) -- (137.3,135.63) -- (137.3,139.63) ;

\draw [fill={rgb, 255:red, 144; green, 19; blue, 254 }  ,fill opacity=1 ][line width=0.75]    (134.2,123) -- (134.25,132.9) ;
\draw [line width=2.25]    (148,143.04) -- (177,143.04) ;
\draw  [fill={rgb, 255:red, 208; green, 2; blue, 27 }  ,fill opacity=1 ] (156.47,135.68) .. controls (156.47,134.42) and (157.49,133.4) .. (158.75,133.4) -- (165.59,133.4) .. controls (166.85,133.4) and (167.87,134.42) .. (167.87,135.68) -- (167.87,150.39) .. controls (167.87,151.65) and (166.85,152.67) .. (165.59,152.67) -- (158.75,152.67) .. controls (157.49,152.67) and (156.47,151.65) .. (156.47,150.39) -- cycle ;
\draw  [fill={rgb, 255:red, 208; green, 2; blue, 27 }  ,fill opacity=1 ][line width=0.75]  (161.3,135.63) -- (165.3,135.63) -- (165.3,139.63) ;

\draw [fill={rgb, 255:red, 144; green, 19; blue, 254 }  ,fill opacity=1 ][line width=0.75]    (162.2,122) -- (162.25,132.9) ;
\draw  [fill={rgb, 255:red, 241; green, 238; blue, 28 }  ,fill opacity=1 ] (105.57,142.8) -- (113.4,134.95) -- (121.25,142.79) -- (113.41,150.64) -- cycle ;
\draw  [fill={rgb, 255:red, 255; green, 255; blue, 255 }  ,fill opacity=1 ] (131.4,120.2) .. controls (131.4,118.65) and (132.65,117.4) .. (134.2,117.4) .. controls (135.75,117.4) and (137,118.65) .. (137,120.2) .. controls (137,121.75) and (135.75,123) .. (134.2,123) .. controls (132.65,123) and (131.4,121.75) .. (131.4,120.2) -- cycle ;
\draw [line width=2.25]    (221,143.54) -- (221,143.54) ;
\draw [line width=2.25]    (261,150.54) -- (290,150.54) ;
\draw [fill={rgb, 255:red, 144; green, 19; blue, 254 }  ,fill opacity=1 ][line width=0.75]    (260.1,132.8) -- (260.06,119.9) ;
\draw  [fill={rgb, 255:red, 241; green, 238; blue, 28 }  ,fill opacity=1 ] (259.57,150.3) -- (267.4,142.45) -- (275.25,150.29) -- (267.41,158.14) -- cycle ;
\draw  [fill={rgb, 255:red, 244; green, 21; blue, 207 }  ,fill opacity=1 ] (263.06,135.57) .. controls (263.08,137.11) and (261.84,138.38) .. (260.29,138.4) .. controls (258.75,138.42) and (257.48,137.19) .. (257.46,135.64) .. controls (257.44,134.09) and (258.67,132.82) .. (260.22,132.8) .. controls (261.77,132.78) and (263.04,134.02) .. (263.06,135.57) -- cycle ;

\draw (186.5,134.66) node [anchor=north west][inner sep=0.75pt]    {$=\sum\limits_{\lambda }\frac{c_{\lambda }}{d_{\lambda }}$};
\draw (267.06,125.3) node [anchor=north west][inner sep=0.75pt]  [font=\footnotesize]  {$\lambda $};
\end{tikzpicture}.
\end{equation}
Here, the colored bullet represents the charge $o_\lambda$, with a coefficient $c_\lambda>0$ and $\sum_\lambda c_\lambda = 1$. We introduce an additional tensor equation to proceed with tensor contractions:
\begin{equation}
\tikzset{every picture/.style={line width=0.75pt}} 
\begin{tikzpicture}[x=0.75pt,y=0.75pt,yscale=-1,xscale=1]

\draw [fill={rgb, 255:red, 144; green, 19; blue, 254 }  ,fill opacity=1 ][line width=0.75]    (195.83,90.51) -- (234.08,90.49) ;
\draw [line width=1.5]    (212.36,70.16) -- (212.26,112.05) ;
\draw  [line width=1.5]  (205.62,70.95) .. controls (205.62,70.92) and (205.62,70.88) .. (205.62,70.85) .. controls (205.6,66.55) and (208.64,63.04) .. (212.42,63.03) .. controls (216.2,63.01) and (219.28,66.49) .. (219.3,70.79) .. controls (219.3,70.83) and (219.3,70.86) .. (219.3,70.9) -- cycle ;
\draw  [color={rgb, 255:red, 0; green, 0; blue, 0 }  ,draw opacity=1 ][fill={rgb, 255:red, 38; green, 173; blue, 95 }  ,fill opacity=1 ][line width=1]  (204.79,89.69) .. controls (204.79,85.71) and (208.22,82.49) .. (212.45,82.49) .. controls (216.68,82.49) and (220.11,85.71) .. (220.11,89.69) .. controls (220.11,93.67) and (216.68,96.89) .. (212.45,96.89) .. controls (208.22,96.89) and (204.79,93.67) .. (204.79,89.69) -- cycle ;
\draw  [fill={rgb, 255:red, 244; green, 21; blue, 207 }  ,fill opacity=1 ] (194.72,87.44) .. controls (196.26,87.44) and (197.51,88.7) .. (197.51,90.25) .. controls (197.5,91.79) and (196.24,93.04) .. (194.7,93.04) .. controls (193.15,93.03) and (191.9,91.78) .. (191.91,90.23) .. controls (191.91,88.68) and (193.17,87.43) .. (194.72,87.44) -- cycle ;
\draw [line width=1.5]    (302.69,95.49) -- (302.6,113.38) ;
\draw  [line width=1.5]  (295.95,96.28) .. controls (295.95,96.25) and (295.95,96.22) .. (295.95,96.18) .. controls (295.93,91.88) and (298.98,88.38) .. (302.76,88.36) .. controls (306.53,88.35) and (309.61,91.82) .. (309.63,96.13) .. controls (309.63,96.16) and (309.63,96.2) .. (309.63,96.23) -- cycle ;
\draw [fill={rgb, 255:red, 144; green, 19; blue, 254 }  ,fill opacity=1 ][line width=0.75]    (287.5,79.51) -- (316.74,79.49) ;
\draw  [fill={rgb, 255:red, 244; green, 21; blue, 207 }  ,fill opacity=1 ] (287.38,76.44) .. controls (288.93,76.44) and (290.18,77.7) .. (290.17,79.25) .. controls (290.17,80.79) and (288.91,82.04) .. (287.37,82.04) .. controls (285.82,82.03) and (284.57,80.78) .. (284.57,79.23) .. controls (284.58,77.68) and (285.84,76.43) .. (287.38,76.44) -- cycle ;

\draw (250.33,89.4) node [anchor=north west][inner sep=0.75pt]    {$=$};

\end{tikzpicture},
\end{equation}
which holds because $[o_\lambda,u(g)]=0$ for any $u(g)\in N$.
Then, we contract the IM from top to bottom:
\begin{equation}
\tikzset{every picture/.style={line width=0.75pt}} 
\begin{tikzpicture}[x=0.75pt,y=0.75pt,yscale=-1,xscale=1]

\draw [line width=2.25]    (47.61,53.67) -- (47.42,138.67) ;
\draw [fill={rgb, 255:red, 144; green, 19; blue, 254 }  ,fill opacity=1 ][line width=0.75]    (54.5,94.18) -- (97.74,94.16) ;
\draw [line width=1.5]    (76.03,73.82) -- (75.93,138.71) ;
\draw  [line width=1.5]  (69.28,74.61) .. controls (69.28,74.58) and (69.28,74.55) .. (69.28,74.52) .. controls (69.26,70.21) and (72.31,66.71) .. (76.09,66.7) .. controls (79.87,66.68) and (82.95,70.16) .. (82.97,74.46) .. controls (82.97,74.49) and (82.97,74.53) .. (82.97,74.56) -- cycle ;
\draw  [color={rgb, 255:red, 0; green, 0; blue, 0 }  ,draw opacity=1 ][fill={rgb, 255:red, 38; green, 173; blue, 95 }  ,fill opacity=1 ][line width=1]  (68.46,93.36) .. controls (68.46,89.38) and (71.89,86.16) .. (76.12,86.16) .. controls (80.35,86.16) and (83.78,89.38) .. (83.78,93.36) .. controls (83.78,97.33) and (80.35,100.56) .. (76.12,100.56) .. controls (71.89,100.56) and (68.46,97.33) .. (68.46,93.36) -- cycle ;
\draw [fill={rgb, 255:red, 144; green, 19; blue, 254 }  ,fill opacity=1 ][line width=0.75]    (54.17,119.98) -- (98.41,119.96) ;
\draw  [fill={rgb, 255:red, 229; green, 126; blue, 33 }  ,fill opacity=1 ][line width=1]  (75.79,110.23) -- (83.78,119.97) -- (75.79,129.71) -- (67.8,119.97) -- cycle ;
\draw  [fill={rgb, 255:red, 208; green, 2; blue, 27 }  ,fill opacity=1 ] (55.68,88.52) .. controls (56.94,88.53) and (57.96,89.56) .. (57.95,90.82) -- (57.9,97.66) .. controls (57.89,98.92) and (56.86,99.93) .. (55.6,99.92) -- (40.89,99.81) .. controls (39.63,99.8) and (38.62,98.77) .. (38.63,97.51) -- (38.68,90.68) .. controls (38.69,89.42) and (39.72,88.4) .. (40.98,88.41) -- cycle ;
\draw  [line width=0.75]  (55.7,93.36) -- (55.67,97.36) -- (51.67,97.33) ;

\draw  [fill={rgb, 255:red, 2; green, 100; blue, 200 }  ,fill opacity=1 ] (55.48,114.02) .. controls (56.74,114.03) and (57.76,115.06) .. (57.75,116.32) -- (57.7,123.16) .. controls (57.69,124.42) and (56.66,125.43) .. (55.4,125.42) -- (40.69,125.31) .. controls (39.44,125.3) and (38.42,124.28) .. (38.43,123.02) -- (38.48,116.18) .. controls (38.49,114.92) and (39.52,113.9) .. (40.78,113.91) -- cycle ;
\draw  [fill={rgb, 255:red, 2; green, 100; blue, 200 }  ,fill opacity=1 ][line width=0.75]  (55.5,118.86) -- (55.47,122.85) -- (51.47,122.83) ;

\draw  [fill={rgb, 255:red, 241; green, 238; blue, 28 }  ,fill opacity=1 ] (39.77,48.83) -- (47.61,40.98) -- (55.45,48.82) -- (47.61,56.67) -- cycle ;
\draw [fill={rgb, 255:red, 144; green, 19; blue, 254 }  ,fill opacity=1 ][line width=0.75]    (61.57,68.02) -- (52.67,67.88) ;
\draw  [fill={rgb, 255:red, 255; green, 255; blue, 255 }  ,fill opacity=1 ] (63.38,65.11) .. controls (64.93,65.11) and (66.18,66.37) .. (66.17,67.91) .. controls (66.17,69.46) and (64.91,70.71) .. (63.37,70.71) .. controls (61.82,70.7) and (60.57,69.44) .. (60.57,67.9) .. controls (60.58,66.35) and (61.84,65.1) .. (63.38,65.11) -- cycle ;
\draw  [fill={rgb, 255:red, 2; green, 100; blue, 200 }  ,fill opacity=1 ] (55.39,62.67) .. controls (56.65,62.67) and (57.67,63.69) .. (57.67,64.95) -- (57.67,71.79) .. controls (57.67,73.05) and (56.65,74.07) .. (55.39,74.07) -- (40.68,74.07) .. controls (39.42,74.07) and (38.4,73.05) .. (38.4,71.79) -- (38.4,64.95) .. controls (38.4,63.69) and (39.42,62.67) .. (40.68,62.67) -- cycle ;
\draw  [fill={rgb, 255:red, 2; green, 100; blue, 200 }  ,fill opacity=1 ][line width=0.75]  (55.44,67.5) -- (55.44,71.5) -- (51.44,71.5) ;

\draw [line width=2.25]    (183.95,103) -- (183.75,135) ;
\draw [fill={rgb, 255:red, 144; green, 19; blue, 254 }  ,fill opacity=1 ][line width=0.75]    (195.83,90.51) -- (234.08,90.49) ;
\draw [line width=1.5]    (212.36,70.16) -- (212.26,135.05) ;
\draw  [line width=1.5]  (205.62,70.95) .. controls (205.62,70.92) and (205.62,70.88) .. (205.62,70.85) .. controls (205.6,66.55) and (208.64,63.04) .. (212.42,63.03) .. controls (216.2,63.01) and (219.28,66.49) .. (219.3,70.79) .. controls (219.3,70.83) and (219.3,70.86) .. (219.3,70.9) -- cycle ;
\draw  [color={rgb, 255:red, 0; green, 0; blue, 0 }  ,draw opacity=1 ][fill={rgb, 255:red, 38; green, 173; blue, 95 }  ,fill opacity=1 ][line width=1]  (204.79,89.69) .. controls (204.79,85.71) and (208.22,82.49) .. (212.45,82.49) .. controls (216.68,82.49) and (220.11,85.71) .. (220.11,89.69) .. controls (220.11,93.67) and (216.68,96.89) .. (212.45,96.89) .. controls (208.22,96.89) and (204.79,93.67) .. (204.79,89.69) -- cycle ;
\draw [fill={rgb, 255:red, 144; green, 19; blue, 254 }  ,fill opacity=1 ][line width=0.75]    (190.5,116.31) -- (234.75,116.3) ;
\draw  [fill={rgb, 255:red, 229; green, 126; blue, 33 }  ,fill opacity=1 ][line width=1]  (212.12,106.56) -- (220.11,116.3) -- (212.12,126.05) -- (204.13,116.3) -- cycle ;
\draw  [fill={rgb, 255:red, 2; green, 100; blue, 200 }  ,fill opacity=1 ] (191.82,110.35) .. controls (193.08,110.36) and (194.09,111.39) .. (194.08,112.65) -- (194.03,119.49) .. controls (194.02,120.75) and (192.99,121.76) .. (191.73,121.75) -- (177.03,121.65) .. controls (175.77,121.64) and (174.76,120.61) .. (174.76,119.35) -- (174.81,112.51) .. controls (174.82,111.25) and (175.85,110.24) .. (177.11,110.25) -- cycle ;
\draw  [fill={rgb, 255:red, 2; green, 100; blue, 200 }  ,fill opacity=1 ][line width=0.75]  (191.83,115.19) -- (191.8,119.19) -- (187.8,119.16) ;

\draw  [fill={rgb, 255:red, 241; green, 238; blue, 28 }  ,fill opacity=1 ] (176.1,98.16) -- (183.94,90.32) -- (191.78,98.16) -- (183.95,106) -- cycle ;
\draw  [fill={rgb, 255:red, 244; green, 21; blue, 207 }  ,fill opacity=1 ] (194.71,87.44) .. controls (196.25,87.44) and (197.51,88.69) .. (197.51,90.24) .. controls (197.51,91.79) and (196.25,93.04) .. (194.71,93.04) .. controls (193.16,93.04) and (191.91,91.79) .. (191.91,90.24) .. controls (191.91,88.69) and (193.16,87.44) .. (194.71,87.44) -- cycle ;
\draw [line width=2.25]    (319.28,94) -- (319.09,129) ;
\draw [line width=1.5]    (352.69,117.16) -- (352.6,129.05) ;
\draw  [line width=1.5]  (345.95,117.95) .. controls (345.95,117.92) and (345.95,117.88) .. (345.95,117.85) .. controls (345.93,113.55) and (348.98,110.04) .. (352.76,110.03) .. controls (356.53,110.01) and (359.61,113.49) .. (359.63,117.79) .. controls (359.63,117.83) and (359.63,117.86) .. (359.63,117.9) -- cycle ;
\draw [fill={rgb, 255:red, 144; green, 19; blue, 254 }  ,fill opacity=1 ][line width=0.75]    (325.83,107.31) -- (335.08,107.3) ;
\draw  [fill={rgb, 255:red, 2; green, 100; blue, 200 }  ,fill opacity=1 ] (327.15,101.35) .. controls (328.41,101.36) and (329.42,102.39) .. (329.41,103.65) -- (329.36,110.49) .. controls (329.35,111.75) and (328.33,112.76) .. (327.07,112.75) -- (312.36,112.65) .. controls (311.1,112.64) and (310.09,111.61) .. (310.1,110.35) -- (310.15,103.51) .. controls (310.16,102.25) and (311.19,101.24) .. (312.44,101.25) -- cycle ;
\draw  [fill={rgb, 255:red, 2; green, 100; blue, 200 }  ,fill opacity=1 ][line width=0.75]  (327.17,106.19) -- (327.14,110.19) -- (323.14,110.16) ;

\draw  [fill={rgb, 255:red, 241; green, 238; blue, 28 }  ,fill opacity=1 ] (311.43,89.16) -- (319.27,81.32) -- (327.12,89.16) -- (319.28,97) -- cycle ;
\draw  [fill={rgb, 255:red, 255; green, 255; blue, 255 }  ,fill opacity=1 ] (337.89,104.5) .. controls (339.44,104.51) and (340.69,105.77) .. (340.68,107.31) .. controls (340.68,108.86) and (339.42,110.11) .. (337.87,110.1) .. controls (336.33,110.1) and (335.08,108.84) .. (335.08,107.3) .. controls (335.09,105.75) and (336.34,104.5) .. (337.89,104.5) -- cycle ;
\draw [fill={rgb, 255:red, 144; green, 19; blue, 254 }  ,fill opacity=1 ][line width=0.75]    (347.5,80.18) -- (363.74,80.16) ;
\draw  [fill={rgb, 255:red, 244; green, 21; blue, 207 }  ,fill opacity=1 ] (347.38,77.11) .. controls (348.93,77.11) and (350.18,78.37) .. (350.17,79.91) .. controls (350.17,81.46) and (348.91,82.71) .. (347.37,82.71) .. controls (345.82,82.7) and (344.57,81.44) .. (344.57,79.9) .. controls (344.58,78.35) and (345.84,77.1) .. (347.38,77.11) -- cycle ;
\draw [fill={rgb, 255:red, 144; green, 19; blue, 254 }  ,fill opacity=1 ][line width=0.75]    (347.5,99.51) -- (363.74,99.49) ;
\draw  [fill={rgb, 255:red, 255; green, 255; blue, 255 }  ,fill opacity=1 ] (347.38,96.44) .. controls (348.93,96.44) and (350.18,97.7) .. (350.17,99.25) .. controls (350.17,100.79) and (348.91,102.04) .. (347.37,102.04) .. controls (345.82,102.03) and (344.57,100.78) .. (344.57,99.23) .. controls (344.58,97.68) and (345.84,96.43) .. (347.38,96.44) -- cycle ;

\draw (108.5,89.66) node [anchor=north west][inner sep=0.75pt]    {$=\sum\limits_{\lambda }\frac{c_{\lambda }}{d_{\lambda }}$};
\draw (244.5,89.66) node [anchor=north west][inner sep=0.75pt]    {$=\sum\limits_{\lambda }\frac{c_{\lambda }}{d_{\lambda }}$};
\draw (186.5,70.66) node [anchor=north west][inner sep=0.75pt]  [font=\footnotesize]  {$\lambda $};
\draw (338.5,61.66) node [anchor=north west][inner sep=0.75pt]  [font=\footnotesize]  {$\lambda $};
\end{tikzpicture}.
\end{equation}
The resulting IM represents a perfectly Markovian bath of a grand canonical ensemble that brings the system qudit to a mixed state diagonal in solitons: $\rho_\text{re} = \sum_\lambda c_\lambda\Pi_\lambda/d_\lambda$.

\section{NEGATIVITY CALCULATIONS}\label{app:negativity}

In this appendix, we show that, for the MPS initial state defined in Eq.~(\ref{eq:MPS_teleportable_entanglement}), the negativity from Eq.~\eqref{eq:negativity} vanishes for $q=2$, while remaining nonzero for typical cases with $q\geq 3$, and we provide an explicit example with nonzero negativity for $q=3$.

Let us start with $q=2$ and write down Eq.~\eqref{eq:rho_explicit_negativity} explicitly:
\begin{align}
   \rho^{{AB}}_\text{eff} = &\frac{1}{4}\Big[ (\ket{0}\bra{0})_{A}\otimes \mathbbm{1}_{B} + (\ket{1}\bra{1})_{A}\otimes \mathbbm{1}_{B}\nonumber\\
   &+(\ket{0}\bra{1})_{A}\otimes (w^0(w^1)^\dagger)_{B} + (\ket{1}\bra{0})_{A}\otimes (w^1(w^0)^\dagger)_{B}\Big].
\end{align}

The partially transposed density matrix is related by a unitary transformation:
\begin{equation}
    (\rho^{{AB}}_\text{eff})^{\mathrm{T}_{{A}}} = \sigma^x_{A}~\rho^{{AB}}_\text{eff}~\sigma^x_{A},
\end{equation}
which is, therefore, a physical density matrix, resulting in exactly zero negativity of $\rho^{{AB}}_\text{eff}$.

\begin{figure}[h]
\hspace{-0.25\textwidth}
\includegraphics[width=0.5\linewidth]{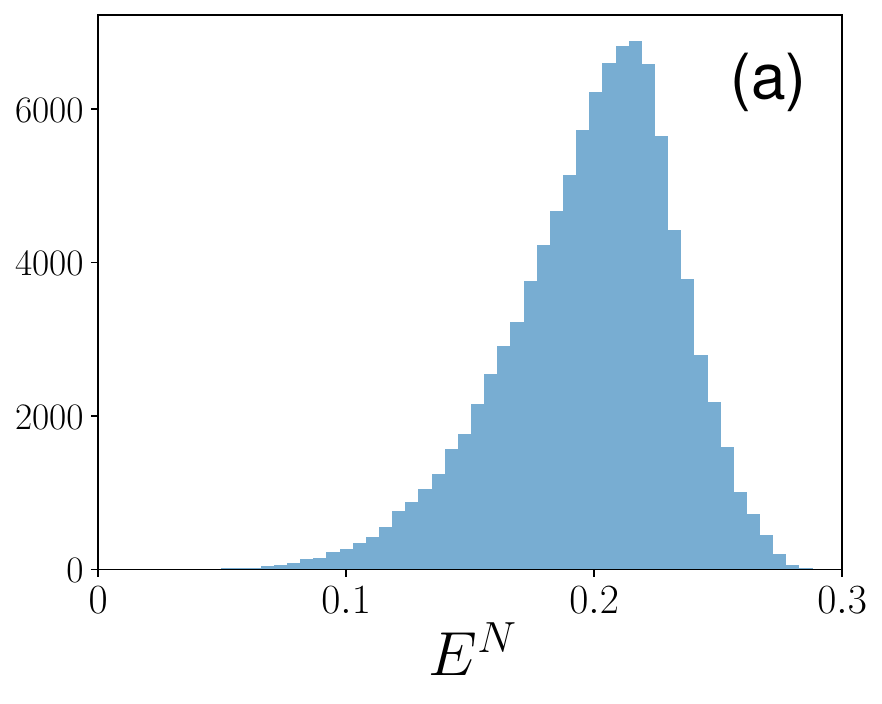}
\includegraphics[width=0.5\linewidth]{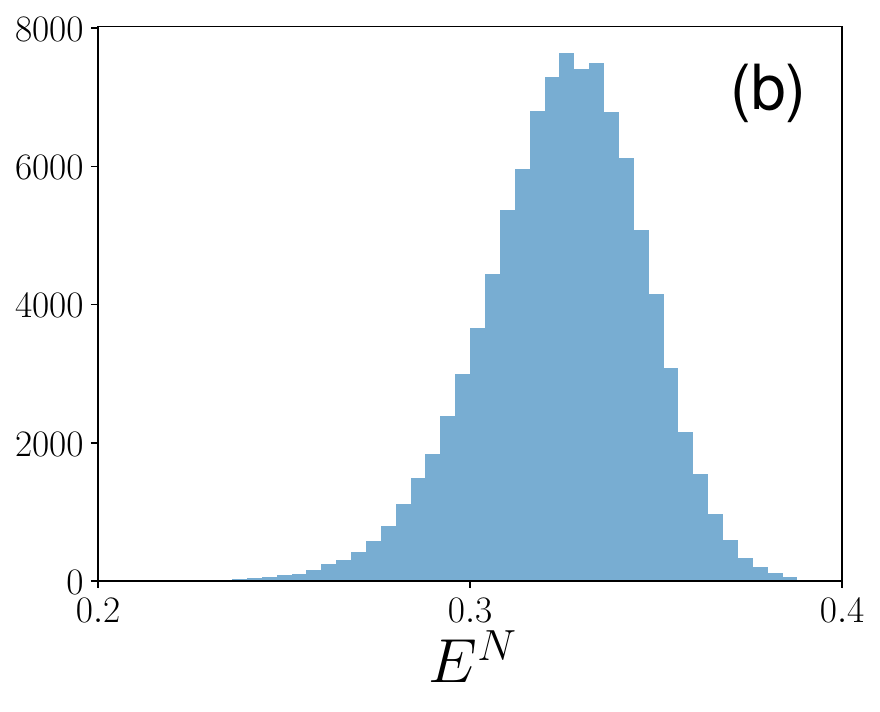} 
\caption{Histograms of negativities for effective two-qudit density matrix from Eq. \eqref{eq:rho_explicit_negativity}, with unitaries $\{w^a\}_{a=0}^{q-1}$ drawn from the Haar ensemble. 
The number of samples is $10^5$. (a) $q=3$. (b) $q=4$.
}
\label{fig:Hist_Negativity}
\end{figure}

Next, let us turn to the case of $q\geq 3$. The unitaries $\{w^a\}_{a=0}^{q-1}$ acting on the auxiliary Hilbert space are generated randomly from the Haar ensemble on $\mathrm{U}(q)$. In Fig.~\ref{fig:Hist_Negativity}, we display the histograms of negativities for $q=3$ and $q=4$. Both show distributions centered at positive values, implying the presence of genuinely quantum correlations in these cases.

\bibliography{Heran_circuit}
\end{document}